\documentclass[letterpaper,12pt]{article}
\pdfoutput=1
\usepackage{jheppub}
\usepackage{amsmath}
\usepackage{graphicx}
\usepackage{caption,subcaption}
\usepackage{cancel}
\usepackage[dvipsnames]{xcolor}
\usepackage{color}
\usepackage{mathrsfs}
\usepackage{amssymb}
\usepackage{epstopdf}
\usepackage{dsfont}
\usepackage{float}
\usepackage{amsthm}

\graphicspath{{./Figures/}}

\newenvironment{emphblock}%
  {\list{}{\leftmargin=0.22in\rightmargin=0.3in}\item[]}%
  {\endlist}

\newcommand*{\hilb}{\mathcal{H}}	
\newcommand*{\hbu}{\mathcal{H}_\text{BU}} 

\DeclareMathOperator{\Tr}{Tr}

\DeclareMathOperator{\Sym}{Sym}

\newcommand*{\id}{\mathds{1}}

\newcommand*{\scri}{\mathscr{I}} 
\newcommand*{\evap}{\mathscr{E}} 
\newcommand*{\swap}{\mathcal{S}} 
\newcommand*{\island}{\mathcal{I}}

\newcommand{\HH}{\mathrm{HH}} 

\addtocontents{toc}{\protect\enlargethispage*{3\baselineskip}} 

\newpage

\title{Observations of Hawking radiation: the Page curve and baby universes}

\author{Donald Marolf}
\author{and Henry Maxfield}

\affiliation{Department of Physics, University of California, Santa Barbara, CA 93106, USA}

\emailAdd{marolf@physics.ucsb.edu}

\emailAdd{hmaxfield@physics.ucsb.edu}

\abstract{
We reformulate recent insights into black hole information in a manner emphasizing operationally-defined notions of entropy, Lorentz-signature descriptions, and asymptotically flat spacetimes. With the help of replica wormholes, we find that experiments of asymptotic observers are consistent with black holes as unitary quantum systems, with density of states given by the Bekenstein-Hawking formula. However, this comes at the cost of superselection sectors associated with the state of baby universes. Spacetimes studied by Polchinski and Strominger in 1994 provide a simple illustration of the associated concepts and techniques, and we argue them to be a natural late-time extrapolation of replica wormholes. The work aims to be self-contained and, in particular, to be accessible to readers who have not yet mastered earlier formulations of the ideas above.}

\begin{document}
\maketitle

\section{Introduction}

Recent work has reinvigorated the idea (see e.g. \cite{Kuchar1981,Caderni:1984pw,Moss:1986yd,
Hawking:1987mz,Giddings:1987cg,Lavrelashvili:1987jg,McGuigan:1988vi,Banks:1988je,Rubakov:1988jf,Hawking:1988ae,Coleman:1988cy,Giddings:1988cx,Giddings:1988wv,Polchinski:1994zs,Louko:1995jw,Maldacena:2001kr,Hawking:2005kf})
that sums over topologies in the gravitational path integral provide missing ingredients necessary to understand black hole information and other issues in gravity and holography. In particular, \cite{Penington:2019kki,Almheiri:2019qdq} built on  \cite{Penington:2019npb,Almheiri:2019psf} to argue that an exchange of dominance between two saddle-point `replica wormhole' geometries resolves a longstanding tension between the perturbative description of black hole evaporation and the interpretation of the black hole's Bekenstein-Hawking entropy as a density of states. Such effects have also been connected \cite{Marolf:2020xie} with the so-called baby universes and with the superselection sectors (`$\alpha$-states') for quantities associated with asymptotic boundaries described in \cite{Coleman:1988cy,Giddings:1988cx,Giddings:1988wv}. See \cite{Almheiri:2020cfm} for a review and references to additional related work.

The bulk of the recent discussions have been couched in terms of Euclidean path integrals. Indeed, even \cite{Almheiri:2019qdq,Hartman:2020swn} which discussed the effect of replica wormholes in Lorentz signature did so by studying Euclidean signature replica wormholes, using them to compute entropies as functions of Euclidean coordinates, and analytically continuing the results to real times.  But it is clearly of interest to understand an \emph{intrinsically} Lorentz-signature description, especially since topology change is generally incompatible with having a smooth Lorentz-signature metric.

In addition, the recent discussions also rely heavily on AdS/CFT duality or related concepts. This was true even for the asymptotically flat analyses of  \cite{Anegawa:2020ezn,Hashimoto:2020cas,Gautason:2020tmk,Krishnan:2020oun} in which arguments were made by analogy with AdS/CFT. But reliance on AdS/CFT presents difficulties as the physics of spacetime wormholes raises the so-called `factorization problem' that calls into question the standard interpretation of AdS/CFT.  As a result, questions have been raised \cite{Giddings:2020yes} as to what physics is really being studied.

Our goal here is to reformulate the recent progress in a manner that i) focusses on operationally defined quantities (the outcomes of `experiments' performed by asymptotic observers),
ii) can be stated and analyzed entirely in Lorentz signature, and iii) emphasizes that the physics described follows directly from having a low energy gravitational path integral that sums over topologies.   While we take the inclusion of this sum over topologies as a fundamental assumption in this work, there will be no explicit input from string theory, holography (AdS/CFT), or any other UV theory of gravity.  To underline the last point, we will work entirely with asymptotically flat spacetimes (though analogous statements apply directly to the asymptotically AdS case as well).  As a result, AdS/CFT is mentioned only briefly in tangential comments.

Nevertheless, the interpretation of the black hole's Bekenstein-Hawking entropy ($S_\mathrm{BH} = \frac{A}{4G} +\, \text{corrections}$) as a density of states will be a common touchpoint throughout our discussion. We do not take this to be a fundamental assumption, but rather a hypothesis to be constantly tested and explored.  Indeed, while to some this interpretation will seem natural due to the success of classical black hole thermodynamics --- or perhaps even required by this success, see e.g.\ \cite{Jacobson:2005kr} ---  it also flies in the face of physics associated with quantum field theory on an evaporating black hole background and perturbative quantum gravity  (see e.g.\ \cite{Hawking:1976ra,Callan:1992rs,Ashtekar:2005cj,Unruh:2017uaw,Rovelli:2017mzl}).

In particular, perturbative quantum gravity would suggest that Hawking radiation is essentially thermal, which is in tension with the statistical interpretation that $S_\mathrm{BH}$ counts black hole states.
Under the standard laws of quantum mechanics, the density of states is an upper bound on the entanglement of any system. Since Hawking evaporation causes $S_\mathrm{BH}$ to decrease over time, the above interpretation thus would appear to force the von Neumann entropy of Hawking radiation to become small in later stages of the evaporation.  As described by Page \cite{Page:1993df}, it would then be natural to expect the von Neumann entropy of radiation from a black hole that forms from rapid collapse to begin at a small value, increase while thermal radiation is produced, but then to `turn over' and decrease once it comes close to saturating this bound, requiring deviations from exact thermality.

\begin{figure}
	\centering
	\includegraphics[width=.6\textwidth]{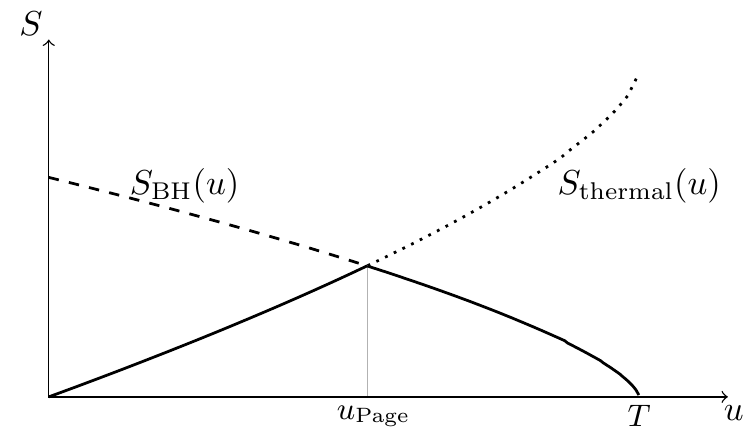}
	\caption{The Page curve for the entropy of Hawking radiation emitted before time $u$ (solid curve). For a while this is increasing, given by the thermal entropy (dotted curve). But under BH unitarity it is bounded by the Bekenstein Hawking entropy $S_{BH}\sim \frac{A}{4G}$ (dashed curve). Consequently, after the Page time $u_\mathrm{Page}$ the entropy must decrease, approximately saturating that bound. \label{fig:Page}}
\end{figure}

The resulting `Page curve' is shown in figure \ref{fig:Page}.  It will feature many times in our discussion below, again as a touchpoint to be compared with various calculations. In particular, the downward sloping part of the Page curve requires information inside the black hole to be returned to the external universe.  The literature on black hole information often describes this as a result of requiring `unitarity'.   But, as noted above, there are rather more assumptions involved than just strict unitary evolution of the full quantum gravity system.  In the present work we will thus instead use the term `Bekenstein-Hawking unitarity' (or BH unitarity) to refer to this suite of ideas, which we summarize as follows:\footnote{The same concept was called `the central dogma' by \cite{Almheiri:2020cfm} in analogy with the term's use in biology. We do not follow this terminology here, so that we might avoid appearing dogmatic.}
\begin{emphblock}
	\textbf{Bekenstein-Hawking unitarity:} in order to describe measurements of distant observers, black holes can be modelled as a quantum system with density of states $e^{S_\mathrm{BH}}$ whose evolution is unitary (up to possible interactions with other quantum systems).
\end{emphblock}
We emphasize that our definition of BH unitarity is operational, referring to observations. In contrast, as we discuss further below, the von Neumann entropy of Hawking radiation is \emph{not} a directly observable quantity; rather, it can only be inferred indirectly from other measurements. Looking ahead, this will be important for our conclusions since it allows BH unitarity to be satisfied despite the fact the the von Neumann entropy may not, strictly speaking, follow the Page curve in figure \ref{fig:Page}.   This discussion has strong overlap with those of \cite{Coleman:1988cy,Giddings:1988cx,Giddings:1988wv,Polchinski:1994zs}.

We will see below that many of the concepts and techniques related to Lorentz-signature spacetime wormholes, baby universes, and the like are well-illustrated by spacetimes described by Polchinski and Strominger in 1994 \cite{Polchinski:1994zs}, which we dub `PS wormholes'.  Indeed, while PS wormholes are not under semi-classical control, and while analyzing them in isolation leads to apparent violations of BH unitary \cite{Polchinski:1994zs}, we will argue them to be a natural late-time extrapolation of the replica wormholes that were shown in \cite{Penington:2019kki,Almheiri:2019qdq} to reproduce the Page curve.  Since this extrapolation turns out to lead to several simplifications, we will devote significant time to discussing PS wormholes in effort to make our treatment as explicit as possible.

Indeed, a final goal of this work is to make the manuscript below accessible to those who have not yet mastered the above references.  Rather than review those works in detail, we instead return to the logical beginning and start in section \ref{sec:HawkingReview} with a brief review of the Hawking effect in a fixed black hole background, but emphasizing both the path integral approach and the in-in formalism that will be useful in later parts of this work.  While none of this material is new, it differs sufficiently from the most common treatments in the literature.  We then use this perspective to discuss the inclusion of semiclassical quantum gravity and perturbative back-reaction in section \ref{sec:semiclassical}.  This sets up the standard challenge for BH unitarity associated with apparent large deviations from the the Page curve, and which is often called `the black hole information problem' \cite{Mathur:2009hf,Harlow:2014yka,Unruh:2017uaw,Marolf:2017jkr}.

The following sections resolve this problem by identifying new saddles for the gravitational path integral.  Some possible effects of new saddles, and especially on \emph{measurements} of entropy by asymptotic observers, are illustrated in section \ref{sec:PS} through the study of PS wormholes.   Although the inclusion of PS wormholes requires assumptions about physics beyond semiclassical control, it provides a simple introduction to ideas that will be of use later in this work.  A key such point is that spacetime wormholes lead to correlations between the outcomes of what might at first appear to be completely independent experiments.  We also discuss challenges for BH unitary raised by PS wormholes alone, setting the stage to introduce and include replica wormholes in section \ref{sec:replicas}.  Doing so resolves the PS challenges and reproduces the Page curve using calculations that are fully under semiclassical control.  We will also see that PS wormholes are a natural late-time extrapolation of replica wormholes.

It then remains to provide a Hilbert space description of the physics of the Page curve, and to characterise the correlations arising from replica wormholes.  This is done in section \ref{sec:BUetc} by slicing open the above path integrals.  We find a `baby universe' Hilbert space of intermediate states which defines superselection sectors associated with the values of asymptotic quantities.  As a result, it leads to an ensemble description of the theory from the viewpoint of asymptotic observers.  Again, the PS wormholes provide a simple illustration.  Section \ref{sec:Disc} concludes with a summary and discussion of open issues.

\section{Hawking radiation and the path integral}\label{sec:HawkingReview}

This section contains a schematic overview of Hawking's original calculation \cite{Hawking:1974sw} of the production radiation using linear quantum fields in a fixed classical spacetime, without back-reaction or evaporation. We also recall how this calculation can be reformulated in terms of a path integral, and how the path integral can be used to compute the R\'enyi entropies of the Hawking radiation. This review lays the groundwork for the semiclassical quantum gravity discussions in section \ref{sec:semiclassical}. In keeping with the general philosophy of this paper, we will emphasise the computation of observables accessible to an asymptotic observer. Readers seeking a more thorough review of the Hawking effect in a fixed background should consult appendix \ref{sec:MoreRev}, the original work \cite{Hawking:1974sw}, or pedagogical introductions such as \cite{Jacobson:2003vx} or \cite{Harlow:2014yka}.

\subsection{Hawking's Heisenberg picture calculation}
 \label{sec:Heisenberg}

\begin{figure}
\centering
\begin{subfigure}[t]{.5\textwidth}
	\includegraphics[width=\textwidth]{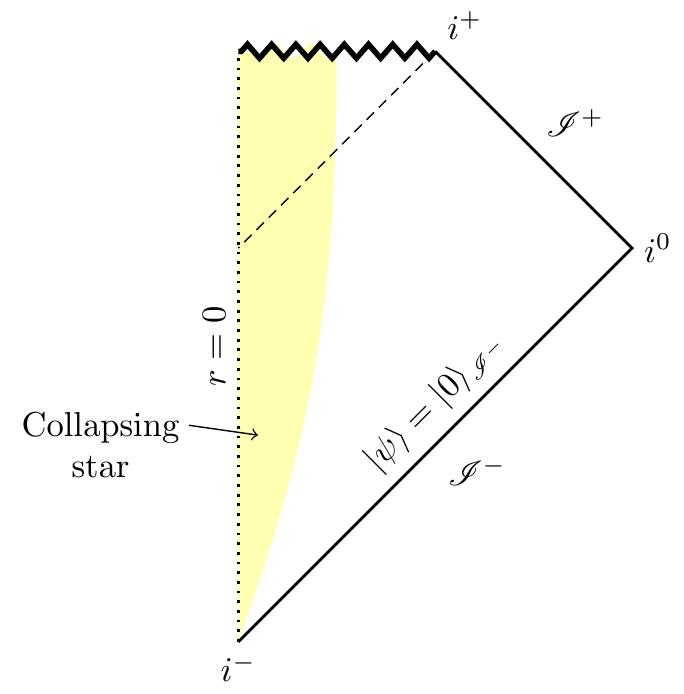}
	\caption{The conformal diagram of a classical spacetime describing collapse of matter (a `star') to form an asymptotically flat black hole with a single asymptotic region.   \label{fig:collapse}}
\end{subfigure}
\hfill
\begin{subfigure}[t]{.4\textwidth}\centering
	\includegraphics[width=.8\textwidth]{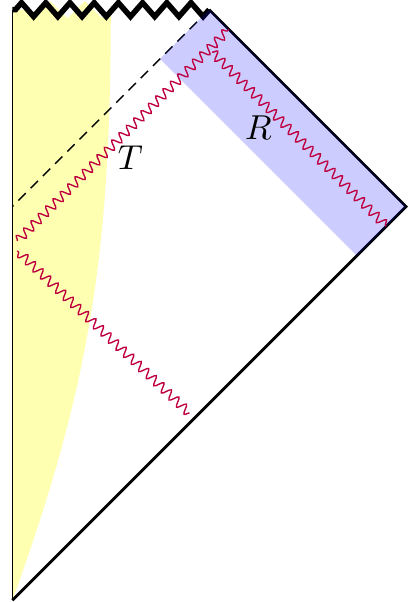}
		\caption{ An illustration of the backwards-propagation of a mode localized at late retarded-time on $\scri^+$.\label{fig:propagate}}
\end{subfigure}
\caption{The final event horizon $\mathcal{H}^+$ is shown as a dashed line, and the singularity as a jagged line.  Future and past null infinity are labelled by $\scri^\pm$. The vertical line marked $r=0$ is a regular origin of spherical polar coordinates. The state is chosen as the vacuum of quantum fields in the flat asymptotic region $\scri^-$, thus $|\psi\rangle  = |0\rangle_{\scri^-}$. In the shaded region of (b) near $\scri^+$, our spacetime is nearly stationary.   There, the backwards-propagation reduces to scattering in a fixed potential and results in a transmitted part $T$ and a reflected part $R$.  For modes localized at late (retarded) times on $\scri^+$, the reflected part $R$ will remain in the nearly stationary region and transmitted part $T$ will be localized very close to $\mathcal{H}^+$.  In particular, the wavelength of $T$ becomes very short in the reference frame shown.   This allows us to complete the backwards-propagation of $T$ from the near-horizon region to $\scri^-$ using geometric optics.}
\end{figure}

The original argument of \cite{Hawking:1974sw} considered a black hole with a single asymptotic region that forms from collapse of matter in an asymptotically flat space. For simplicity we consider a spherically symmetric collapse of uncharged matter so that the final black hole is Schwarzschild.   A conformal diagram for such a spacetime is shown in figure \ref{fig:collapse} below.  And for further simplicity, we follow \cite{Hawking:1974sw} in taking the quantum fields to be massless so that their initial data is specified at past null infinity $\scri^-$.  There the spacetime is completely flat, and the state $|\psi\rangle$ of the quantum fields is taken to coincide with the Minkowski vacuum on $\scri^-$.

We are interested in the predictions of observations made at future null infinity $\scri^+$. In particular, we would like to compute the expectation values $\langle\psi|\mathcal{O}(\scri^+)|\psi\rangle$ of operators $\mathcal{O}(\scri^+)$ defined at $\scri^+$. Following \cite{Hawking:1974sw}, we work in the Heisenberg picture. We thus evolve the operators $\mathcal{O}(\scri^+)$ backwards in time to write them in terms of operators at $\scri^-$.
Since the Hilbert space at $\scri^+$ can be described as a Fock space of  `out' scattering states, we can build all operators at $\scri^+$ from creation and annihilation operators $a_m^\dagger(\scri^+)$, $a_m(\scri^+)$, labelled by some complete orthonormal set of modes indexed by $m$. Using the Heisenberg evolution back to $\scri^-$, we can write $a_m(\scri^+)$ in terms of corresponding operators $a_n^\dagger(\scri^-)$, $a_n(\scri^-)$ acting on the Fock space of `in' states, and similarly for $a_m^\dagger(\scri^+)$. Since we took the initial state at $\scri^-$ to be the vacuum $|0\rangle_{\scri^-}$ annihilated by all $a_n(\scri^-)$, this rewriting allows us to compute all observables at $\scri^+$.

For a free quantum field theory, the relationship between creation and annihilation operators at $\scri^+$ and those at $\scri^-$ is linear. The Heisenberg evolution is thus given by a Bogoliubov transformation
\begin{equation}\label{eq:Bogoliubov}
	a_m(\scri^+) = \sum_{n} \left(\alpha_{mn} \, a_n(\scri^-) + \beta_{mn} \, a_n^\dagger(\scri^-)\right) ,
\end{equation}
for some coefficients $\alpha_{mn},\beta_{mn}$. For example, if we compute the expectation value of an occupation number $N_m(\scri^+) = a_m^\dagger(\scri^+)a_m(\scri^+)$ of a mode at $\scri^+$, we find
\begin{equation}
	\langle\psi|N_m(\scri^+)|\psi\rangle = \sum_n |\beta_{mn}|^2.
\end{equation}
Black holes radiate as a simple consequence of the fact that $\beta_{mn}$ is nonzero, so the outgoing occupation numbers are positive despite choosing an ingoing vacuum.

At least for operators associated with field modes $m$ that are localized at late retarded times (large affine parameter $u$ along $\scri^+$), it is straightforward to compute the Bogoliubov transformation \eqref{eq:Bogoliubov} using two facts.  The first is that, in the region close to $\scri^+$, the spacetime is well-approximated by that of a stationary black hole.  Mode propagation in this region thus reduces to solving a Schr\"odinger-type problem.  The second important fact is that, once the mode is propagated backward into the near-horizon region, it becomes localized very close to the horizon.  In particular, as a result of the second property we may use the WKB approximation to justify either the use of geometric optics in further propagating the mode back to $\scri^-$ \cite{Hawking:1974sw}, or the use of the adiabatic approximation to evaluate correlators without explicitly completing the backwards propagation to $\scri^-$ \cite{Unruh:1977ga,Fredenhagen:1989kr,Jacobson:1993hn,Jacobson:2003vx}.  These features are illustrated in figure \ref{fig:propagate}.  When combined, they establish the familiar result that the occupation numbers $N_m(\scri^+)$ of such late-time modes are thermally distributed, with grey-body factors appropriate to the black hole. Interactions do not change this qualitative picture. The details of this argument are not relevant to our presentation below, but we include a brief summary in appendix \ref{sec:MoreRev} for readers wishing to review them.  Readers seeking a more thorough discussion should consult the original paper  \cite{Hawking:1974sw} or reviews such as \cite{Jacobson:2003vx,Harlow:2014yka}.

In the above discussion we formulated Hawking's calculation as the computation of expectation values of all possible operators on $\scri^+$. This is equivalent to describing the state of quantum fields on $\scri^+$. Indeed, one way to define the density matrix of a region is as the linear functional that maps operators on that region to their expectation values. Connecting to the usual Hilbert space language, there is a unique $\rho$ such that this functional acts as $\mathcal{O}\mapsto \Tr(\rho\mathcal{O})$. We can recover matrix elements $\rho_{ij}$ of $\rho$ explicitly from expectation values by choosing $\mathcal{O}=|j\rangle\langle i|$, where the states $|i\rangle$, $|j\rangle$ are chosen from a complete basis of pure states on $\scri^+$.

Famously, despite choosing a pure state on $\scri^-$, the state $\rho$ on $\scri^+$ is not pure; that is, it cannot be written as $|\psi\rangle\langle\psi|$ for any $|\psi\rangle$.  This impurity arises for the simple reason that $\scri^+$ is not a Cauchy surface, as Cauchy surfaces must reach the regular origin shown as a vertical black line in figures \ref{fig:collapse}, \ref{fig:propagate}. Equivalently, while we can perform Heisenberg evolution of operators from $\scri^+$ back to $\scri^-$, we cannot do the reverse, since the operator resulting from forward evolution will have support on the black hole interior.

\subsection{Path integral version}
\label{sec:PIV}

We now recall how the computation outlined in section \ref{sec:Heisenberg} can be formulated as a path integral over quantum fields\footnote{This differs from the Hartle-Hawking derivation of Hawking radiation \cite{Hartle:1976tp}, which considered the worldline path integral over trajectories of a particle.}.  In this description, the actual computation of the effect is somewhat more cumbersome.  However, as we will see in the remaining sections below, the path integral framework allows us to straightforwardly incorporate both perturbative back-reaction and certain non-perturbative quantum gravity effects.

In our experience, most textbook treatments of path integrals work in the Schr\"odinger picture and emphasize the co-called `in-out' formulation.  In particular, the latter is naturally associated with computations of transition amplitudes.  However, since our discussion will continue to emphasize expectation values, we will instead focus on the `in-in' formulation of path integrals below.  We will also  continue to use the Heisenberg picture as in section \ref{sec:Heisenberg} above.  Both choices will simplify the discussion of various issues in the sections that follow. But the departure from standard textbook treatments suggests that we proceed slowly for the moment.  We will thus first review various general features of in-in Heisenberg-picture path integrals in section \ref{sec:PIprelim} before returning to Hawking emission in section \ref{sec:PIHawk}.

\subsubsection{Path integral preliminaries}
\label{sec:PIprelim}

Before turning to expectation values, we begin by considering the path integral between initial and final Cauchy surfaces $\Sigma_\pm$.  We use $\phi$ to denote the set of local bulk fields over which we integrate. The corresponding Heisenberg-picture operators $\hat \phi$ are defined by insertions of the field $\phi$ (or more general functionals of $\phi$) into the path integral.  We first consider a path integral with boundary conditions specifying that the fields on $\Sigma_\pm$ take definite values $\phi_{\pm}$. These boundary conditions correspond to eigenstates of the field operators on $\Sigma_\pm$ with eigenvalues $\phi_\pm$, and this path integral computes the inner product ${}_+ \langle \phi_+|\phi_-\rangle_-$:
\begin{equation}\label{eq:PIU}
	{}_+ \langle \phi_+|\phi_-\rangle_- \propto \int\displaylimits_{\left.\phi\right|_{\Sigma_\pm} = \phi_\pm} \mathcal{D}\phi \; e^{i I[\phi]}.
\end{equation}
  There is of course a choice of phases to be made in defining such eigenstates, and this choice is associated with the choice of possible boundary terms in the path integral action $I[\phi]$ (and with the fact that such boundary terms can change under canonical transformations). In addition, it can be difficult to keep track of normalisations in the path integral, so we should ultimately consider normalization-independent ratios. %

Since $|\phi_\pm\rangle_\pm$ are defined as eigenstates of different sets of field operators, on $\Sigma_+$ or on $\Sigma_-$, they give different bases for the Hilbert space. The inner products ${}_+ \langle \phi_+|\phi_-\rangle_-$ give the change of basis matrix. These may be thought of as the matrix elements of the time-evolution operator $U = P\exp\left(-i\int dt H(t)\right)$ with a time-dependent Hamiltonian $H(t)$, so we will loosely use $U$ to indicate the path integral \eqref{eq:PIU}.

\begin{figure}[h!]
\centering
	\includegraphics[width=.9\textwidth]{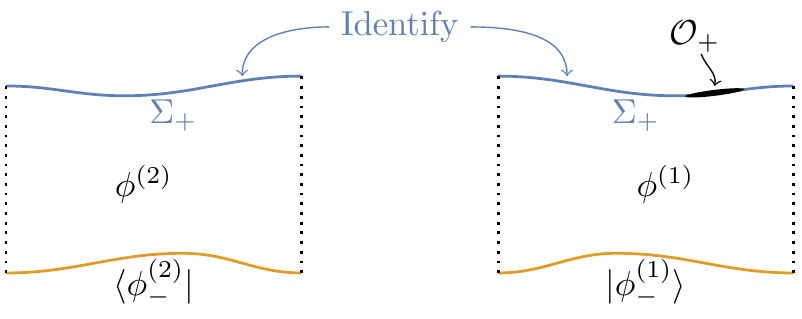}
	\caption{A path integral that computes the matrix elements $\langle\phi^{(2)}_-|\mathcal{O}_+|\phi^{(1)}_- \rangle$. The right copy of the spacetime contains fields $\phi^{(1)}$ and is weighted by $e^{iI[\phi^{(1)}]}$, while the left copy contains fields $\phi^{(2)}$ and is weighted by $e^{-iI[\phi^{(2)}]}$ (or more generally by the CPT conjugate of the action on the left copy).   This conjugation is associated with the fact that the initial conditions for the right copy (fixing the field on $\Sigma_-$) are defined by the ket-state $|\phi^{(1)}_- \rangle$ while those for the left copy are defined by the bra-state $\langle \phi^{(2)}_- |$.\label{fig:ininPI}}
\end{figure}

Given an operator $\hat{\mathcal{O}}_+$ defined in terms of fields on $\Sigma_+$, we can describe its Heisenberg evolution back to $\Sigma_-$ by computing its matrix elements ${}_-\langle \phi^{(2)}_-|\hat{\mathcal{O}}_+|\phi^{(1)}_-\rangle_-$ between the pair of field eigenstates $|\phi^{(1,2)}_-\rangle_-$ on $\Sigma_-$. To do this, we can insert a complete sets of states $|\phi_+\rangle_+$ on which $\hat{\mathcal{O}}_+$ takes definite values $\mathcal{O}_+(\phi_+)$. This leaves us to compute the two overlaps $\langle\phi_+|\phi^{(1)}_-\rangle$ and $\langle\phi^{(2)}_-|\phi_+\rangle$ before integrating over $\phi_+$. Since there are two such overlaps to compute, we have a doubled set of fields $\phi^{(1,2)}$ in the path integral, though these sets must be identified at $\Sigma_+$:
\begin{equation}
	\langle \phi^{(2)}_-|\hat{\mathcal{O}}_+|\phi^{(1)}_-\rangle = \int\displaylimits_{\substack{ {\left.\phi^{(1,2)}\right|_{\Sigma_-} = \phi^{(1,2)}_-} \\ {\left.\phi^{(1)}\right|_{\Sigma_+} = \left.\phi^{(2)}\right|_{\Sigma_+} = \phi_+}}} \mathcal{D}\phi^{(1)} \mathcal{D}\phi^{(2)} \;e^{i I[\phi^{(1)}]-i I[\phi^{(2)}]}\;\mathcal{O}_+(\phi_+)
\end{equation}
We may equivalently think of doubling not the fields on a given spacetime, but the spacetime itself. The doubled spacetime then has two branches which are glued to each other on $\Sigma_+$; see figure \ref{fig:ininPI}. This perspective becomes particularly natural once we incorporate quantum gravity effects, since the geometry can fluctuate independently on each branch of the spacetime. The first branch (which provides a home for the field $\phi^{(1)}$) begins at the initial `ket' state $|\phi^{(1)}_-\rangle$ and describes a forward time-evolution computing $U$. We then insert the operator $\hat{\mathcal{O}}_+$ before passing to the second branch of the spacetime.  The field $\phi^{(2)}$ lives on this second branch, and the associated path integral computes the backward evolution $U^\dag$. The combination of these gives the familiar Heisenberg evolution of the operator. The distinction between forward and backward evolution is implemented in the path integral by the relative sign between $I[\phi^{(1)}]$ and $I[\phi^{(2)}]$ --- or, more generally, by CPT conjugation which may also act nontrivially on fields.

Because our quantum field theory is unitary, if we happen to consider a trivial operator for which $\mathcal{O}_+(\phi_+)$ is independent of $\phi_+$ then the backwards and forwards evolutions will cancel.  In that case the result is clearly independent of the choice of slice $\Sigma_+$ on which the two spacetime branches are joined.  More generally, so long as we interpret
$\mathcal{O}_+(\phi_+)$ as being evaluated on one of the two branches, we may choose the two spacetime branches to be glued along an arbitrary Cauchy surface $\Sigma$, as long as the support of $\hat{\mathcal{O}}_+$ lies in the past of $\Sigma$.
This slicing-independence will prove useful in our discussions below.

The eigenstates $|\phi_-\rangle_-$ of field configurations on the initial slice $\Sigma_-$ are typically not of direct physical interest.  But other boundary conditions can be described by integrating over field configurations on $\Sigma_-$ with some choice of weighting. This corresponds to allowing a general state, written as a superposition of eigenstates $|\phi_-\rangle_-$ as defined by its wavefunction.  Now, since it is usually inconvenient to specify states of interest through their explicit wavefunction, we may instead choose to describe them by introducing further path integrals. For example, in our Hawking effect problem, we might specify the initial Minkowski vacuum state $|0\rangle_{\scri^-}$ by inserting a path integral over semi-infinite flat Euclidean space and connecting it to the real Lorentz-signature path integral computing $U$.

\begin{figure}[h!]
\centering
	\includegraphics[width=.5\textwidth]{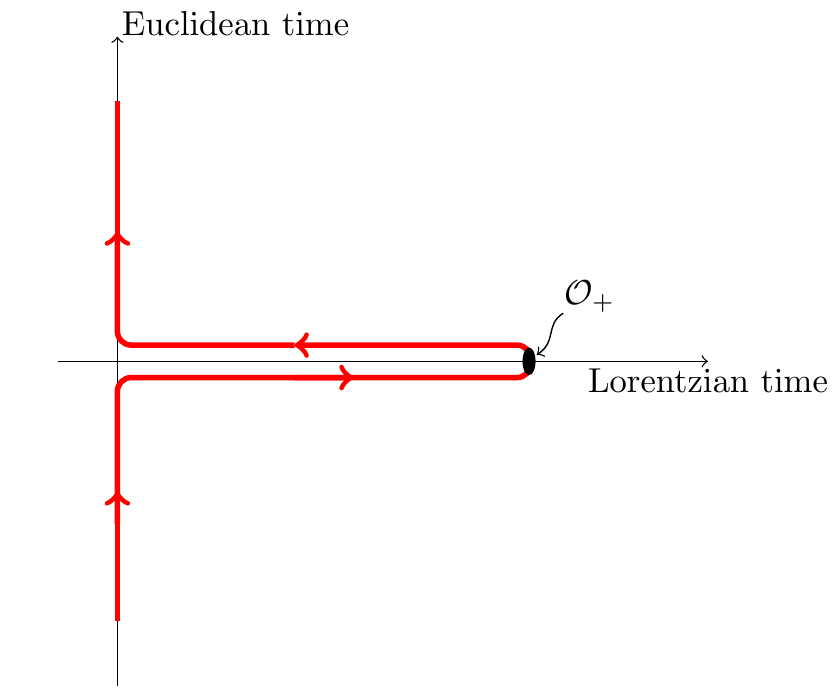}
	\caption{An in-in (or Schwinger-Keldysh) contour in the complex time-plane that computes the expectation value of $\mathcal{O}_+$ at Lorentzian time $t$ in the vacuum state $|0\rangle$. The contour begins at negative infinite Euclidean time and follows the Euclidean axis to the origin.  This part of the contour computes $|0\rangle$ in terms of fields at $t=0$.  The contour then proceeds along the Lorentzian axis (this part of the contour corresponding to the right spacetime of figure \ref{fig:ininPI}) until $\mathcal{O}_+$ is inserted at $t$, whence it returns to the origin (the left spacetime of figure \ref{fig:ininPI}).  Finally, it proceeds from the origin to positive infinite Euclidean time to compute $\langle 0|$.  For clarity, the various parts of the contour have been slightly displaced from the axes in the figure. \label{fig:SK}}
\end{figure}

We can now assemble these ingredients: an initial `ket' state prepared (perhaps) by a Euclidean path integral, a Lorentzian path integral performing forward time evolution, insertion of the operator of interest, backward time evolution, and finally the preparation of the initial `bra' state. The resulting spacetime on which we perform the path integral (see figure \ref{fig:SK}) is the `in-in' or Schwinger-Keldysh contour, and encodes the natural formulation of dynamics when we do not wish to specify a final state \cite{Schwinger:1960qe,Bakshi:1962dv,Keldysh:1964ud}.

\subsubsection{The in-in formulation of Hawking emission}
\label{sec:PIHawk}

\begin{figure}[hb]
\centering
\begin{subfigure}[b]{.48\textwidth}\centering
	\includegraphics[width=\textwidth]{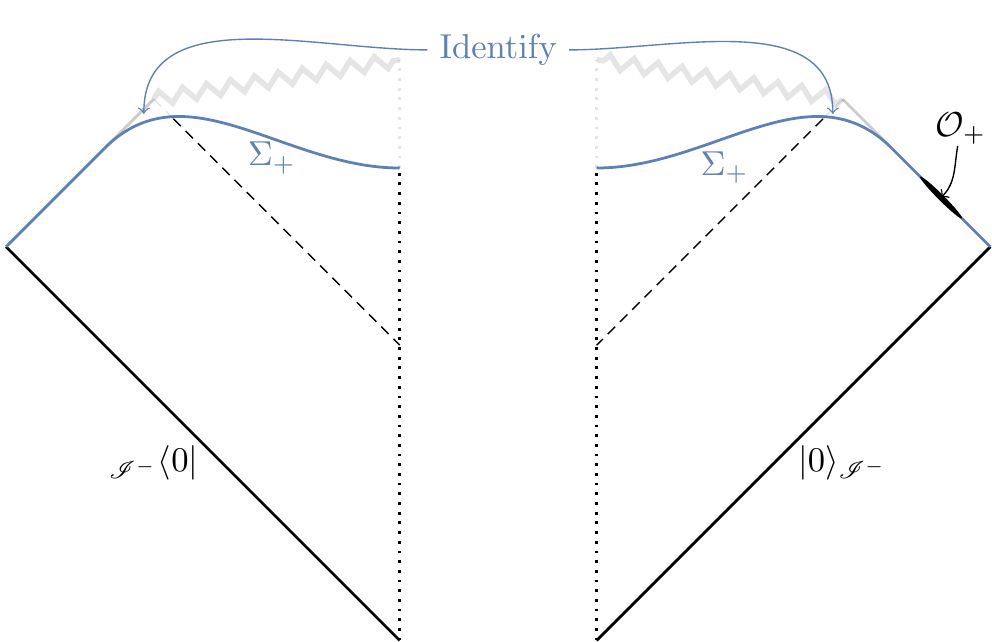}
	\caption{The path integral which computes the expectation value of an operator $\mathcal{O}_+$ on $\scri^+$. The right and left copies of the spacetime perform forward and backward time-evolution respectively. They are glued together along a Cauchy surface $\Sigma_+$, which must coincide with $\scri^+$ in the region where $\mathcal{O}_+$ is supported (denoted by the black blob) but which is otherwise arbitrary. The region to the future of $\Sigma+$ is not part of the spacetime on which our path integral is performed. \label{fig:HawkingO}}
\end{subfigure}
\hfill
\begin{subfigure}[b]{.48\textwidth}\centering
	\includegraphics[width=\textwidth]{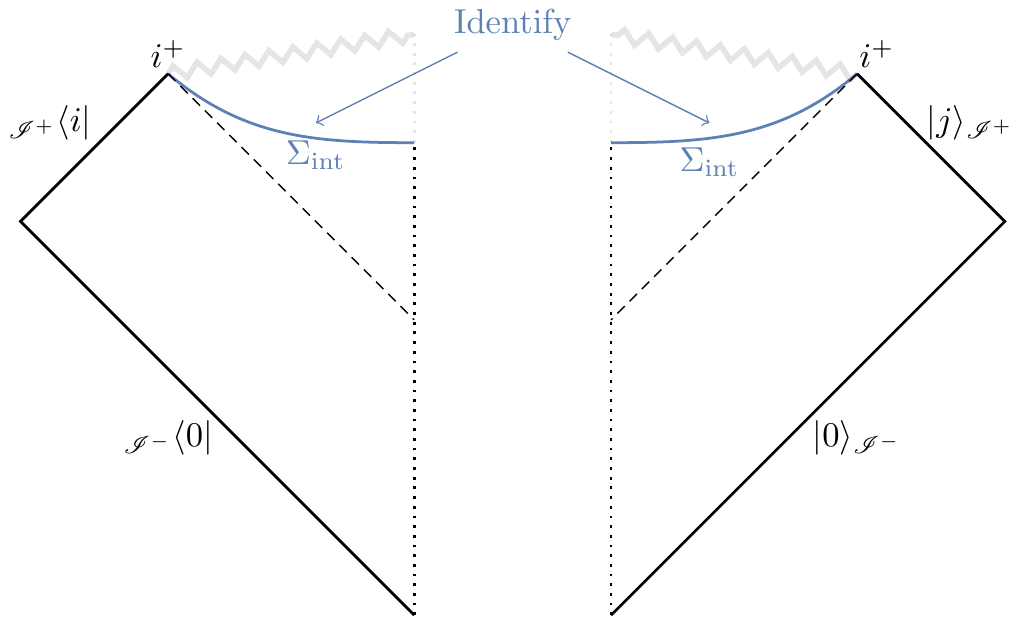}
		\caption{The path integral which computes matrix elements  of the density matrix on $\scri^+$. Along the two copies of $\scri^+$, we impose boundary conditions which weight field configurations according to the wavefunctions of states $|i\rangle_{\scri^+}$, $|j\rangle_{\scri^+}$.  If $\scri^+$ were a Cauchy surface, this would cause the path integral to fall into two disconnected pieces, indicating that the state is pure. Here, this does not happen since the two branches remain joined along $\Sigma_\mathrm{int}$, which is a Cauchy surface for the black hole interior.\label{fig:HawkingRho}}
\end{subfigure}
\caption{Path integrals computing (a) the expectation value of an operator at $\scri^+$ and (b) components of the density matrix on $\scri^+$.}
\label{fig:HawkPI}
\end{figure}

Let us now apply the above general description of quantum fields in curved spacetime to the problem at hand.  The resulting path integral is shown in figure \ref{fig:HawkingO}. We are interested in the expectation values of an operator $\mathcal{O}_+$ located on $\scri^+$, so we should take our future boundary $\Sigma_+$ to lie in the far future and to coincide with $\scri^+$ in the region where $\mathcal{O}_+$ is supported. Away from our operator insertion, we our free to extend $\Sigma_+$ to a complete Cauchy surface in any way we please.  Furthermore,  the slicing independence described above guarantees the final result to be independent of such choices. The path integral is then performed on two copies of the spacetime, but only in the region to the past of the Cauchy surface $\Sigma_+$. These two copies are identified along $\Sigma_+$, where we also insert a weighting corresponding to our operator $\mathcal{O}_+$. Since these insertions are restricted to $\scri^+$, the identification effectively performs a partial trace over the interior part of $\Sigma_+$.

As discussed at the end of section \ref{sec:Heisenberg}, computing expectation values of all operators on $\scri^+$ is equivalent to describing the state there. In particular, we can compute components $\rho_{ij}$ of the density matrix on $\scri^+$ by choosing our operator $\mathcal{O}_+$ to be $|j\rangle_{\scri^+}{}_{\scri^+}\langle i|$ for pure states $|i\rangle_{\scri^+}$, $|j\rangle_{\scri^+}$ on $\scri^+$. We depict this in figure \ref{fig:HawkingRho}. This operator insertion corresponds to a boundary condition that weights field configurations on the two branches of the Schwinger-Keldysh contour  independently, so the branches are no longer meaningfully joined along $\scri^+$; in operator terms, this says that our $\mathcal{O}_+$ has rank one. If $\scri^+$ were a Cauchy surface, then this boundary condition would cause the path integral to split into two disconnected pieces. Our $\rho_{ij}$ would then become a product of (conjugate) functions of $i$ and $j$ alone, and hence a rank one matrix describing a pure state. However, this does not occur because any Cauchy surface $\Sigma_+$ must include a piece $\Sigma_\mathrm{int}$ covering the interior of the black hole as well as a piece running along $\scri^+$. The two branches of the contour remain connected through $\Sigma_\mathrm{int}$, and the state on $\scri^+$ is mixed. This joining of the two branches is the path integral implementation of what is often called `tracing out' the interior state living on $\Sigma_\mathrm{int}$.

In practice the simplest way to evaluate the above path integrals may well be to relate it to the Heisenberg-picture computation of section \ref{sec:Heisenberg} and to use the results computed there.  Nevertheless, the formulation in terms of the path integrals of figure \ref{fig:HawkPI} will prove useful in our quantum gravity discussions below.

\subsection{Entropies from the Hawking path integral}
\label{sec:HawkingEnt}

We now conclude our review of the Hawking effect on a fixed background with a discussion of entropies.  The main point will be to review how path integrals may be used to study the R\'enyi entropies of subsets of the Hawking radiation at $\scri^+$, quantifying the tension between the original Hawking calculation and BH unitarity via the Page curve in figure \ref{fig:Page}.

\begin{figure}
	\centering
	\includegraphics[width=.7\textwidth]{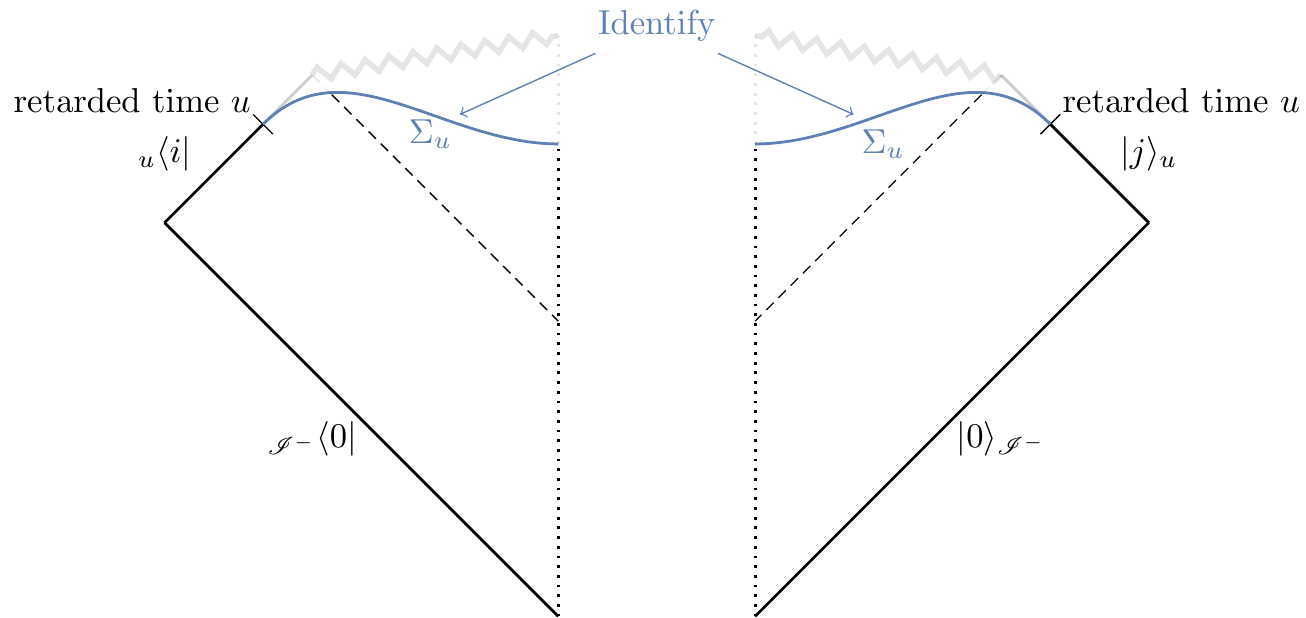}
	\caption{The path integral on this geometry computes matrix elements ${}_u\langle j |\rho_{u} | i \rangle_{u}$ of the density matrix $\rho_{u}$ describing Hawking radiation in the piece $\scri_u$ of $\scri^+$ before retarded time $u$. Two copies of the original black hole spacetime have been glued together along a surface $\Sigma_u$, which defines a Cauchy surface when joined to $\scri_u$.  We impose boundary conditions on $\scri_u$ corresponding to the states $|i \rangle_{u}$, $|j \rangle_{u}$.\label{fig:HawkingRhou}}
\end{figure}

First, we must slightly generalize the above discussion to compute the density matrix $\rho_{u}$ associated not with the entirety of $\scri^+$, but only with the Hawking radiation that reaches the subset  $\scri_u \subset \scri^+$ of points at retarded times $u' < u$. To compute matrix elements of $\rho_{u}$, we simply modify the discussion above as depicted in figure \ref{fig:HawkingRhou}.
On $\scri_u$, we fix boundary conditions according to the desired matrix elements.  We then  join the two branches of the path integral along a partial Cauchy surface $\Sigma_u$ that reaches $\scri^+$ at $u$ (rather than joining them on some $\Sigma_\mathrm{int}$ that reaches $\scri^+$ only at its future endpoint $i^+$).

Next recall that we are interested in R\'enyi entropies.
The $n$th R\'enyi entropy of a density matrix $\rho$ is defined by
\begin{equation}
\label{eq:Renyis}
S_n(\rho) = -\frac{1}{n-1} \log \left(\frac{\Tr(\rho^n)}{\left( \Tr \rho \right)^n} \right),
\end{equation}
 where we have allowed for the possibility  that $\rho$ has not yet been normalised (i.e., that it may not have unit trace).
As noted above, in path integral constructions it is typically simpler to work with unnormalized states than to keep track of all normalizations.

\begin{figure}
	\centering
	\raisebox{110pt}{\scalebox{1.5}{$\Tr(\rho(u)^2) =$}}
	\includegraphics[width=.4\textwidth]{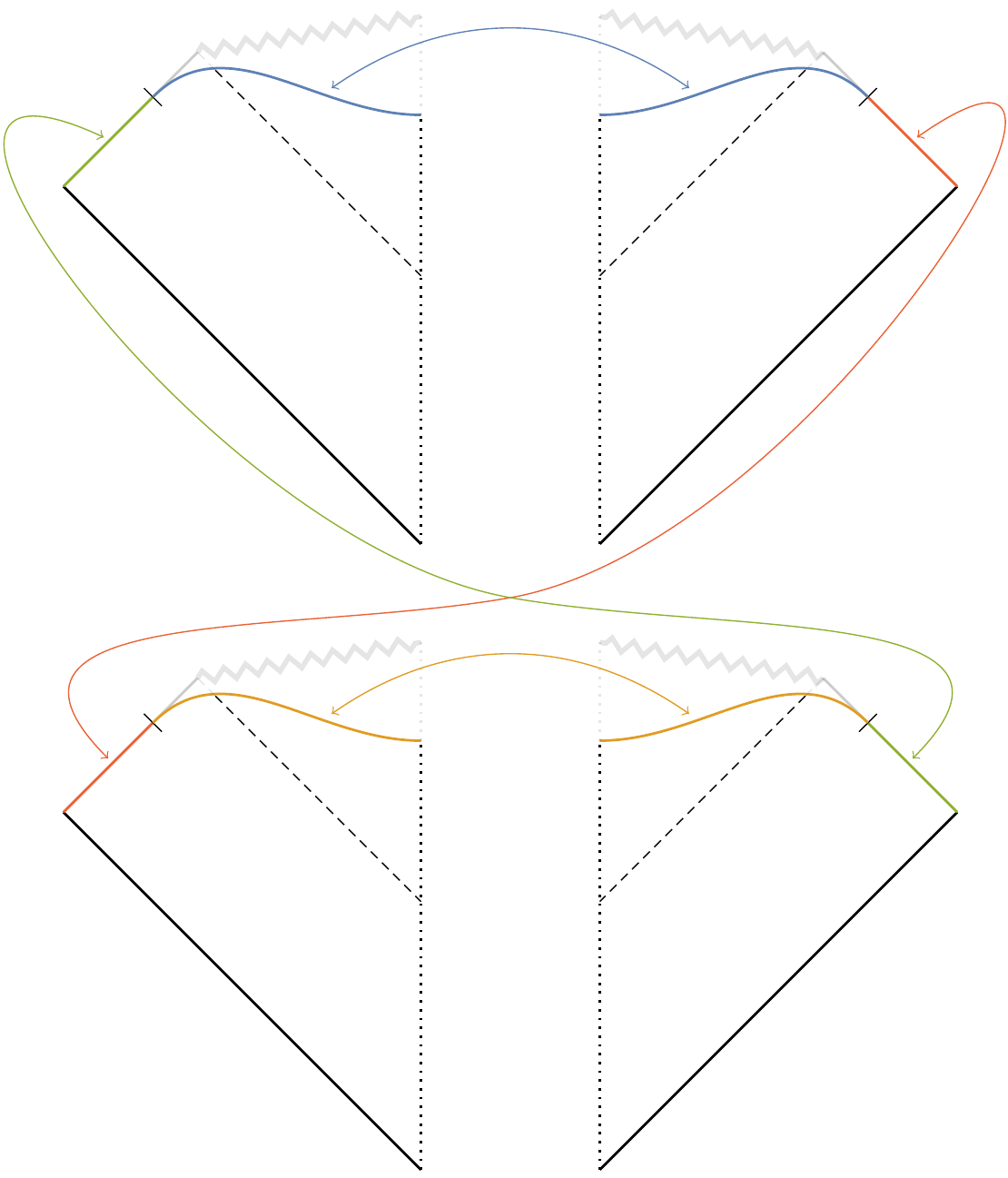}
		\caption{To compute $\Tr(\rho(u)^2)$, we perform the path integral on the geometry built from two replicas of figure \ref{fig:HawkingRhou}, identified as shown.\label{fig:HawkingRenyi}}
\end{figure}

To compute $\Tr(\rho(u)^n)$ from the path integral, we start with $n$ copies of the spacetime depicted in figure \ref{fig:HawkingRhou} to construct $n$ replicas of $\rho(u)$. We then sew these replicas together as instructed by the matrix products and trace in $\Tr(\rho^n)$. Specifically, the `ket' boundary labelled by the state $|j_r\rangle_u$ on the $r$th replica becomes identified with the `bra' boundary labelled by ${}_u\langle i_{r+1}|$ on the $(r+1)$th replica since the insertion of complete sets of states amounts to setting $i_{r+1}=j_r$ and then summing over a complete set of such wavefunctions. The trace completes this pattern cyclically. The result is shown in figure \ref{fig:HawkingRenyi} for the case $n=2$.  It is often of interest to compute (or to imagine computing) the R\'enyi entropies for all integers $n\ge 2$, studying an appropriate analytic continuation\footnote{Carlson's theorem from complex analysis states that any analytic function $f(z)$ that agrees with given values on the positive integers and satisfies the bounds $|f(z)| \le C e^{\tau |z|}$ for some real $C, \tau$ (for all complex $z$) and $|f(iy)| \le C e^{c|z|}$ for real $y$ and some $c<\pi$.  For systems with finite-dimensional Hilbert spaces, the R\'enyis always satisfy such conditions.  In practice, the same seems to hold for physically-interesting states on infinite-dimensional Hilbert spaces.}, and taking the limit $n \rightarrow 1$ which defines the von Neumann entropy $S(\rho)$.


For any $n$, the resulting R\'enyi entropy will be infinite due to high-frequency modes at the `entangling surface' where $\Sigma_u$ meets $\scri^+$ at retarded time $u$. This divergence is local at the entangling surface and is state-independent, so it is not related to the physics of interest (in particular, it is independent of $u$). We will subsequently assume that some regulator has been chosen, for example subtraction of the Minkowski vacuum result, and implicitly discuss the resulting finite quantity throughout.

Since Hawking radiation rapidly becomes thermal at $\scri^+$, after some brief transient behavior any correlation functions on $\scri^+$ decay rapidly when clusters of points are separated by more than a thermal retarded time.  As a result, over large stretches of time the density matrix on $\scri^+$ may be thought of as a tensor product of thermal density matrices (with appropriate grey-body factors) associated with smaller pieces of $\scri^+$.   As a result, all R\'enyi entropies $S_n$ and the von Neumann entropy $S$ will increase linearly with $u$ at large $u$.  As noted in the introduction, this behavior is inconsistent with BH unitarity which would require $S$ to be bounded by the Bekenstein-Hawking entropy $S_{BH}$ defined by the Bondi mass at each retarded time $u$.

\section{Semiclassical path integrals and back-reaction}
\label{sec:semiclassical}

For our review of Hawking's calculation in section \ref{sec:HawkingReview}, we treated spacetime as a background field with a fixed nondynamical metric, and  we integrated only over matter fields. We now wish to incorporate gravitational dynamics by integrating over metrics. Of course, outside of simple toy models it will be difficult to perform (or even define) the gravitational path integral exactly. Instead, we will treat the path integral as a weak-coupling expansion in a nonlinear effective theory. In practice, this means that we look for saddle-point configurations for which the classical (or, perhaps, the quantum-corrected) effective action is stationary under variations of the metric and other fields, and we then integrate over fluctuations around these saddles.

We will thus need to specify boundary conditions for the metric. It is natural to impose boundary conditions in asymptotic regions of spacetime where gravity becomes weak, in analogy with scattering problems in quantum field theory.  We will integrate over asymptotically flat metrics, and choose in-states and out-states for gravitons (along with matter fields) on $\scri^\pm$. Alternatively, following the review of section \ref{sec:HawkingReview}, we may use boundary conditions that do not completely specify a final state, and we may instead compute an asymptotic observable using an in-in formalism. In either case, we specify the metric and states only in the asymptotic region.  We will place no restrictions on the metric deep in the spacetime interior.  We will thus include contributions from any saddle-point metric matching the specified asymptotics.  In particular, we allow all spacetime topologies.

To describe perturbative quantum effects, it will be convenient for us to treat the metric separately from matter fields, and begin by `integrating out' the matter. For a given spacetime with metric $g$, we can use ideas reviewed in section \ref{sec:HawkingReview} to perform the matter path integral as a QFT on the fixed background, which we can write as a quantum effective action:
\begin{equation}\label{eq:Ieff}
	e^{iI_\text{eff}[g]} := \int \mathcal{D}\phi \; e^{iI_\text{matter}[\phi,g]} \,.
\end{equation}
To incorporate perturbative effects from the fluctuations of the metric itself, such as black hole evaporation by emission of gravitons, this `matter' effective action should also incorporate a one-loop effective action from integrating out linearised metric perturbations; see e.g. \cite{Vilkovisky:1984st,DeWitt:1985sg}. A saddle-point in the integral over metrics $g$ is then a stationary point of the combined gravitational (Einstein-Hilbert) action and matter effective action $I_\text{EH}[g]+I_\text{eff}[g]$.

\subsection{Incorporating back-reaction}\label{sec:hawking}

We now have everything we need to begin making predictions using semiclassical gravity. We first adapt the calculations of section \ref{sec:PIV} to incorporate a  dynamical metric, preparing an initial state of matter at $\scri^-$ to form a black hole, and asking for the expectation value of some observable at $\scri^+$. The relevant boundary conditions are similar to the situation pictured in figure \ref{fig:HawkingO}, with the two branches of the in-in contour joined at a future boundary. But thus far the metric has been specified only asymptotically at $\scri^\pm$, and in the interior we sum over allowed possible metrics. As already noted above, in practice this means that we will proceed by studying saddle points, where here we explicitly mean saddle points of  $I_\text{EH}[g]+I_\text{eff}[g]$.

Finding saddles can be construed as solving the associated equations of motion.  However, one should realize that this is not a standard Cauchy evolution problem for two reasons.  The first is that the quantum-corrected effective action is generally non-local.  The second is that we impose boundary conditions at both copies of $\scri^-$ and also at both copies of $\scri^+$, rather than imposing two conditions (on fields and on their derivatives) on a single Cauchy slice.  As a result, there can be multiple saddles that contribute to a given path integral, and it can be challenging to determine whether one has in fact found all of the relevant ones.    One is thus often left with simply searching for saddles and seeing what physics they entail.  If one later finds additional saddles, one will need to correct the original calculation to take the new saddles into account.

It is natural to begin by assuming quantum effects to be small and treating $I_\text{eff}$ as a small correction to $I_\text{EH}[g]$.  In particular, the latter includes a factor of the inverse Newton constant $1/G$, and is thus very large in the semiclassical gravity limit $G \rightarrow 0$.  The most obvious saddle for our path integrals is thus given by starting with the classical collapsing black hole solution that was used as a fixed background in section \ref{sec:Heisenberg} and including perturbative corrections from $I_\text{eff}$.  Note that the variation of $I_\text{eff}$ with respect to the metric is precisely the expectation value of the stress tensor of the quantum matter fields\footnote{And analogous corrections to the equations of motion built from linearized gravitons.} over which we have already integrated in the initial state $|0\rangle_{\scri^-}$, up to effects associated with post-selection when the state is also (partially) specified at $\scri^+$.  So this indeed incorporates back-reaction from the Hawking radiation described earlier. We shall focus on this saddle below, turning to other possible saddles only in sections \ref{sec:PS} and \ref{sec:replicas}.

Let us begin by ignoring post-selection at $\scri^+$, so that back-reaction is precisely given by the expected stress-energy tensor in the state $|0\rangle_{\scri^-}$.  As is well known, this tensor carries a flux of positive energy to infinity and a flux of negative energy into the black hole.  The flux is small, so significant changes to the background occur only when they can build up over long times, or over large affine parameters.

Now, in the original classical solution of figure \ref{fig:collapse}, the only null geodesics that extend to infinite affine parameters toward the future are those that lie entirely outside the event horizon.  As a result, any additional null geodesic that extends to large affine parameter must be confined to the region close to the original event horizon.  We thus conclude that there is a large region inside the black hole where perturbative corrections give little change in the physics, and where the spacetime continues to collapse at least until such time as the curvatures become large (which presumeably means Plank scale).  For simplicity, we will continue to call this large-curvature a singularity and to indicate it by a jagged line on spacetime diagrams.  This is consistent with our current ignorance and lack of control over Planck-scale physics, though we do not rule out the possibility that a better description may become available in the future.

On the other hand, back-reaction can be significant when one follows a null geodesic that lies just inside the event horizon of the original background.  Congruences of such geodesics can be studied using the Raychaudhuri equation (see e.g.\ \cite{Wald:1984rg}).  In particular, while they begin with a slight negative expansion, if this initial negative value is sufficiently small (i.e., for congruences close enough to the event horizon of the original background) the incoming flux of negative energy causes the expansion to evolve through zero and to eventually become positive.  This indicates that such congruences in fact escape to $\scri^+$.  Taking a one-parameter family of such congruences and using the cuts on which the expansions vanish to define an apparent horizon, the fact that each successive congruence must begin with a more and more negative expansion means that this apparent horizon must shrink.  And again, this description must continue to hold until the curvature becomes Planck scale, at which point the apparent horizon is also correspondingly small.  We denote this locus $\evap$ and refer to it as the `endpoint' of Hawking evaporation in the expectation that little more of interest can happen after this point\footnote{This expectation will become an explicit assumption for the purposes of section \ref{sec:PS}.  However, our goal in this work is to avoid sensitivity to effects that are not under semiclassical control.  A critical point is thus that no such assumptions are needed for the replica wormhole derivation of the Page curve that will be reviewed in section \ref{sec:replicas}.}.  We will idealize $\evap$ as a codimension-2 surface, though it reality it describes a region of small but finite size. We define the `evaporation time' $u_\evap$ to be the retarded time of the past boundary of $\evap$; that is, the time at which Planckian curvatures are first visible asymptotically.

\begin{figure}
\centering
	\includegraphics[width=.35\textwidth]{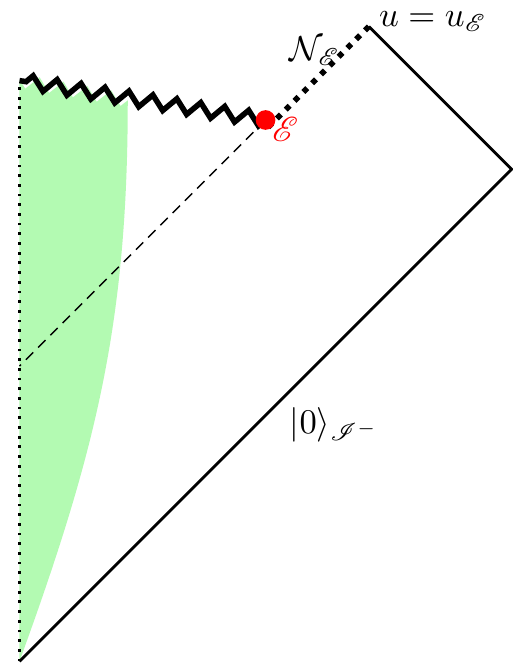}
\caption{The region under semiclassical control in an evaporating black hole spacetime.  In the far past, the diagram coincides with \ref{fig:collapse} and the black hole forms from collapse of matter.  This region is bounded by the jagged line (called the `singularity'), its endpoint $\evap$, and the outgoing null congruence $\mathcal{N}_\evap$ (dotted line).  Planck scale physics becomes important at the singularity and $\evap$, and may influence further evolution of the spacetime.   The event horizon $\mathcal{H}^+$ (dashed line) is defined to be the boundary of the past domain of dependence of the singularity and $\evap$, and we refer to this past domain of dependence as the black hole interior.\label{fig:EvaporatingSC}}
\end{figure}

Without a better understanding of Planck scale physics, it is impossible to say whether and how the singularity and $\evap$ influence other parts of the spacetime.  But there is a unique perturbatively-semicalssical evolution in regions of spacetime from which they are causally separated, and of course also in the region to the past of the singularity and $\evap$.  This region of semiclassical control is shown  figure \ref{fig:EvaporatingSC}.  It is not geodesically complete, and does not contain a complete $\scri^+$.  Instead, it has a future boundary defined by
the singularity, $\evap$, and (using our spherical symmetry to rule out caustics and the like) the outgoing null congruence $\mathcal{N}_\evap$ from $\evap$ (dotted line in figure \ref{fig:EvaporatingSC}) at retarded time $u_\evap$.
However, it can be used to study black hole evaporation so long as we do not ask about what occurs beyond $\mathcal{N}_\evap$.

In particular,
let us now use the spacetime of figure \ref{fig:EvaporatingSC} to construct back-reacted saddles for the density matrix $\rho(u)$ on a region $\scri_u\subset \scri^+$ that is expected to be under semiclassical control.  We thus wish to find a back-reacted analogue of figure \ref{fig:HawkingRhou}.  The one issue we must consider is post-selection at $\scri_u$, as this can modify the stress-energy fluxes to $\scri^+$ and across ${\cal H}^+$.  However, as typical states at $\scri^+$ have stress-energy fluxed close to the mean, such effects are typically small.  And even when they are large, they make little impact on qualitative features of figure \ref{fig:EvaporatingSC}.

\begin{figure}
\centering
	\includegraphics[width=.5\textwidth]{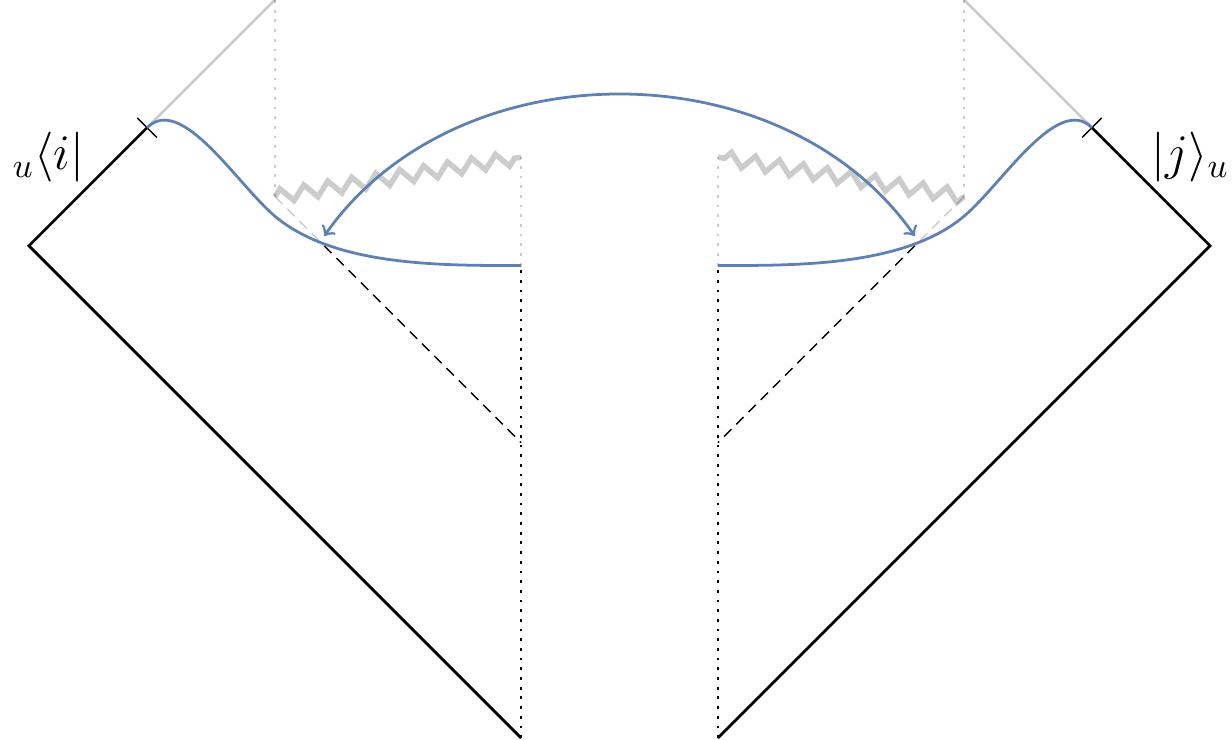}
\caption{The saddle-point spacetime for computing the density matrix of an evaporating black hole. The future of the blue slice $\Sigma_u$ where the identification occurs is not part of the configuration, so the spacetime is weakly curved everywhere, and in particular excludes the singularity.\label{fig:rhoEvaporatingSC}}
\end{figure}

We may thus construct a saddle for $\rho(u)$ in direct analogy with figure \ref{fig:HawkingRhou}, and in particular by sewing two copies of figure \ref{fig:EvaporatingSC} to each other along a partial Cauchy surface $\Sigma_u$ that runs from the regular origin at the center of the collapsing matter to retarded time $u$ at $\scri^+$ as shown in figure \ref{fig:rhoEvaporatingSC}.  The only difference from working on a fixed background is that gravity dynamically determines the spacetime away from the boundaries. The contribution of this saddle to the path integral is independent of the choice of $\Sigma_u$, since the phases in the classical action from the two branches of the contour cancel, and the matter evolves unitarily on a fixed background.  We see that the entire calculation is under semiclassical control and makes no reference to strong curvature regions.

For future reference, and because it involves essentially the same physics as Hawking's original calculation \cite{Hawking:1974sw}, we refer to the density matrix defined by saddles of the form shown in figure \ref{fig:rhoEvaporatingSC} as the Hawking density matrix:
\begin{equation}\label{eq:rhoHawking}
	\rho(u) \approx \rho_\text{Hawking}(u).
\end{equation}
Because back-reaction is small, $\rho_\text{Hawking}(u)$ is essentially a thermal state with a temperature that varies slowly with retarded time $u$.

Since the predictions for any experiment are encoded in the density matrix, we see that  perturbatively-semiclassical gravity suffices to make probabilistic predictions for any measurement of the Hawking radiation that avoids particularly late retarded times (at which the black hole has become Planck scale). Since these predictions are encoded in the highly mixed and quasi-thermal density matrix $\rho_\text{Hawking}(u)$, they violate BH unitarity and indicate the black hole density of states to be unrelated to the Bekenstein-Hawking entropy.  Indeed, by the usual argument that starting with an arbitrarily large black hole leads to arbitrarily large entropy on $\scri_u$
even when the Bondi mass at $u$ is held fixed, it suggests the actual black hole density of states to be infinite.

However, with access only to the Hawking radiation produced in a single black hole evaporation, we cannot operationally verify that the state is mixed. It turns out that this critical fact provides interesting room for further physics.  The remainder of this paper is largely devoted to this point.  In order to describe such possibilities without yet delving into the technical complications of replica wormholes, and to make connections with the historical literature, section \ref{sec:PS} will use the crutch of making assumptions about physics that is beyond semiclassical control.  But we will see in sections \ref{sec:replicas} and \ref{sec:BUetc} that this crutch can be discarded, and that semiclassical gravitational physics {\it does} predict physics consistent with BH unitarity.

\section{Entropy measurements and potential new saddles}
\label{sec:PS}

Our calculation in the previous section has led us to suspect that the Hawking radiation is in a highly mixed state on $\scri_u$, which in particular violates BH unitarity. Continuing with our philosophy of concentrating on the predictions for asymptotic observers, we might like to imagine performing an experiment to directly verify such violations. But this is impossible without access to several copies of the state.  Indeed, as an immediate consequence of the familiar fact that a mixed state is equivalent to an ensemble of pure states, no measurement on a single copy can help us to distinguish a mixed state from an unknown pure state.

We must therefore form several black holes, taking care to prepare them in identical initial states, and collect their decay products. We end up with $n$ sets of Hawking radiation, presumably in $n$ identical copies of the same state since they were all prepared in the same way. With $n$ identical copies in hand, it is a straightforward task to test whether a state is pure or highly mixed.  For example, one may use the swap test of \cite{buhrman2001quantum,Hayden:2007cs} which we will describe below.

Now, what does semiclassical gravity predict for the state $\rho^{(n)} (u)$ of our $n$ sets of radiation on $\scri_u$?   At first sight, this may appear to be a frivolous question; surely it is trivially $n$ copies of the result already obtained in \eqref{eq:rhoHawking},
\begin{equation}\label{eq:rhoHawkingn}
	\rho^{(n)}(u) \approx \left[\rho_\text{Hawking}(u)\right]^{\otimes n}\; \raisebox{-2pt}{\text{\Large ?}}
\end{equation}
However, as observed by Polchinski and Strominger \cite{Polchinski:1994zs}, this conclusion is too hasty.  While it is true that our saddle-point computation of $\rho^{(1)} \approx \rho_\text{Hawking}$ immediately leads to a saddle that would give \eqref{eq:rhoHawkingn},
considering $n$ copies of the state together turns out to allow potential new saddle points.\footnote{In this section and the next, we simply observe this phenomenon and study its implications.  Interpretations of the new saddles and discussions of the underlying physics they represent will be deferred to section \ref{sec:BUetc}.}

The purpose of this section to describe path integrals that predict experimental measurements of entropy and to connect them with the potential new saddles discussed in \cite{Polchinski:1994zs}.  Before doing so, we will admit to the reader that the potential new saddles advocated in  \cite{Polchinski:1994zs} involve physics that is {\it not} under semiclassical control.  It is thus important that they will not form the basis of any analyses in section \ref{sec:replicas} or \ref{sec:BUetc}, or for the final conclusions of this work.   We nevertheless review this proposal here for three other reasons.  The first is that it serves as a pedagogical tool to explain the idea of new saddles without yet delving into the technical complications of replica wormholes.  The second is that this helps to place recent developments in an appropriate historical context, as proposal of \cite{Polchinski:1994zs} turns out to have many similarities to the replica wormholes of section \ref{sec:replicas}.  And the third is that it suggests some of the physics that may in fact lie behind the semiclassical replica wormholes of section \ref{sec:replicas}.

We thus dedicate section \ref{sec:PSP} to reviewing the proposal of \cite{Polchinski:1994zs},  recasting the discussion in terms of experimental measurements at infinity. This is followed by a short aside in section \ref{sec:SW}, which describes how the black hole information problem is related to the lack of factorization of quantum gravity amplitudes.   Experiments that involve only some $\scri_u\subset \scri^+$ are introduced in section \ref{sec:PSu}, and section \ref{sec:PSchallenges} then describes shortcomings of the Polchinski-Strominger proposal, all of which will be resolved by replica wormholes in section \ref{sec:replicas}.

Before diving in, we should remark that the Polchinski-Strominger work \cite{Polchinski:1994zs} was largely described in terms of two-dimensional models of gravity inspired by analogy with the string worldsheet. We interpret their proposal more broadly, applying it to more general theories of gravity in any dimension. In particular, much of \cite{Polchinski:1994zs} was concerned with the physics of the endpoint of evaporation $\evap$, the details of which will be unimportant for our considerations.

\subsection{Polchinki and Strominger's proposal}
\label{sec:PSP}

To understand how considering $n>1$ black holes can lead to new saddles, let us first construct the boundary conditions appropriate for such multi-black-hole experiments. For the purposes of the current section, we take our experimenter to collect all of the Hawking radiation emitted to $\scri^+$ for all times, deferring discussion of subsets $\scri_u$ to section \ref{sec:PSu}.  This will necessarily involve making assumptions about physics that is not under semiclassical control.\footnote{As described in section \ref{sec:PSu}, in the Polchinski-Strominger context this issue will \emph{not} be resolved just by considering the subsets $\scri_u$.  But replica wormholes will offer a resolution in section \ref{sec:replicas}.}

We will treat each black hole as if it is formed and decays in its own separate asymptotic region.  As a result, our boundary conditions will be precisely $n$ copies of the boundary conditions of figure \ref{fig:rhoEvaporatingSC} in the limit $u\rightarrow \infty$ or, equivalently, extended from $\scri_u$ to all of $\scri^+$.  Placing each black hole in its own asymptotic region is a convenient abstraction, though the conclusions should be equally valid for $n$ black holes in a single asymptotic region, so long as we prepare black holes which are sufficiently well-separated in time or space.\footnote{This can be thought of as a version of the cluster decomposition principle.} The boundary conditions for computing the components of the $n$-evaporation density matrix $\langle i_1,\ldots,i_n|\rho^{(n)}|j_1,\ldots,j_n\rangle$ thus involve $2n$ separate asymptotic boundaries, $n$ with boundary conditions at $\scri^+$ specifying a `ket' state $|j_r\rangle$, and $n$ conjugate copies specifying a `bra' state $\langle i_r|$.

For $n=1$, we expect a saddle given by extending figure \ref{fig:rhoEvaporatingSC} to $u = \infty$.  As noted above, this extension must involve assumptions about effects in the strong curvature region.
Roughly speaking, our interpretation of the assumption of \cite{Polchinski:1994zs} is that the black hole evaporates completely, but that information in the black hole interior does \emph{not} emerge at $\scri^+$.  Indeed, Polchinski and Strominger describe information reaching the singularity as being transferred to a `baby universe' that branches off from the parent universe and does not return, a perspective which we will explore further in section \ref{sec:BUetc}.  For our purposes, we can cleanly state the required assumption as follows:\footnote{See e.g.\ \cite{Giddings:1992hh,Ashtekar:2005cj,Ashtekar:2008jd,Bianchi:2018mml,DAmbrosio:2020mut}    for other scenarios for late-time quantum gravity effects.}%

\begin{figure}
\centering
	\includegraphics[width=.3\textwidth]{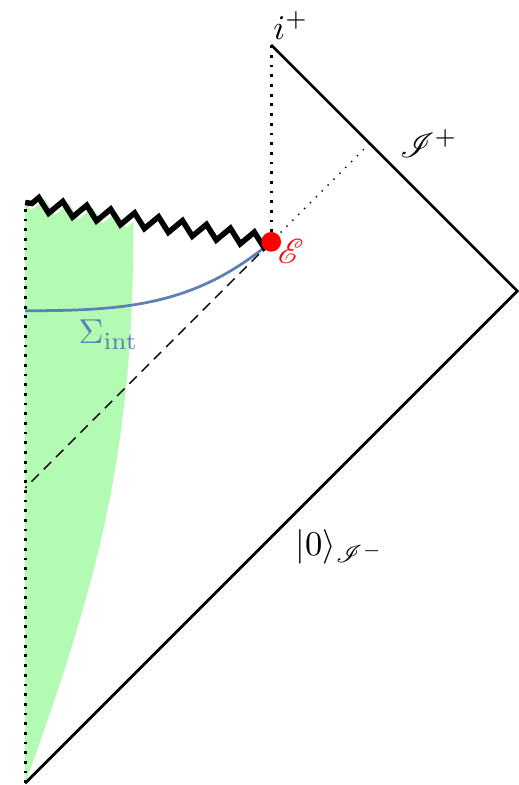}
\caption{An extension of the spacetime of figure \ref{fig:EvaporatingSC} to larger $u$ under the PS assumption. We have a complete $\scri^+$ with Minkowski-like future timelike infinity $i^+$, and may treat $\scri^+ \cup \Sigma_\mathrm{int}$ as a Cauchy surface whenever  $\Sigma_\mathrm{int}$ is Cauchy in the black hole interior.  \label{fig:Evaporating}}
\end{figure}

\begin{emphblock}
	\textbf{PS assumption:} The extension of any evaporating black hole spacetime beyond the region of semiclassical control shown in figure \ref{fig:EvaporatingSC} is such that (1) the spacetime is empty near future timelike infinity $i^+$, so that this region resembles that of Minkowski space; and (2) for any Cauchy surface $\Sigma_\mathrm{int}$ of the black hole interior, we may treat $\scri^+ \cup \Sigma_\mathrm{int}$ as a (disconnected) Cauchy surface for the full spacetime.%
\end{emphblock}
We depict the evaporating black hole spacetime under this assumption in figure \ref{fig:Evaporating}.
We note that the PS assumption requires that the physics of $\evap$ is appropriately local: in particular, the state of any radiation emitted to $\scri^+$ after the black hole becomes Planckian will be independent of the history of the black hole, such as the state on $\Sigma_\mathrm{int}$ away from the strongly curved region $\evap$.

The PS assumption immediately allows us to sew together two copies of figure \ref{fig:Evaporating} to define a back-reacted saddle for the density matrix $\rho = \lim_{u\rightarrow \infty}\rho(u)$ on all of $\scri^+$; see either top or bottom of figure \ref{fig:PSrho2a} below. The result satisfies the definition of a spacetime wormhole given in the introduction, since the boundary consists of two complete and disconnected copies of $\scri^+$. For this reason, and because spacetimes like that of figure \ref{fig:Evaporating} were often championed by Hawking,  we refer to this spacetime as the Hawking wormhole.

For $n>1$ replicas, the spacetime which gives rise to the na\"ive result \eqref{eq:rhoHawkingn} for the $n$-evaporation density matrix $\rho^{(n)}$ is then simply $n$ copies of the Hawking wormhole with boundary conditions $|j_r\rangle$ and $\langle i_r|$ for $r=1,2,...n$; see figure \ref{fig:PSrho2a} for $n=2$. But since the boundary conditions are invariant under independent permutations of bras and kets, it is clear that we can then build further wormholes with identical boundary conditions by simply pairing `bra' and `ket' boundaries in different ways. This construction defines $n!$ distinct wormholes over which our path integral must sum, one for each permutation of the $n$ kets relative to the $n$ bras.   We refer to the doubled-spacetimes defined by the $n!-1$ non-trivial pairings as PS wormholes.  The single PS wormhole for the $n=2$ case is shown in figure \ref{fig:PSrho2b}.   Note that, although each wormhole involves $\evap$ and its future (and thus leaves the domain of semiclassical control), since all $n!-1$ PS wormholes are diffeomorphic to $n$-copies of the Hawking wormhole of figure \ref{fig:PSrho2a}, they also have precisely the same validity as the Hawking wormhole to be interpreted potential saddles.
\begin{figure}
\centering
\begin{subfigure}[t]{.48\textwidth}\centering
	\includegraphics[width=.8\textwidth]{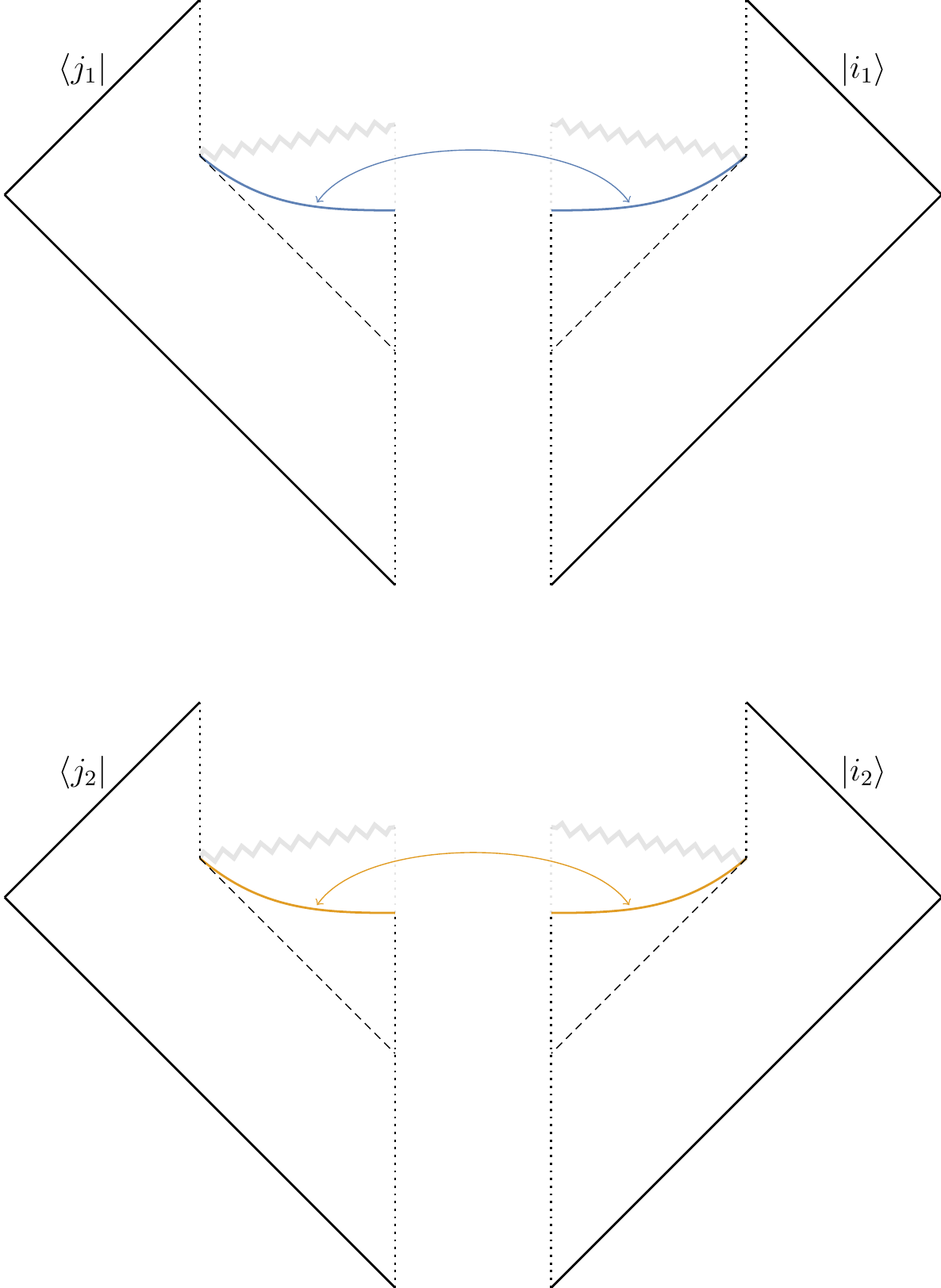}
	\caption{Extending figure \ref{fig:rhoEvaporatingSC} as in figure \ref{fig:Evaporating} gives a Hawking wormhole. Two copies of this wormhole are shown.}
\label{fig:PSrho2a}
\end{subfigure}
\hfill
\begin{subfigure}[t]{.48\textwidth}\centering
	\includegraphics[width=.8\textwidth]{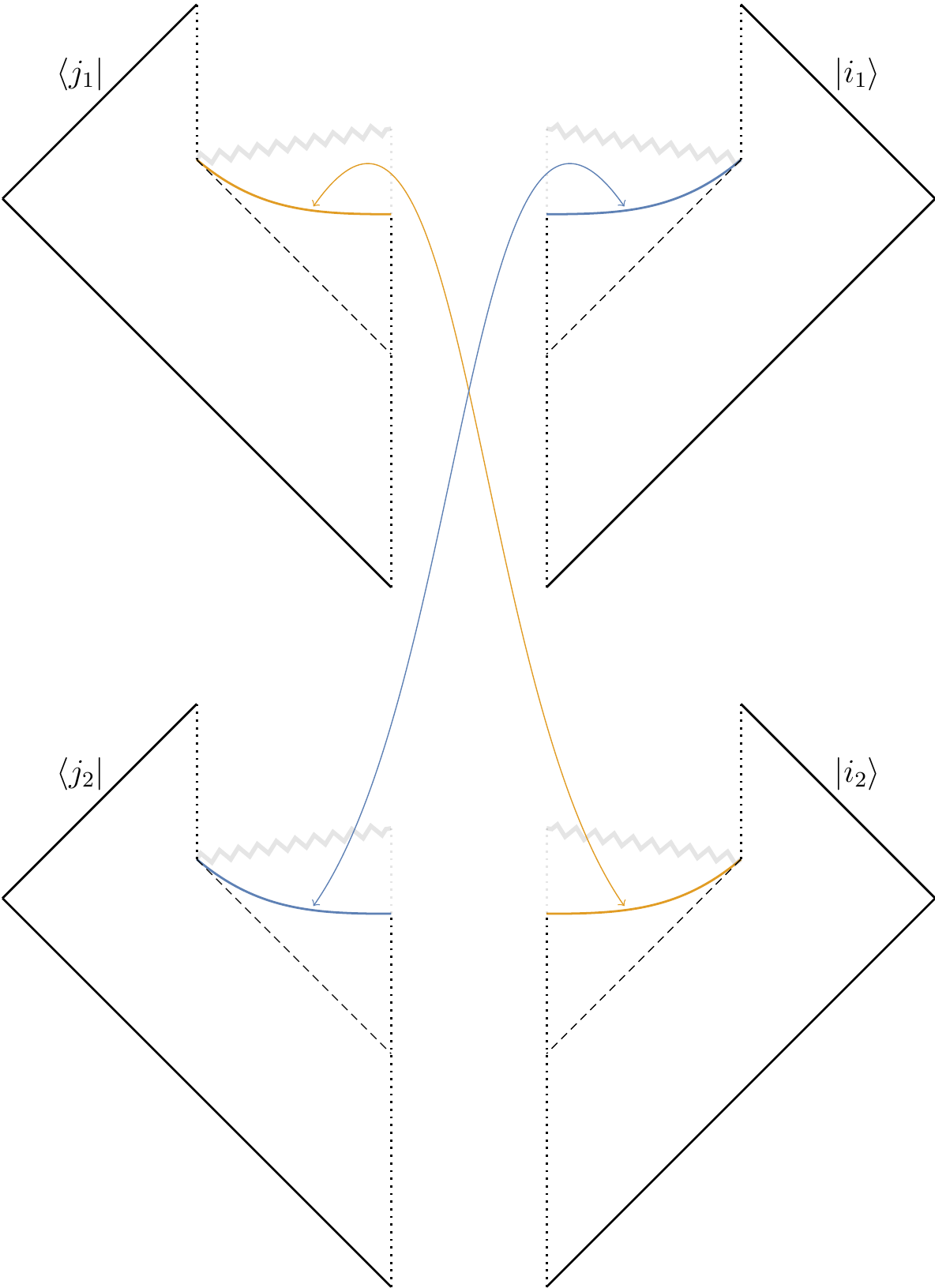}
		\caption{Another contribution to $\rho^{(2)}$ with the same boundary conditions, for which the identifications between black hole interiors have been swapped.}
\label{fig:PSrho2b}
\end{subfigure}
\caption{The Hawking (a) and Polchinski-Strominger (b) wormholes contributing to the density matrix $\rho^{(2)}$ describing the decay products at $\scri_+$ of two identically-prepared black holes.\label{fig:PSrho2}}
\end{figure}

Indeed,
the fact that all $n!$ saddles are diffeomorphic also requires them to contribute precisely the same weight to the path integral.
We therefore find the components of our density matrix to be given by a sum over all permutations $\pi\in \Sym(n)$, where $\Sym(n)$ denotes the symmetric group on $n$ indices:
\begin{equation}\label{eq:rhon1}
	\langle i_1,\ldots,i_n|\rho^{(n)}|j_1,\ldots,j_n\rangle = \sum_{\pi\in \Sym(n)} \langle i_1|\rho_\text{Hawking}|j_{\pi(1)}\rangle \cdots \langle i_n|\rho_\text{Hawking}|j_{\pi(n)}\rangle + \cdots,
\end{equation}
and where we have not normalised the state.  The ellipsis ($+\cdots$) in \eqref{eq:rhon1} indicates various potential corrections, including any that from possible further saddles that have not yet been identified. We will assume such corrections to be negligible for the rest of section \ref{sec:PS}.

As a result of \eqref{eq:rhon1}, the rules for our semiclassical path integral, while treating PS wormholes as saddles, imply that the state $\rho^{(n)}$ of the $n$-evaporation Hawking radiation collected by our experimenter  differs significantly from the state $\rho_\text{Hawking}^{\otimes n}$ that would describe $n$ identical independent copies of the mixed state $\rho_\text{Hawking}$ that she would collect from a single evaporation.

Since this may at first seem surprising, it is useful to note that \eqref{eq:rhon1} admits a natural Hilbert space interpretation. After we collapse $n$ black holes and allow them to evaporate, we must trace out the $n$ interiors. But once evaporation has proceeded to completion, we see from figure \ref{fig:PSrho2b} that the interiors are no longer attached to a corresponding external spacetime.  As a result, there is no longer anything to distinguish them. The sum in \eqref{eq:rhon1} treats the $n$ interiors as indistinguishable objects obeying Bose statistics. We could say that each black hole interior is like a Bosonic particle, carrying many internal degrees of freedom to describe the state of the matter that formed the black hole and the ingoing Hawking partners. When we trace these out, having several interiors in the same quantum state means that we must include a symmetrisation as is familiar from Bosonic Fock spaces.  This then leads to  \eqref{eq:rhon1}. We will explore the Hilbert space interpretation in more detail in section \ref{sec:BUetc}.

To understand the implications of \eqref{eq:rhon1}, it is useful to introduce a unitary operator $U_\pi$ for each permutation $\pi$ in the symmetric group $\Sym(n)$, where the $U_\pi$ act to permute states among the $n$ collections of Hawking radiation:
\begin{equation}
	U_\pi \left(|i_1\rangle\otimes\cdots\otimes |i_n\rangle\right) = |i_{\pi(1)}\rangle\otimes\cdots\otimes |i_{\pi(n)}\rangle.
\end{equation}
We can equivalently think of $U_\pi$ as a geometric symmetry operator acting on $n$ copies of $\scri^+$ by the diffeomorphism which permutes them.
Momentarily dropping the $\cdots$ in \eqref{eq:rhon1}, we find
\begin{equation}\label{eq:rhon2}
	\rho^{(n)} = \sum_{\pi\in \Sym(n)} U_\pi\, \rho_\text{Hawking}^{\otimes n}\propto P_\text{Sym}\, \rho_\text{Hawking}^{\otimes n},
\end{equation}
where $P_\text{Sym} = \frac{1}{n!}\sum_{\pi\in S_n} U_\pi$ is a projection onto the completely symmetric subspace that is invariant under all permutations.

We can now ask what our experimentalist should expect when she tries to verify that the radiation is mixed. For a simple concrete example, we take the case of $n=2$ copies and perform the swap test \cite{buhrman2001quantum,Hayden:2007cs}. This means that we simply measure the swap operator $\swap$, which acts to exchange the two copies of the radiation.  In terms of our previous notation, this operator is $\swap=U_\pi$ where $\pi$ is the nontrivial permutation in $\Sym(2)$.  Such measurements have two possible outcomes $\pm 1$ corresponding to the eigenvalues of $\swap$. For swap measurements performed on two uncorrelated copies of a single (normalised) density matrix $\rho$, the expectation value of such outcomes is
\begin{equation}\label{eq:swapTest}
	\Tr(\swap \rho\otimes \rho) = \Tr(\rho^2) = e^{-S_2(\rho)}.
\end{equation}
The quantity \eqref{eq:swapTest} is known as the `purity' of $\rho$, and the last equality relates it to the second R\'enyi entropy $S_2(\rho)$ as defined in \eqref{eq:Renyis}.  For a highly mixed state such as $\rho_\text{Hawking}$ (which has $S_2$ of order $G_N^{-1}$) the expectation value is very close to zero.  It is thus essentially equally likely that the measurement gives $+1$ as $-1$. On the other hand, for pure $\rho$ it is guaranteed to obtain $+1$. We can therefore perform only a handful of measurements and distinguish reliably between the two cases.

Now, from \eqref{eq:rhon2} it is manifest that $\rho^{(2)}$ is invariant under the action of $\swap$. We thus find $\Tr(\swap\rho^{(2)})=\Tr(\rho^{(2)})=1$, and we predict that our experimenter will {\it always} obtain the result $+1$ from measurement of $\swap$. In other words, if
we are inspired by \eqref{eq:swapTest} to summarize her observations by defining the `swap (R\'enyi) entropy'
\begin{equation}
\label{eq:SwapEnt}
S_2^{\text{swap}} : = - \log \Tr \left( \swap \rho^{(2)} \right),
\end{equation}
then this swap entropy vanishes. This can be generalised to an $n$th swap R\'enyi entropy, defined through the the expectation value of a permutation operator acting on $n$ copies of the radiation as
\begin{equation}
\label{eq:SwapEntn}
S_n^{\text{swap}} : = -\frac{1}{n-1} \log \Tr \left( U_\tau \rho^{(n)} \right),
\end{equation}
where $\tau$ is a cyclic permutation of the $n$ copies,\footnote{For $n\neq 2$, $U_\pi$ is not Hermitian, but it can still be measured since it is normal (commutes with its adjoint). This is equivalent to measuring both its Hermitian and anti-Hermitian parts, which are commuting Hermitian operators.}
\begin{equation}\label{eq:tau}
	\tau = (1\,2\,\cdots\,n)\in\Sym(n).
\end{equation}
 We leave the $n$ in the definition of $\tau$ implicit, since it will be clear from context. Once again, from \eqref{eq:rhon2} it is manifest that $\rho^{(n)}$ is invariant under $U_\tau$, so the outcome of such a measurement will always be unity, and $S_n^{\text{swap}}=0$.

More generally, any measurement (of more complicated permutations for example, or even complete tomography to obtain the density matrix) will reproduce the expectations from a pure state, as will be made more manifest in section \ref{sec:BUetc}.

\begin{figure}
\centering
\begin{subfigure}[t]{.48\textwidth}\centering
	\includegraphics[width=.9\textwidth]{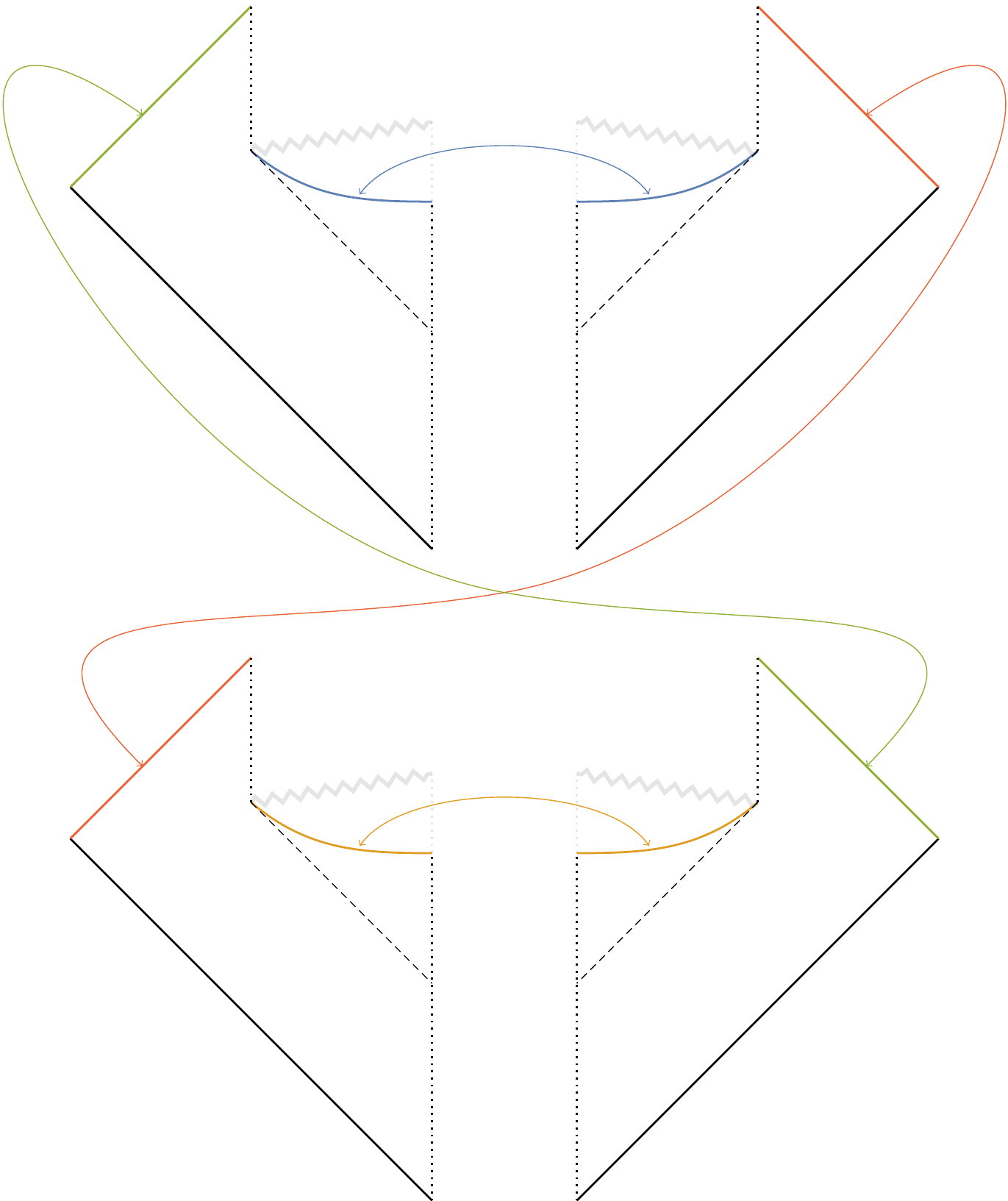}
	\caption{A swap test saddle with na\"ive connections in the bulk.}
\label{fig:PSswapa}
\end{subfigure}
\hfill
\begin{subfigure}[t]{.48\textwidth}\centering
	\includegraphics[width=.9\textwidth]{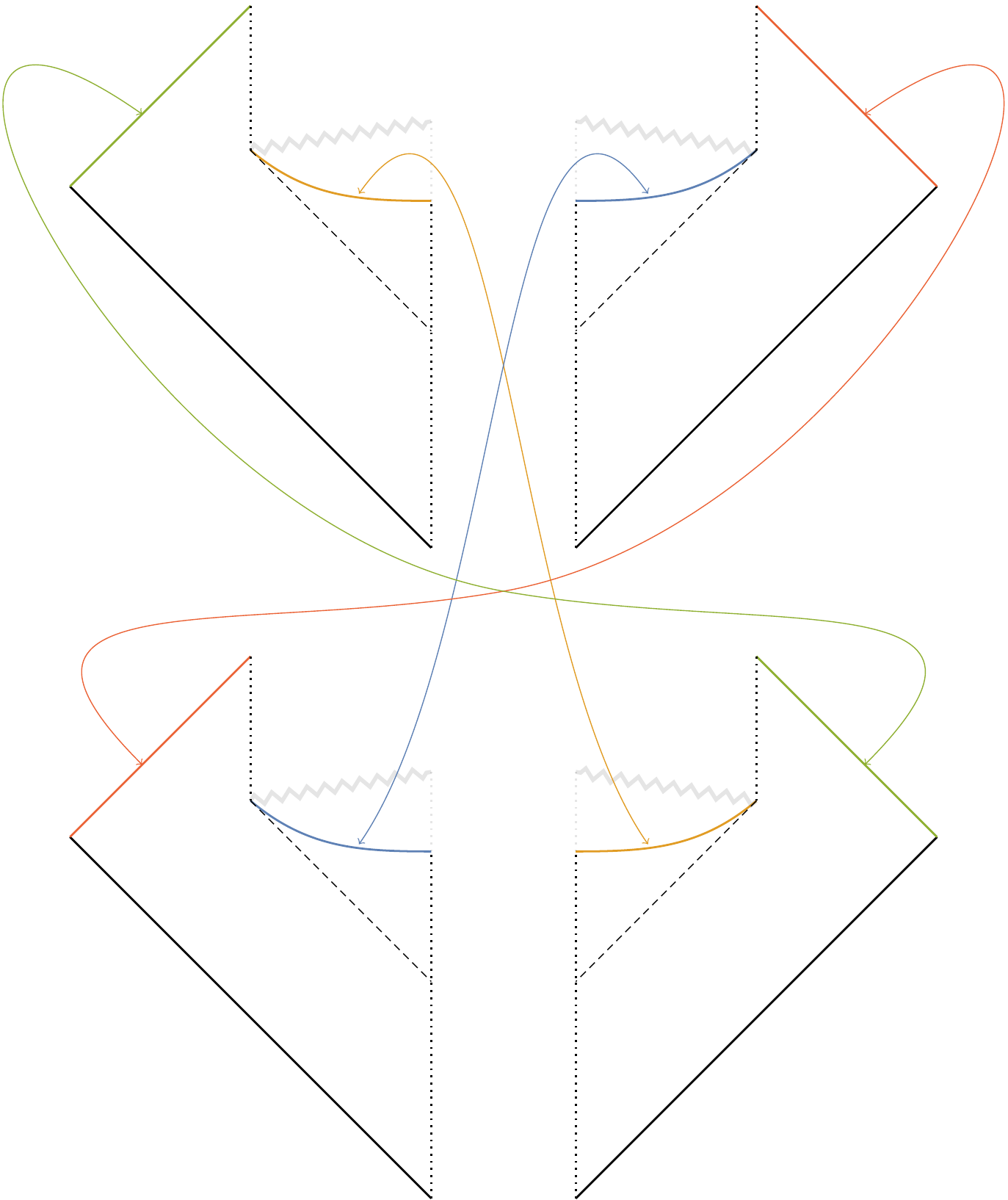}
		\caption{A saddle in which the swap on the boundary is effectively cancelled by an additional dynamical swap in the bulk.}
\label{fig:PSswapb}
\end{subfigure}
\caption{\label{fig:PSswap}}
\end{figure}

For future reference, we note that the expectation value of $\swap$ for radiation collected from two identically-prepared evaporating black holes can be directly formulated as a gravitational path integral. Two saddle points satisfying these boundary conditions are shown in figure \ref{fig:PSswap}.  These are essentially the same saddles pictured in figure \ref{fig:PSrho2}, where the identification of the black hole interiors can be either `unswapped' or `swapped', but now with boundary conditions appropriate to our swap test expectation value. The point is that summing these saddles gives precisely the same result as taking the trace of the saddles in figure \ref{fig:PSrho2} since \ref{fig:PSswapa} is diffeomorphic to the spacetime defined by taking the trace of \ref{fig:PSrho2b} and \ref{fig:PSswapb} is diffeomorphic to that for the trace of \ref{fig:PSrho2a} .
We have attempted to swap the radiation to check whether the state is mixed, but the gravitational path integral has dynamically hidden this from us by performing a matching swap of black hole interiors.

\subsection{Wormholes and factorization}
\label{sec:SW}

As noted above,  the path integral boundary conditions  required to compute the density matrix on $\scri^+$ involves two disconnected copies of $\scri^+ \cup \scri^-$, and thus two disconnected boundaries. So despite the fact that it involves only one density matrix, we may characterize this argument as a `two-replica' calculation.  Furthermore, as discussed above, the Hawking result $\rho_\text{Hawking}$ is obtained from a spacetime wormhole, in the sense that disconnected boundaries become connected through the dynamical bulk.  In the Hawking wormhole this happens due the two copies of figure \ref{fig:Evaporating} being joined along the slice $\Sigma_\mathrm{int}$, which does not reach the asymptotic boundary.  Both features are closely associated with the failure of BH unitarity due to the large entropy of $\rho_{\text{Hawking}}$ on $\scri^+$.

If one believes in BH unitarity, it might thus seem natural to seek a one-replica calculation that describes black hole evaporation.
Rather than concentrating on observables, we might try to compute components of the S-matrix directly, or equivalently the wavefunction of the Hawking radiation at $\scri^+$ for a given initial state at $\scri^-$. For this, we would like to compute the path integral with boundary conditions on a single copy of $\scri^+ \cup \scri^-$.

However,  there is no clear way to compute this path integral semiclassically, even after making assumptions about the endpoint of evaporation $\evap$. We might attempt to proceed by using a single copy of the evaporating black hole geometry of figure \ref{fig:Evaporating}, and then perform the path integral of the quantum fields on this background with appropriate initial and final boundary conditions. But we run into difficulty due to the presence of the future singularity (the jagged line in figure \ref{fig:Evaporating}). First, we do not expect that our low-energy effective theory will be valid in the high-curvature regions near the singularity. Second, there is no obvious prescription for the boundary conditions or measure that we should apply when we integrate over quantum fields at the singularity, and the spacetime we have chosen may not be a stationary point of the action depending on what variations are allowed by the boundary conditions.  This is a more severe problem than the one encountered at the endpoint of evaporation $\evap$ when studying the path integral of figure \ref{fig:PSrho2}, since the current problem affects all of the interior Hawking partners and scales with a positive power of the black hole's  initial size. Resolving this by choosing a prescription to replace the singularity with a boundary condition is equivalent to the black hole final state proposal of \cite{Horowitz:2003he}. We will instead take the more conservative point of view that semiclassical gravity simply does not offer an answer to this question.

On the other hand, we have seen above that the Polchinski-Strominger proposal gives operationally-defined entropies indicating the final state to be pure.  As a result, it is natural to expect whatever physics lies behind this operational purity to also enable calculations of the above S-matrix components.  At least in some sense, it should then cause the `two-replica' Hawking wormhole calculation of $\rho$ to factorize into a product of `one-replica' S-matrices.  Aficionados of the AdS/CFT correspondence will thus recognize that the black hole information problem is a special case of the so-called `factorization problem' of AdS/CFT \cite{Rey:1998yx,Maldacena:2004rf,ArkaniHamed:2007js}.  We shall return to this issue in the discussion of section \ref{sec:discussions}.

\subsection{Experiments on part of the radiation}
\label{sec:PSu}

Section \ref{sec:PSP} discussed predictions for the swap test as applied to the entirety of radiation on $\scri^+$, and found that they are consistent with a pure state. Here, we will generalise this to ask for the predictions of the PS proposal when we measure only the radiation on the part $\scri_u$ of $\scri^+$ to the past of some retarded time $u$. We postpone interpretation of the results to section \ref{sec:Violate}, where (along with other difficulties) we will discover them to be inconsistent with BH unitarity.  Nevertheless, this calculation will be a helpful warm-up for the replica wormholes introduced in section \ref{sec:replicas}.

From the PS proposal \eqref{eq:rhon2}, the expectation value of an operator $\mathcal{O}^{(n)}$ acting on $n$ sets of Hawking radiation is given by
\begin{equation}
\label{eq:PSsumperm}
	\Tr\left(\mathcal{O}^{(n)}\rho^{(n)}\right) = \sum_{\pi \in \Sym(n)}  \Tr\left(  \mathcal{O}^{(n)} U_\pi(\scri^+) \rho_\text{Hawking}^{\otimes n}\right)\; ,
\end{equation}
where we have here used the more explicit notation $U_\pi(\scri^+)$ to include the region $\scri^+$ on which the permutation operator acts. Strictly speaking we should divide by a normalisation factor determined by setting $\mathcal{O}^{(n)} =\id $.  However,
except for the term defined by the identity permutation $\pi = \id$,
all terms in this normalization factor are exponentially small.  Thus  the resulting corrections are negligible.

We will ask for predictions when we measure a swap operator $\swap(\scri_u)$ (or more generally $U_\tau(\scri_u)$ for the cyclic permutation $\tau$ from equation \eqref{eq:tau}), but now acting only on $\scri_u$, capturing the Hawking radiation that emerges before the retarded time $u$. As before, we encode the result in a `swap R\'enyi entropy'
\begin{equation}
\label{eq:swapEntu}
S_n^{\text{swap}}(u) : = -\frac{1}{n-1} \log \Tr \left( U_\tau(\scri_u) \rho^{(n)} \right)
\end{equation}
generalising \eqref{eq:SwapEnt2}.

We begin with the case $n=2$, where there are two terms:
\begin{equation}\label{eq:PSswapu}
	\Tr \left( \swap(\scri_u) \rho^{(2)} \right) = \Tr\left(  \swap(\scri_u) \rho_\text{Hawking}^{\otimes 2}\right)   +\Tr\left(  \swap(\scri_u) \swap(\scri^+) \rho_\text{Hawking}^{\otimes 2}\right).
\end{equation}
The first term is the expectation value of the swap operator $\swap(\scri_u)$ in the tensor product state $\rho_\text{Hawking} \otimes \rho_\text{Hawking}$. From \eqref{eq:swapTest}, this yields $e^{-S^{\text{Hawking}}_2(u)}$, where $S^\text{Hawking}_2(u)$ is the second R\'enyi entropy of the part ($\scri_u$) that is swapped. To understand the contribution of the second term, note that the product of two swap operators is again a swap operator: $\swap(\scri_u) \swap(\scri^+) = \swap(\overline{\scri_u})$, where $\overline{\scri_u}$ is the complement of $\scri_u$ in $\scri^+$.  As a result, the contribution of the second term to $\Tr \left( \swap(\scri_u) \rho^{(2)} \right)$ is of precisely the same form as the first, but with
$S^\text{Hawking}_2(u)$ replaced with the R\'enyi entropy $\bar S_2^\text{Hawking}(u)$ of the radiation on $\overline{\scri_u}$ associated with the Hawking state.  Thus we find
\begin{equation}
\begin{aligned}
	S^\text{swap}_2(u) &\sim -\log \left[ e^{-S_2^\text{Hawking}(u)} + e^{-\bar{S}_2^\text{Hawking}(u)}  \right] \\
	&\sim \min \left\{ S_2^\text{Hawking}(u),\bar{S}_2^\text{Hawking}(u) \right\},\label{eq:PSswapent}
\end{aligned}
\end{equation}
where we may approximate the function as a minimum of the two terms because $S_2^\text{Hawking}(u)$, $\bar{S}_2^\text{Hawking}(u)$ are both very large.
\begin{figure}
\centering
\begin{subfigure}[t]{.48\textwidth}\centering
	\includegraphics[width=.9\textwidth]{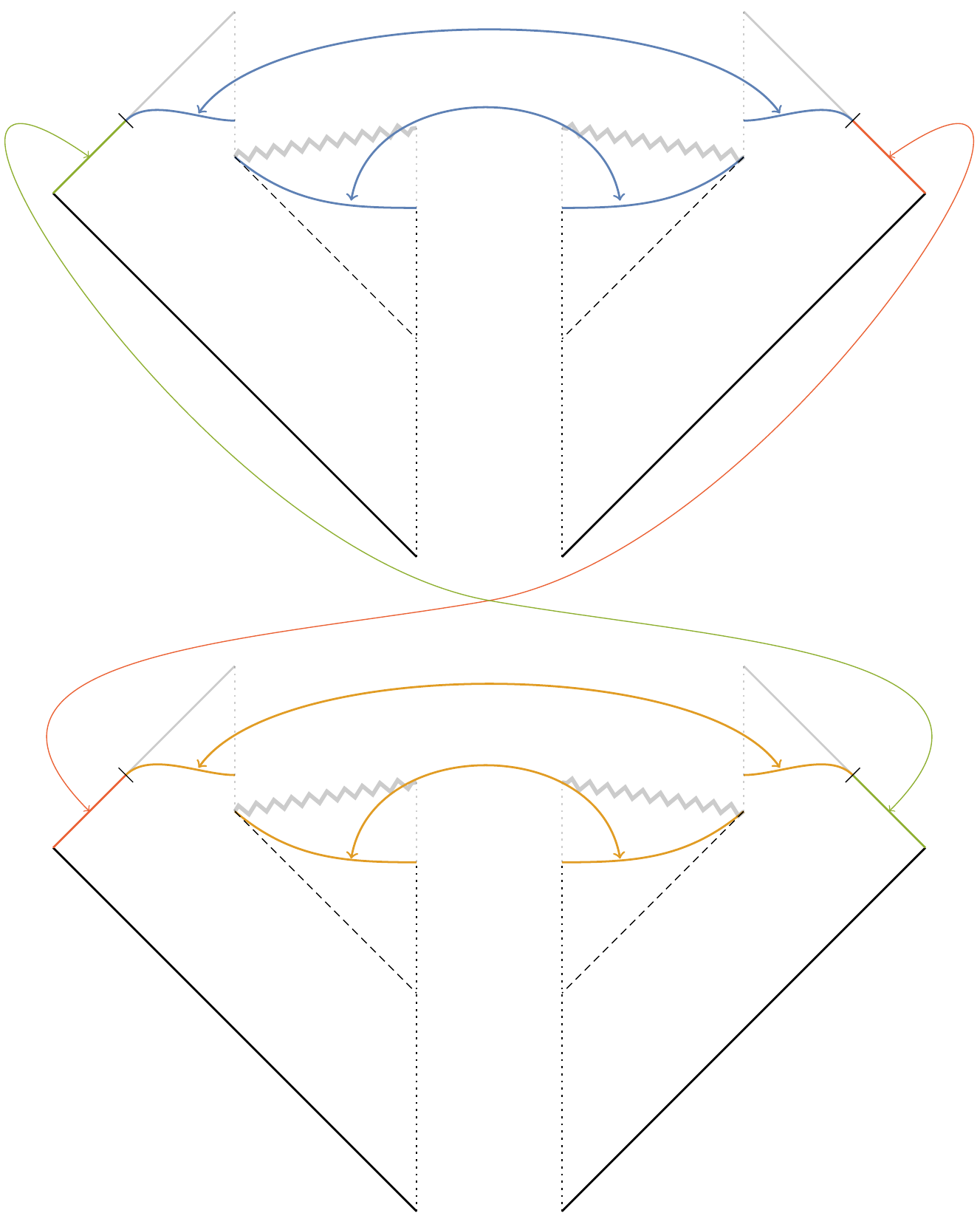}
	\caption{The geometry giving rise to the first term in \eqref{eq:PSswapu} and  \eqref{eq:PSswapent}.}
\label{fig:PSswapua}
\end{subfigure}
\hfill
\begin{subfigure}[t]{.48\textwidth}\centering
	\includegraphics[width=.9\textwidth]{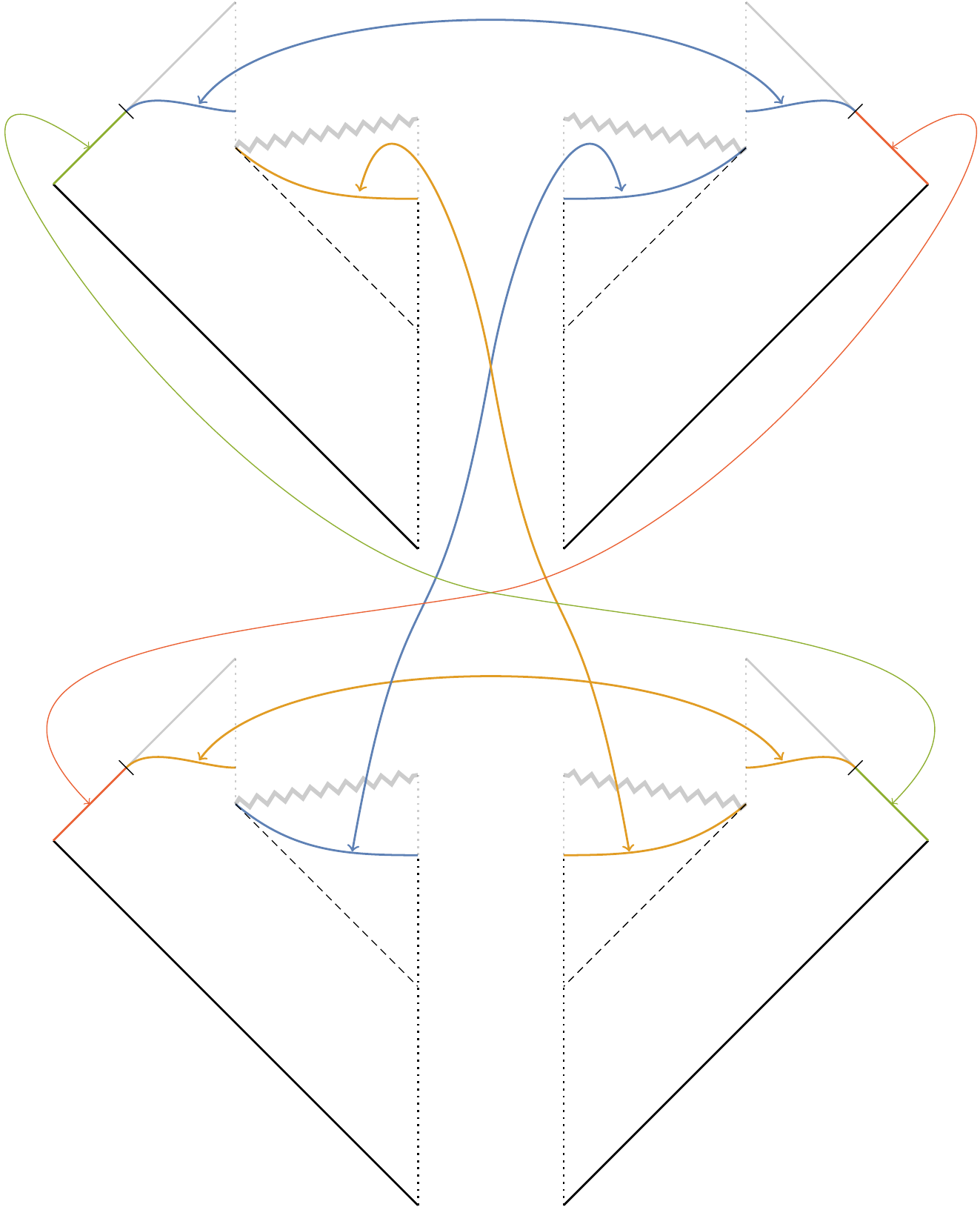}
		\caption{The geometry with swapped interiors gives the second term in \eqref{eq:PSswapu} and  \eqref{eq:PSswapent}.}
\label{fig:PSswapub}
\end{subfigure}
\caption{The two PS wormholes contributing to the computation of $\Tr \left( \swap(\scri_u) \rho^{(2)} \right)$, the expectation value of a swap operator acting on $\scri_u$ for two sets of radiation. \label{fig:PSswapu}}
\end{figure}
The two geometries of the path integral corresponding to the computation \eqref{eq:PSswapent} are shown in \ref{fig:PSswapu}. The minimum in \eqref{eq:PSswapent} comes from choosing only the dominant saddle.

Generalising this to cyclic permutation on $n$ sets of radiation, there are $n!$ terms, but only two terms are important, the identity permutation $\id$ and the inverse $\tau^{-1}$ of the cyclic permutation we are measuring. Other PS wormholes are exponentially suppressed relative to at least one of the two included terms.\footnote{There is an exception when both terms are comparable $S_n^\text{Hawking}(u)\approx \bar{S}_n^\text{Hawking}(u)$, in which case additional permutations give further interesting corrections: see footnote \ref{foot:transitions}.} In analogy with \eqref{eq:PSswapent}, the two terms give
\begin{equation}
	S^\text{swap}_n(u) \sim \min \left\{ S_n^\text{Hawking}(u),\bar{S}_n^\text{Hawking}(u) \right\},\label{eq:PSswapent2}
\end{equation}
with the second coming from the relation $U_{\tau}(\scri_u) U_{\tau^{-1}}(\scri^+) = U_{\tau^{-1}}(\overline{\scri_u})$.

\subsection{Challenges for the Polchinski-Strominger proposal}
\label{sec:PSchallenges}

The observations of section \ref{sec:PSP} may suggest that the semiclassical gravity predictions for an asymptotic observer always conspire to produce results consistent with BH unitarity.  However, if the only relevant contributions from the path integral are those discussed above, with further consideration one still finds serious problems.

These problems are described below.
Using arguments related to the problem that will be described in \ref{sec:Violate}, \cite{Polchinski:1994zs} concluded in their context that black holes in fact violate BH unitarity and instead described black holes as `long-lived remnants'.\footnote{Here the term `remnant' means an object with unbounded entropy (that is, infinitely many internal states) below a fixed mass.}
These difficulties will all be resolved in section \ref{sec:replicas} by appealing to the recently-discovered replica wormholes of \cite{Almheiri:2019qdq,Penington:2019kki}. Nonetheless, we will first discuss the issues in more detail so we can better appreciate this resolution.

\subsubsection{What happens at the endpoint of evaporation $\evap$?}\label{sec:endpoint}

As observed above, we lose semiclassical control near the endpoint of evaporation $\evap$ once the black hole is of Planckian size.
We have thus far followed \cite{Polchinski:1994zs} in making the PS assumption, but it would be a great improvement if we were able to arrive at the same conclusions without such assumptions, and with the semiclassical approximation justified throughout the calculation.

\subsubsection{Violations of BH Unitarity}
\label{sec:Violate}

We now discuss the result of section \ref{sec:PSu}, where we computed the expectation value of a cyclic permutation acting on the radiation arriving at $\scri^+$ before retarded time $u$. Since von Neumann entropies are more familiar and more physical than R\'enyis, we will phrase the calculation in terms of the `swap von Neumann entropy' $S^\text{swap}(u)$ obtained by formally taking the $n\to 1$ limit of \eqref{eq:PSswapent2},
\begin{equation}\label{eq:PSswapvN}
	S^\text{swap}(u) \sim \min \left\{ S^\text{Hawking}(u),\bar{S}^\text{Hawking}(u) \right\}.
\end{equation}
However, the same considerations apply directly to R\'enyi entropies as well. We interpret $S^\text{swap}(u)$ as a prediction for the von Neumann entropy that an asymptotic observer would deduce by performing measurements on many copies of the Hawking radiation emitted before time $u$.

To understand these quantities, we must simply note that Hawking's state does not contain significant long-range correlations, so can be regarded as a product of uncorrelated thermal states emitted at different times. This means that $S^\text{Hawking}(u)$ and $\bar{S}^\text{Hawking}(u)$ are well-approximated by the thermal entropy of Hawking radiation emitted before and after the time $u$ respectively. In particular, the sum $S^\text{Hawking}(u) + \bar S^\text{Hawking}(u)$ gives the total entropy $S^\text{Hawking}(\infty)$ of all radiation at $\scri^+$ in the Hawking saddle, up to order one corrections from the vicinity of the boundary between $\scri_u$ and $\overline{\scri_u}$. In particular, $S^\text{Hawking}(u)$ monotonically increases from zero to $S^\text{Hawking}(\infty)$, while $\bar{S}^\text{Hawking}(u)$ monotonically decreases between the same values.

At early times we have $S^\text{Hawking}(u) < \bar{S}^\text{Hawking}(u)$, so \eqref{eq:PSswapvN} is dominated by the first term, corresponding to the saddle-point in figure \ref{fig:PSswapua}.  The swap entropy $S^\text{swap}(u)$ thus increases until $S^\text{Hawking}(u) = \bar S^\text{Hawking}(u)$, at which point there is a first order phase  transition, the second saddle-point in figure \ref{fig:PSswapub} becomes dominant, and $S^\text{swap}(u)$ decreases back to zero. While this is qualitatively very much like the Page curve in figure \ref{fig:Page}, it disagrees quantitatively and we find a result which is incompatible with BH unitarity. The key point is that the entropy $S^\text{Hawking}(\infty)$ on $\scri^+$ in the Hawking saddle exceeds the Bekenstein-Hawking entropy $S_{\text{BH}}$ of the initial black hole by a factor of order one.  This discrepancy occurs because black hole evaporation is thermodynamically irreversible and hence produces thermal entropy; the generalized second law is not saturated by evaporation in the Hawking saddle.

\begin{figure}
	\centering
	\includegraphics[width=.6\textwidth]{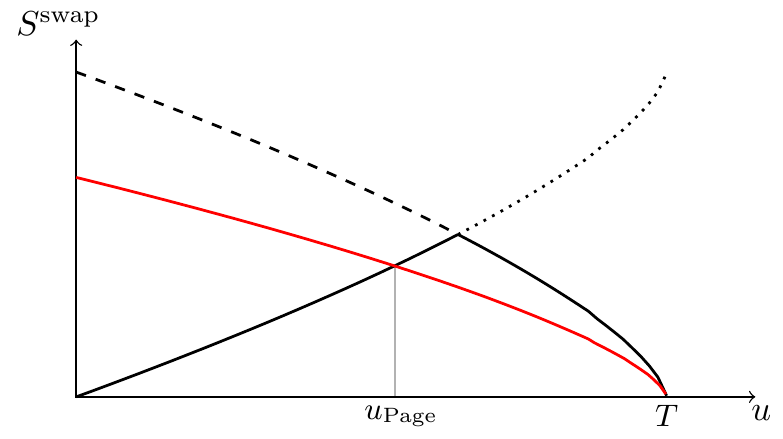}
	\caption{The $u$-dependent swap entropy computed from PS saddles (solid black curve) rapidly transitions from agreement at small $u$  with $S^\text{Hawking}(u)$ (increasing black curve) to agreement at large $u$ with $\bar S^\text{Hawking}(u)$ (decreasing black curve).
The transition occurs at the $u_0$ for which $S^\text{Hawking}(u_0) = \bar S^\text{Hawking}(u_0)$.  However, because $S^\text{Hawking}(\infty) > S_\text{BH}(0)$, we have  $S^\text{Hawking}(0) > S_\text{BH}(0)$ and in fact also at all $u$.  In particular, $S^\text{Hawking}(u_0) = \bar S^\text{Hawking}(u_0)$ exceeds $S_\text{BH}(u_0)$, violating BH unitarity. \label{fig:splot}}
\end{figure}

We can be very explicit for Schwarzschild black holes evaporating by production of massless particles, since the various entropies are determined as a function of time by dimensional analysis. In particular, the production of Hawking radiation is determined by the geometry, which provides the only length scale $R$. The emitted power (energy per unit time) is therefore proportional to $R^{-2}$, and thermal entropy is produced at a rate per unit time proportional to $R^{-1}$. Meanwhile, in $D$ spacetime dimensions the black hole mass $M$ is proportional to $G^{-1}R^{D-3}$, and $S_\text{BH}$ is proportional to $G^{-1}R^{D-2}$. From this, we can solve everything up to a few unknown dimensionless constants to find\footnote{We can define $S_\mathrm{BH}(u)$ as the Bekenstein-Hawking of a black hole with mass given by the Bondi mass at time $u$. Equivalently, this is entropy of the black hole when the radiation arriving at $\scri^+$ at time $u$ was emitted, where the precise definition of emission time is not important since the evaporation timescale $u_\evap$ is a positive power of $G_N^{-1}$.}
\begin{align*}
	S_\text{BH}(u) &= S_\text{BH}(0) \left(1-\frac{u}{u_\evap}\right)^\frac{D-2}{D-1}\,, \\
	S^\text{Hawking}(u) &= S^\text{Hawking}(\infty)\left[1- \left(1-\frac{u}{u_\evap}\right)^\frac{D-2}{D-1}\right]\,, \\
	\bar S^\text{Hawking}(u) &= S^\text{Hawking}(\infty)\left(1-\frac{u}{u_\evap}\right)^\frac{D-2}{D-1}\,,
\end{align*}
where $D$ is the spacetime dimension and $u_\evap$ is the time taken for complete evaporation.  The only undetermined parameter relevant for us is the (constant) ratio
\begin{equation}
	r = \frac{\bar S^\text{Hawking}}{S_\text{BH}} =  \frac{d S^\text{Hawking}}{du}\Big/ \left|\frac{d S_\text{BH}}{du}\right| =  \frac{S^\text{Hawking}(\infty)}{S_\text{BH}(0)} >1,
\end{equation}
which depends in detail on the dynamics through greybody factors. However, the only point that is important for us is that it is greater than one.  Indeed, as shown in figure \ref{fig:splot} one finds a violation of BH unitarity by a factor of $r$.  For four-dimensional black holes, Page has computed the corresponding ratio for von Neumann entropies in various cases \cite{Page:1976df,Page:2013dx}; for example, for Schwarzschild black holes radiating by emission of gravitons and photons he computed $r\approx 1.48$.\footnote{The total entropy of Hawking radiation $S^\text{Hawking}(\infty)$ depends on details of the endpoint of evaporation beyond semiclassical physics. We can safely ignore these details, since the effect on the entropy does not (by our PS assumption) scale with the original size of the black hole.}

The above paragraphs describe a problematic violation of entropy bounds, but only by an order one ratio. However, as is familiar from other discussions, we can magnify the problem by refusing to let the black hole evaporate freely and instead feeding it with matter so that it remains at a given size for as long as we desire (perhaps even eternally as in \cite{Almheiri:2019yqk}). If this time is very long, then in the middle of this period $S_2^\text{Hawking}(u)$ and $\bar S_2^\text{Hawking}(u)$ will both become very large, so $S^\text{swap}_2(u)$ is also very large.  But the Bekenstein-Hawking entropy $S_\text{BH}$ is fixed by the current mass of the black hole.  So from this analysis it would appear that black holes have an unbounded number of internal states below any given mass, a serious failure of BH unitarity.

\subsubsection{Violations of causality}
\label{sec:PSCausality}

Perhaps an even greater problem than the failure of BH unitarity is the observation that  \eqref{eq:PSswapent} entails a possible violation of causality.  In particular, since it involves the entropy $\bar S_2^\text{Hawking}(u)$ on $\overline{\scri_u}$, it predicts the swap entropy measured by our experimenter at a finite time $u$ to depend on the entire future of the black hole!  This is particularly sharp if we imagine first performing this measurement at a finite distance from the black hole (or at an AdS boundary), whence we can subsequently throw matter into the black hole depending on the swap entropy obtained.  Such violations of causality appear large enough to even throw the consistency of above calculations into doubt. We take this to suggest that a consistent framework will require additional corrections to the swap entropy at finite $u$; such further corrections will be explored in the next section.

\section{Replica wormholes}
\label{sec:replicas}

It is natural to ask if the above challenges might be resolved by finding further new saddles. Similar ideas have been investigated in various forms for many years; see e.g.\ \cite{Hawking:1987mz,Giddings:1987cg,Hawking:1988ae,Maldacena:2001kr,Hawking:2005kf}. We are now able to make this more concrete, since in the past year a new class of saddles  have been argued to exist. These are known as replica wormholes for reasons that will shortly become clear. They were discovered as contributions to path integrals of the form studied in section \ref{sec:PSu} above, in our context giving the expectation value of the cyclic permutation operators $U_\tau(\scri_u)$ acting on $n$ copies of a subset of Hawking radiation. As we review below, the replica wormholes reproduce the expectations from the Page curve quantitatively, via a path integral over spacetimes where the semiclassical approximation can be trusted everywhere. This implies that the replica wormhole geometries must also contribute to other observables, and in general to the components of the $n$-evaporation density matrix $\rho^{(n)}$, which we explore in section \ref{sec:RWother}.

\subsection{Replica wormhole spacetimes}
\label{sec:repQES}

In a sense, replica wormholes are a generalisation of PS wormholes studied in section \ref{sec:PS}, so we first revisit these in a way that is suggestive of the required generalisation. Specifically, we will reconsider the swap entropies of the Hawking radiation that emerges before some finite retarded time $u$, as discussed in section \ref{sec:PSu}. Recall that this is an expression for the expectation value of the cyclic permutation $U_\tau(\scri_u)$ applied to $n$ copies of Hawking radiation on $\scri_u$.

The PS wormholes for this amplitude (pictured in figure \ref{fig:PSswapub} for $n=2$) are built from $2n$ pieces, consisting of $n$ `ket' replicas $\mathcal{M}_r$ of the evaporating black hole spacetime and $n$ conjugate `bra' replicas $\bar{\mathcal{M}}_r$, labelled by a replica index $r=1,\ldots,n$. These spacetimes terminate at a future Cauchy surface $\Sigma$ where they are sewn together. The surface $\Sigma$ is divided into three pieces, with a different rule for sewing replicas along each piece. First, we have a region $\scri_u$ on $\scri^+$, where the boundary conditions require us to join spacetimes with the cyclic permutation $\tau$, so $\mathcal{M}_r$ joins to $\bar{\mathcal{M}}_{\tau(r)}$. Next, we have an exterior piece $\Sigma_{\mathrm{ext}}$ stretching from retarded time $u$ on $\scri^+$ to the regular origin $r=0$ that is expected to emerge after the final evaporation of the black hole (we can take $\Sigma_{\mathrm{ext}}=\scri^+$ if we like). In this region, we sew without permutation, so $\mathcal{M}_r$ joins to $\bar{\mathcal{M}}_r$; this is also fixed by the boundary conditions on $\scri^+$, which require such an identification in a neighborhood to the future of retarded time $u$, where $\Sigma_\mathrm{ext}$ begins. Finally, we have a Cauchy surface for the black hole interior $\Sigma_{\mathrm{int}}$, reaching from the original regular origin (before the black hole evaporates) to the evaporation endpoint $\evap$. Here, the boundary conditions do not uniquely specify any sewing rule, and we can join $\mathcal{M}_r$ to $\bar{\mathcal{M}}_{r'}$ along $\scri_\mathrm{int}$ with any choice of permutation we desire. The path integral includes a sum over all possibilities, and the dominant permutation for a given calculation is dynamically determined. In particular, the interesting new contribution to the swap entropies \eqref{eq:PSswapent2} arose from choosing the permutation on $\Sigma_\mathrm{int}$ to match the permutation $\tau$ on $\scri_u$ imposed by the boundary conditions.

\begin{figure}
\centering
\begin{subfigure}[t]{.49\textwidth}\centering
	\includegraphics[width=.8\textwidth]{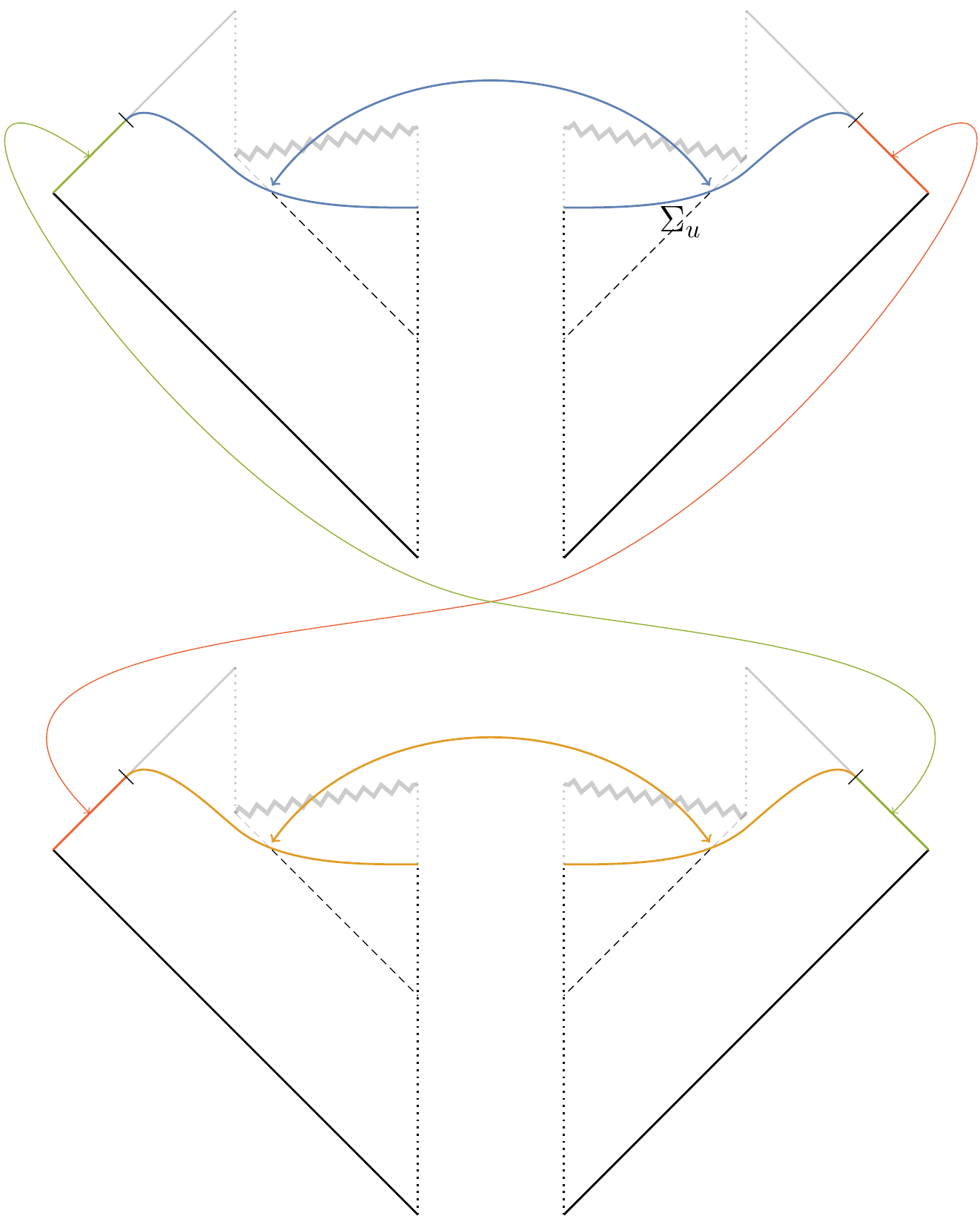}
	\caption{The `trivial' geometry with the boundary conditions appropriate for the swap operator acting on $\scri_u$. This arises from two copies of the Hawking wormhole in figure \ref{fig:rhoEvaporatingSC}.}
\label{fig:replicarho2a}
\end{subfigure}
\hfill
\begin{subfigure}[t]{.49\textwidth}\centering
	\includegraphics[width=.8\textwidth]{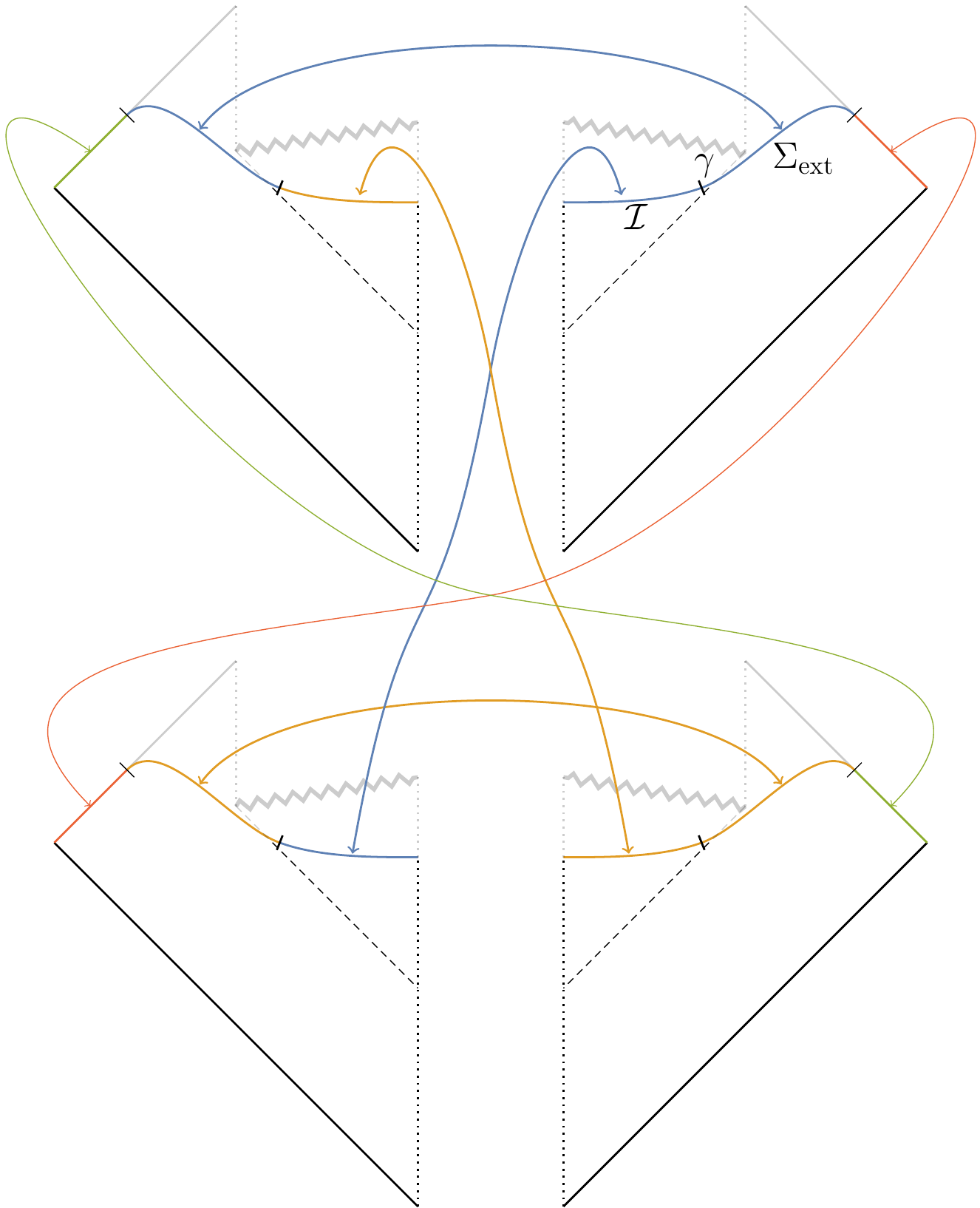}
		\caption{A replica wormhole geometry with the same boundary conditions, obtained from (a) by changing the identifications along the `island' $\island$.}
\label{fig:replicarho2b}
\end{subfigure}
\caption{Two geometries in the path integral contributing to the expectation value of the swap operator acting on $\scri_u$ for two sets of Hawking radiation. The right spacetime is an $n=2$ replica wormhole. To obtain this, we divide the partial Cauchy surface $\Sigma_u$ into two pieces $\island$ and $\Sigma_\mathrm{ext}$ along the codimension-2 surface $\gamma$.  The connections along $\Sigma_\mathrm{ext}$ are the same as in (a), but are swapped along the island $\island$. The configuration shown is not a saddle as it does not incorporate back-reaction from the structure near the special surface $\gamma$, and incorporating such back-reaction will make the spacetime metric complex. That is, passing the contour of integration through the desired saddle requires deforming it away from real Lorentzian metrics.  However, the replica wormhole saddle will coincide with the (real) Hawking saddle in the formal limit $n\rightarrow 1$ of the replica number $n$.\label{fig:replicarho2}}
\end{figure}

This description also applies to replica wormholes, but generalised to allow a more general choice of Cauchy surface $\Sigma$ where we sew the replicas, and to also allow a more general splitting of this surface into pieces. The region $\scri_u$ is fixed by the boundary conditions, so must remain unchanged, but we are free to choose how the remainder $\Sigma_u$ of the Cauchy surface is split into two pieces: a partial Cauchy surface $\island$ (the `island') generalising $\Sigma_\mathrm{int}$ in the discussion above, and its complement in $\Sigma_u$ which we continue to call $\Sigma_\mathrm{ext}$. The exterior surface $\Sigma_{\mathrm{ext}}$ extends to meet $\scri^+$ at retarded time $u$, where the boundary conditions specify that bra and ket spacetimes are connected in the trivial way, but we sew along the interior island pieces $\island$ with a nontrivial permutation. For the boundary conditions computing the expectation value of $U_\tau(\scri_u)$, the most interesting possibility again arises when we take the sewing permutation on $\island$ to match the cyclic permutation $\tau$ which acts on $\scri_u$. The novelty of the replica wormholes is that we take the Cauchy surface $\Sigma_u$ to be connected, so that $\island$ and $\Sigma_{\mathrm{ext}}$ meet at a common codimension-2 boundary $\gamma = \partial\island$. Indeed, for Lorentz-signature spacetimes of this form, the causal structure must have an interesting singularity:  points on $\gamma$ will have several past light cones, one for each bra spacetime that meets at $\gamma$ (and also one for each ket spacetime).  This is an important feature, but we will treat it only briefly below, referring the reader to \cite{Louko:1995jw}, \cite{Dong:2016hjy}, and \cite{CDMRW} for further details and deferring discussion of further implications to section \ref{sec:SingCausal}. The resulting spacetime is depicted in figure \ref{fig:replicarho2} for $n=2$.

We can already see why such spacetimes might avoid the PS-wormhole's dependence on physics near $\evap$ and the resulting loss of semiclassical control discussed in section \ref{sec:endpoint}.  By joining replicas along a Cauchy slice $\Sigma_u$ which stays far from regions of strong curvature, the entire singularity --- and in particular the endpoint $\evap$ --- is excluded from the spacetime under consideration, just as for the Hawking wormhole in figure \ref{fig:rhoEvaporatingSC}. We will see that such replica wormholes exist for all times $u<u_\evap$ lying to the past of the future lightcone of $\evap$ (after the black hole forms), and thus remove the dependence on UV physics until the black hole reaches Planckian dimensions.

The matter path integral in this replica wormhole spacetime is a Schwinger-Keldysh path integral on an $n$-sheeted spacetime which includes the insertion of a permutation operator $U_\tau(\island)$ acting on the island $\island$, as well as the operator $U_\tau(\scri_u)$ imposed by the boundary conditions. In principle, we should compute this for every such replica wormhole spacetime, in particular for all choices of $\island$, and then perform the integral over metrics. Different choices of $\gamma$ in a given single-sheeted spacetime result in different geometries for the $n$-sheeted whole, so our gravitational path integral integrates over all inequivalent choices of $\Sigma_\mathrm{int}$ (and we might also sum over nontrivial permutations $\pi$ acting on the island).  If saddle-points exist, the location of $\gamma$ in the resulting geometry will be determined dynamically by extremizing an appropriate action.

\subsection{Quantum extremal surfaces}\label{sec:QES}

The interesting question now is whether this replica wormhole topology can yield a new semiclassical saddle for given boundary conditions at some replica number $n$.  The general case for integer replica number $n >1$ is still under exploration.\footnote{See \cite{Faulkner:2013yia,Hartman:2013mia,Lewkowycz:2013nqa,Penington:2019kki,Mirbabayi:2020fyk} for related constructions in Euclidean signature and \cite{CDMRW} for saddles with Lorentz-signature boundary conditions analogous to those considered here.}  However, we are able to make more progress by considering a formal analytic continuation of the calculation to non-integer $n$, studying the problem for $n-1 \to 0^+$ to first order in $(n-1)$. This will not only be convenient, but also physically interesting, since the corresponding limit of R\'enyi entropies gives the von Neumann entropy. Specifically, we will first compute the same observables as section \ref{sec:PSu}, studying the path integrals with boundary conditions appropriate for computing the expectation value of a cyclic permutation $\tau$ acting on $n$ copies of the radiation emitted before retarded time $u$, encoded in the `swap entropy'
\begin{equation}
\label{eq:SwapEnt2}
S_n^{\text{swap}}(u) : = -\frac{1}{n-1} \log \Tr \left( U_\tau(\scri_u) \rho^{(n)} \right).
\end{equation}
Continuing this to non-integer $n$ and taking the $n\to 1$ limit defines the `swap (von Neumann) entropy' $S^{\text{swap}}(u) := \lim_{n\to 1} S_n^{\text{swap}}(u)$.

In section \ref{sec:PSu}, we found a new interesting contribution to this path integral arising when we chose to join the replicas along the black hole interiors $\Sigma_\mathrm{int}$ by the same permutation $\tau$ as we apply on $\scri_u$. Our strategy will be to emulate this for replica wormholes as described above, replacing $\Sigma_\mathrm{int}$ by a general partial Cauchy surface $\island$. We will reformulate the calculation of the path integral on such geometries in such a way that $n$ need not be an integer. For $n=1$ exactly the permutation group $\Sym(n=1)$ is trivial and there is only the original saddle for $\Tr\rho(u)$ that computes the normalization of the state. Nonetheless, by continuing the problem to study a neighbourhood of $n=1$  we introduce nontrivial dependence on the choice of $\island$, but can still state the calculation in terms of the $n=1$ geometry and associated matter state. As pointed out in \cite{Almheiri:2019qdq,Penington:2019kki}, the condition for a saddle to exist at order $(n-1)$ was found some time ago: see \cite{Dong:2017xht}, building on \cite{Faulkner:2013ana,Dong:2016hjy}.  The condition is that the splitting surface $\gamma=\partial\island$ is a quantum extremal surface (QES) \cite{Engelhardt:2014gca}. See also \cite{CDMRW} for discussion of saddles  for real-time path integrals when $n-1$ is not infinitesimal.

Before reviewing the argument, we recall the definition of a QES. This is a `quantum version' of an extremal surface, which is a stationary point of the area functional $A[\gamma]$. To go from `classical' to `quantum' extremal surface, we simply replace the area function with a quantum corrected version, the generalised entropy:\footnote{We have written $S_\mathrm{gen}$ as a functional of the partial Cauchy surface $\island$ (up to equivalence under changes that leave the domain of dependence invariant), rather than its bounding surface $\gamma$. These data are equivalent unless our spacetime includes a closed universe component (that is, a partial Cauchy slice with empty boundary).}
\begin{equation}\label{eq:Sgen}
	S_\mathrm{gen}(\island;u) = \frac{A[\partial\island]}{4G_N} + S_\mathrm{matter}(\island\cup\scri_u).
\end{equation}
Since the matter fields are pure on a full Cauchy surface, the second term is also the matter entropy on the partial Cauchy surface $\Sigma_\mathrm{ext}$ bounded by $\gamma$ and by $\scri^+$ at retarded time $u$.  This is the more standard way of describing a generalized entropy.
The final argument $u$ in $S_\mathrm{gen}$ reminds the reader that this matter entropy term depends on where we choose this partial Cauchy surface to meet $\scri^+$.

The definition of a QES $\gamma=\partial\island$ is that $S_\mathrm{gen}$ is stationary to first order variations of $\gamma$. In the definition \eqref{eq:Sgen}, $S_\mathrm{matter}(\island\cup\scri_u)$ is the von Neumann entropy of matter fields on $\island\cup\scri_u$ in the state under consideration. For a matter QFT, this entropy is divergent and depends on the choice of UV cutoff. Nevertheless, there is strong evidence  \cite{Susskind:1994sm,Jacobson:1994iw,Frolov:1996aj} (see also the appendix of \cite{Bousso:2015mna}) that the combination $S_\mathrm{gen}$ is finite and not UV sensitive, since matter fields give an equal and opposite infinite renormalisation to $G_N^{-1}$ (using the `bare' value of $G_N$ at the EFT cutoff in \eqref{eq:Sgen}). Relatedly, if the theory has higher derivative terms or non-minimal couplings to gravity (perhaps induced by quantum effects) then the $\frac{A[\gamma]}{4G_N}$ term should be replaced by the corresponding notion of geometric entropy \cite{Dong:2013qoa,Camps:2013zua,Miao:2014nxa}. These features are not special to replica wormholes, but are familiar from quantum corrections to black hole thermodynamics, for example. Operationally, it suffices to evaluate \eqref{eq:Sgen} using the finite IR value of $G_N$ and a finite subtracted expression for $S_\mathrm{matter}$ using some convenient regulator which is local at $\gamma$.

We now sketch how the generalised entropy functional and the QES prescription arise from the path integral, by looking for replica wormhole saddle-points with boundary conditions for computing $\Tr \left( U_\tau(\scri_u) \rho^{(n)} \right)$.\footnote{Here we depart from the historical presentation of the arguments in order to simplify the discussion.} These spacetimes are \emph{replica symmetric}: that is, the geometry respects the $n$-fold cyclic symmetry possessed by the boundary conditions, as well as the two-fold symmetry swapping `bra' and `ket' branches of the Schwinger-Keldysh contour. We are thus considering metrics that are obtained by taking $2n$ replicas of a single spacetime ($n$ of them after CPT conjugation) to the past of a Cauchy surface $\island\cup \Sigma_\mathrm{ext}\cup\scri_u$, and gluing them along that Cauchy surface, though on the $\island$ and $\scri_u$ portions of this Cauchy surface we perform this gluing using a cyclic permutation $\tau$ between replicas. It suffices to check that we have a saddle-point varying only amongst such replica symmetric configurations, since the symmetry ensures stationarity to variations which break this symmetry. This will enable us to describe the problem in terms of a single copy.

First, we compute the matter path integral on such a geometry. As noted above, the replica wormhole geometry is the Schwinger-Keldysh contour giving the expectation value of $U_\tau(\scri_u\cup \island)$ for the matter state on the final Cauchy surface $\Sigma = \island \cup \Sigma_\mathrm{ext}\cup\scri_u$ produced by unitary evolution from the initial conditions on $\scri^-$. We can express this in terms of the R\'enyi entropy of the matter reduced density matrix $\rho_{\island\cup\scri_u}$. This is much the same as the discussion in section \ref{sec:HawkingEnt}, except we now are computing the R\'enyi entropy on $\island\cup\scri_u$, not just on $\scri_u$. We can thus write the $n$-replica matter path integral as
\begin{equation}\label{eq:Znmatter}
	Z^{(n)}_\text{matter} = \Tr(\rho_{\island\cup\scri_u}^n) = \left(Z^{(1)}_\text{matter}\right)^n e^{-(n-1)S_n(\island\cup\scri_u)}.
\end{equation}
The factor $\left(Z^{(1)}_\text{matter}\right)^n =\Tr(\rho_{\island\cup\scri_u})^n$ gives the normalisation of the state on the unreplicated geometry, and is independent of $\gamma$. The matter effective action is thus given by $n$ times the $n=1$ effective action, plus a term from the R\'enyi entropy on $\island\cup\scri^+$:
\begin{equation}
	\log Z^{(n)}_\text{matter} = n \log Z^{(1)}_\text{matter}  - (n-1)S_n(\island\cup\scri_u) .
\end{equation}
This extra term will become the matter von Neumann entropy $S_\mathrm{matter}(\island\cup\scri^+)$ appearing in the generalised entropy \eqref{eq:Sgen} when we continue this close to $n=1$. In particular, since the R\'enyi entropy is defined for any $n\geq 1$, we have succeeded in describing the matter integral in such a way that $n$ is not restricted to be an integer. We now do the same for the integral over replica-symmetric metrics.\footnote{What we have called the matter path integral should include linearised metric fluctuations as explained after equation \ref{eq:Ieff}, here computing the entropy of gravitons. In particular, the path integral thus incorporates small deviations from replica-symmetric metrics. In practice, this is rather subtle, but the subtleties are local at $\gamma$ and so do not accumulate to become significant.}

Since the Einstein-Hilbert action is local, it is tempting to say that the action on our $n$ replicas is simply $n$ times the action on a single copy. This is almost true, but as described in \cite{Dong:2016hjy} (following similar Euclidean observations in \cite{Lewkowycz:2013nqa}) there is an additional local contribution at the surface $\gamma$. To understand this, it is helpful to imagine deforming the Schwinger-Keldysh contour to pass through our final Cauchy surface in an imaginary time direction, so that we can think of the geometry as being Euclidean (at least locally, very close to $\gamma$).
\begin{figure}
\centering
	\includegraphics[width=.8\textwidth]{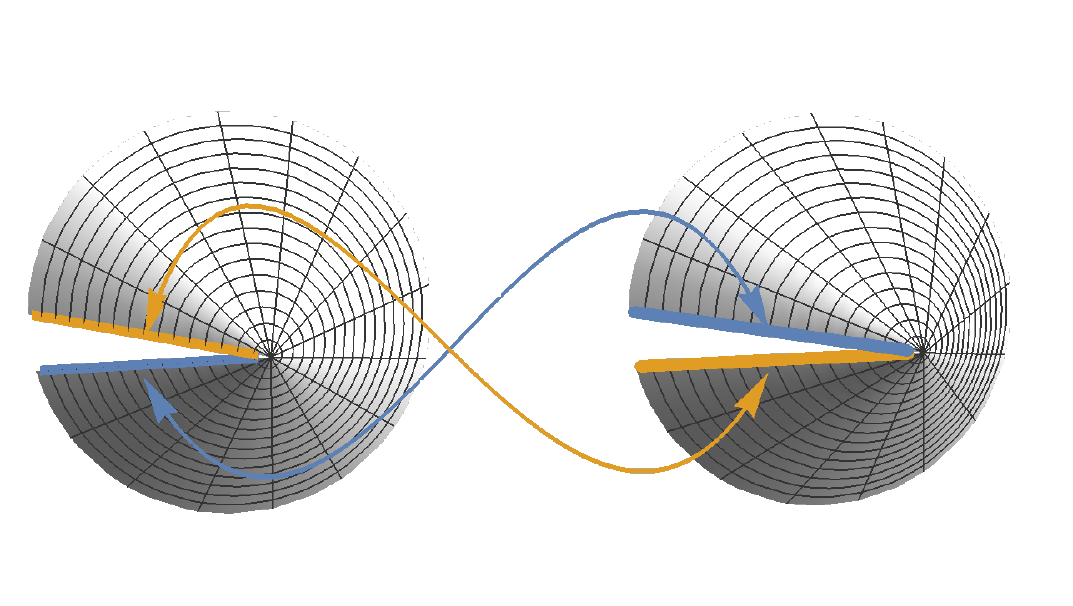}
\caption{A sketch of the $n=2$ replica wormhole geometry near the splitting surface $\gamma$, which lies at the tip of the cones. Comparing to \ref{fig:replicarho2b}, one cone corresponds to the top two spacetimes (one `ket' and one `bra') and the other to the bottom two, identified along $\Sigma_\mathrm{ext}$, leaving a cut in the cone along $\island$. The two cones are joined along $\island$ with a swap (or more generally, a cyclic permutation) as indicated by the colours and arrows. The conical defect in each replica is required for the resulting replica wormhole spacetime to be smooth at $\gamma$. \label{fig:cones}}
\end{figure}
In this vicinity, the $n$-sheeted geometry (sketched in figure \ref{fig:cones} for $n=2$) is obtained from the metric on a single copy by slicing the $n$ replicas along the surface $\island$ emanating from $\gamma$, and joining them back together with cyclic identifications. Our $n$-copy geometry must be smooth at $\gamma$ so that we satisfy the equations of motion there, but this implies that each single geometry is not smooth: it has a conical defect at $\gamma$ with opening angle $\frac{2\pi}{n}$. In particular, this requires back-reaction that will modify the geometry on each replica in some $n$-dependent way. However, we can find a saddle-point by solving the equations of motion from varying the metric on a single copy while imposing the $\frac{2\pi}{n}$ defect boundary condition at $\gamma$, which is a problem which can be continued in $n$, and in the $n\to 1$ limit we return to the original smooth geometry. Now, if we were simply to evaluate the gravitational action on this single-copy singular configuration, we would find a contribution $(1-n^{-1})\frac{A[\gamma]}{4G_N}$ from curvature with delta-function support at $\gamma$.\footnote{For one way to see this, we can split the action $\frac{1}{16\pi G_N}\int \mathcal{R}$ into an integral in the directions parallel to $\gamma$, which gives the area, and a transverse two-dimensional integral. We can evaluate the latter integral on a small circle centred on $\gamma$ using the Gauss-Bonnet theorem. See \cite{Louko:1995jw} and \cite{CDMRW} for details of this procedure in Lorentz signature near singularities in the causal structure of the form associated with $\gamma$.} But this contribution should not be there, since we are really evaluating the action on the $n$-copy metric, which is smooth and so has no such singular piece. To make up for this difference between the correct $n$-replica action and the singular Einstein-Hilbert action, we must subtract this `by hand'. The gravitational action on the replica manifold is therefore given \cite{Lewkowycz:2013nqa,Dong:2016hjy} by\footnote{It is also true that \eqref{eq:replicaEH} defines a good variational principle on the singular single copy defined by taking the $n$-fold quotient of the $n$-replica manifold.  See \cite{Dong:2019piw} for a full discussion of the Euclidean case.  The Lorentz signature case follows by analytic continuation; see also \cite{CDMRW}.}
\begin{equation}\label{eq:replicaEH}
	i S_\mathrm{EH}^{(n)} = n i S_\mathrm{EH}^{(1)} - (1-n^{-1})\frac{A[\gamma]}{4G_N}.
\end{equation}
 In the $n\to 1$ limit, the area term above gives the geometric term in the generalised entropy \eqref{eq:Sgen}. Note that this is a real contribution when the action is evaluated in Euclidean signature, so despite the Lorentzian setting it weights the path integral by the exponential $e^{-(1-n^{-1})\frac{A[\gamma]}{4G_N}}$, and not by a phase.  The same basic phenomenon was observed long ago in \cite{Louko:1995jw}.

We now have a description for the path integral over replica symmetric configurations that we can nicely continue in $n$, and which we can study for small values of $(n-1)$. There are two types of term appearing in the action which weights this integral. First, we have terms which are independent of $\gamma$, namely the local gravitational action $S_\mathrm{EH}^{(1)}$  in \eqref{eq:replicaEH} and the normalising factor $\left(Z_\mathrm{matter}^{(1)}\right)^n$ from \eqref{eq:Znmatter}. Together, these just give $n$ times what we will call the single-copy action (that is, the gravitational action --- including a contribution from the singularity for $n\neq 1$ --- plus matter effective action). Secondly, we have the two terms making up the generalised entropy, namely the area term in \eqref{eq:replicaEH} and the matter entropy in \eqref{eq:Znmatter}. For $n- 1\ll 1$, the second class of terms give a small correction so we can ignore them at first, obtaining simply the path integral that computes the norm of the state.  Since we will also need to divide by this result to get our final expectation value, such contributions cancel completely in the final expression.

The above considerations fix a saddle-point geometry on a single replica.  However, there remains a residual integral over codimension two surfaces $\gamma$ within that geometry.  Note that we require a saddle point for this integral as well if we are to specify a saddle for the full $n$-fold replicated geometry.

For the integral over $\gamma$, the weighting is provided by the second set of terms:
\begin{align}
	\Tr \left( U_\pi(\scri_u) \rho^{(n)} \right) &\longrightarrow \int \mathcal{D}\gamma \; e^{-(n-1)S_\mathrm{gen}(\island;u)} \qquad (n-1 \ll 1) \label{eq:gammaint} \\
	&\sim \sum_{\gamma=\partial\island \ \mathrm{QES}} e^{-(n-1)S_\mathrm{gen}(\island;u)}, \label{eq:QESsum}
\end{align}
where the last step indicates a saddle-point evaluation of the integral over surfaces. Now, in principle we should also realize that for $n>1$  the singularity at $\gamma$ will back-react on the single-replica metric and thus change the value of the single-replica action $S_{EH}^{(1)}$.  But since at $n=1$ we work at are at stationary point for $S_{EH}^{(1)}$, this effect is quadratic in $(n-1)$ and can thus be ignored for the purpose of computing our swap von Neumann entropy; see \cite{CDMRW} for further discussion discussion of back-reaction at finite $n-1$ in saddles for real-time path integrals.

The saddle-points of \eqref{eq:gammaint} are precisely the quantum extremal surfaces, since these are the points at which $S_\mathrm{gen}$ is stationary. We may attempt to approximate this by including only the dominant saddle-point and noting that for $n$ near $1$ the dominant saddle is given by the term in which $S_\mathrm{gen}$ takes the smallest value.\footnote{This is not always sufficient.  As described in \cite{Penington:2019kki,Dong:2020iod,Akers:2020pmf}, other saddles can sometimes play important roles --  especially when two QESs has similar values of $S_\mathrm{gen}$. But their inclusion appears to only strengthen the arguments presented here. See also
\cite{Vidmar:2017pak,Murthy:2019qvb,MarolfWangWang2020} for related comments on corrections near transitions in which saddles exchange dominance.\label{foot:transitions}}
Expressing this in terms of the swap entropy \eqref{eq:SwapEnt2}, we can summarise the resulting replica wormhole contribution by the simple formula
\begin{equation}\label{eq:RT}
	S^{\text{swap}}(u) \sim \min_{\gamma=\partial\island \text{ QES}} S_\mathrm{gen}(\island;u).
\end{equation}
That is, we evaluate $S_\mathrm{gen}$ for island bounded by surfaces $\gamma$ such that it is stationary to first order variations, and if there are multiple such surfaces we choose the smallest result.

The result \eqref{eq:RT} is a version of the Ryu-Takayanagi formula \cite{Ryu:2006bv,Ryu:2006ef} first stated by Engelhardt and Wall \cite{Engelhardt:2014gca}, following generalisations to time-dependent situations \cite{Hubeny:2007xt} and inclusions of quantum corrections \cite{Faulkner:2013ana}. In this context where the quantum extremal surface $\gamma$ is compact and hence bounds an island $\island$, it has become known as the `island formula' \cite{Almheiri:2019hni}. These were all originally stated in the context of holographic duality, with the result interpreted as a von Neumann entropy of a dual quantum system. Here, we do not assume any such dual description so our interpretation is rather different, instead predicting the outcome of `swap' experiments performed on multiple sets of Hawking radiation.

Incidentally, the argument reveals why we should expect that $S_\mathrm{gen}$ is finite and not UV sensitive. We obtain $S_\mathrm{gen}$ as a limit of partition functions over replica manifolds which are smooth, with no singular features at the surface $\gamma$. These features of $S_\mathrm{gen}$ are ensured if we have a sensible effective theory.

The first term in $S_\mathrm{gen}(\gamma;u)= \frac{A[\gamma]}{4G} + S_\mathrm{ext}(\Sigma_\mathrm{int}\cup \scri_u)$ is naturally of order $G_N^{-1}$, while the second matter entropy term will typically be a small correction of order one. This means that in most circumstances, a QES will be close to a classical extremal surface. However, this in not always the case. In particular, for evaporating black holes there may be no nontrivial classical extremal surface  but, as we will see presently, due to the parametrically long times involved there is nonetheless a QES.

\subsection{Contributions from replica wormholes}
\label{sec:replicaCon}

The discussion of section \ref{sec:QES} reduced the study of replica wormholes near $n=1$ to the study of quantum extremal surfaces in the original semiclassical $n=1$ saddle.  Recall that  for us this is the `Hawking wormhole' in figure \ref{fig:rhoEvaporatingSC}. A trivial case is when the island $\island$ and hence the QES $\gamma=\partial\island$ are empty, in which case we obtain the original Hawking result $S^\mathrm{Hawking}(u) = S_\mathrm{matter}(\scri_u)$ for the swap entropy. It remains to ask whether there might also be a nontrivial QES in this spacetime.

\begin{figure}
\centering
	\includegraphics[width=.4\textwidth]{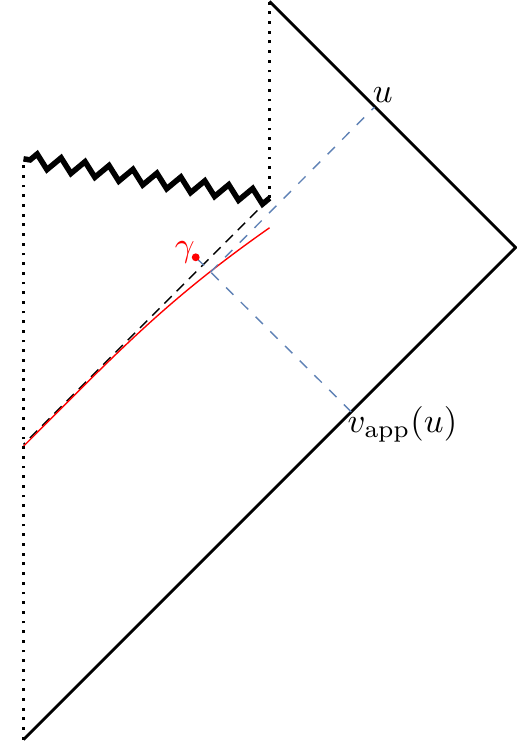}
\caption{A sketch of the location of the nontrivial QES in an evaporating black hole. The red curve is the location of the apparent horizon, where the area is stationary to variations in the outward null direction. The function $v_\mathrm{app}(u)$ is defined as the ingoing time which intersects the apparent horizon at outgoing time $u$, as shown by the dashed blue null lines. The QES $\gamma$ for $S^\mathrm{swap}(u)$ lies at ingoing time  $v_\mathrm{app}(u)$, just behind the event horizon.\label{fig:EvaporatingQES}}
\end{figure}

This is precisely the question that was studied in references \cite{Penington:2019npb,Almheiri:2019psf}.  Those works showed that a non-trivial QES $\gamma$ exists soon after the black hole forms (after roughly a scrambling time, which is logarithmic in $G_N$). To locate the QES, we first define the function $v_\mathrm{app}(u)$ so that for a given outgoing time $u$, the apparent horizon of the black hole lies at ingoing time $v=v_\mathrm{app}(u)$. Given our spherical symmetry, we may define the apparent horizon  as the (spherical) surface on which the area of the transverse sphere is stationary under variations in the outgoing null direction. This surface is slightly outside the event horizon since the black hole is evaporating, so the function $v_\mathrm{app}(u)$ is well-defined for times $u$ soon after formation of the black hole, up until the evaporation time $u_\evap$. The works \cite{Penington:2019npb,Almheiri:2019psf} showed that a QES computing $S^\mathrm{swap}(u)$ exists very close to the event horizon at advanced time close to $v_\mathrm{app}(u)$, with the corrections to this advanced time being of order the inverse black hole temperature $\beta$. This is sketched in figure \ref{fig:EvaporatingQES}.

The generalised entropy of $\gamma$ is dominated by the area term,  so $S_\mathrm{gen}(\island;u)$ is close to the Bekenstein-Hawking entropy $S_\mathrm{BH}(u)$. This QES thus becomes dominant after the Page time and causes $S^\mathrm{swap}(u)$ to follow the Page curve:
\begin{equation}\label{eq:Page}
	S^\mathrm{swap} \sim \min\left\{ S^\mathrm{Hawking}(u) , S_\mathrm{BH}(u) \right\}\, .
\end{equation}

The physics that allows such a QES to exist is rather generic, and in particular is independent of the dimension or asymptotics of the spacetime. Using spherical symmetry, it is sufficient to argue that $S_\mathrm{gen}$ is stationary to variations in ingoing and outgoing null directions. The outgoing variation of the area vanishes on the apparent horizon, so it is unsurprising that the outgoing variation of $S_\mathrm{gen}$ can vanish on a nearby surface $\gamma$. The ingoing variation is more subtle, requiring a balance between quantum entropy and classical area terms. This is possible due to the exponential divergence of outgoing geodesics near to the event horizon, producing a logarithmically growing contribution to the entropy. For detailed arguments, we refer the reader to the original references \cite{Penington:2019npb,Almheiri:2019psf} with AdS asymptotics (though for similar calculations with flat asymptotics see \cite{Hartman:2020swn,Gautason:2020tmk,Anegawa:2020ezn,Krishnan:2020oun,Hashimoto:2020cas}). We emphasise that there is no classical extremal surface close to $\gamma$ at which the $\frac{A[\gamma ]}{4G}$ term would be stationary on its own. The entropy term is thus critically important for the extremisation, with large gradients in entropy arising from the large relative boost between the near-horizon and asymptotic region. As a result, the corresponding replica wormholes are not related to any saddle of the classical action, but only exist as saddles of the quantum-corrected effective action as discussed above.

To make contact with the discussion of section \ref{sec:PS}, we can think of the PS wormholes as spacetimes of the above form in which we simply take the island to be the entire black hole interior, $\island = \Sigma_\mathrm{int}$, so that $\gamma=\evap$. If we choose the area term from $\evap$ to be zero, the corresponding generalised entropy is  $S_\mathrm{gen}(\Sigma_\mathrm{int};u) = \bar{S}^\mathrm{Hawking}(u)$, giving a third term over which we should minimise in \eqref{eq:Page}. Since this term arises from saddle-points which are not under semiclassical control, it is unclear whether or not it should really be allowed. But in the presence of the new QES and the accompanying replica wormholes, we see that it is in any case irrelevant. For any time $u< u_\evap$, the replica wormhole generalized entropy $S_\mathrm{gen}(\island;u)\sim S_\mathrm{BH}(u)$ is smaller than $\bar{S}^\mathrm{Hawking}(u)$ by  at least a factor of order one.  Since these quantities are both large, the difference is also large. At follows that PS wormholes never dominate, and in fact can only provide at most an exponentially small correction that we ignore.

We thus no longer require any input beyond semiclassical physics or assumptions about $\evap$, resolving the problem of section \ref{sec:endpoint}. Furthermore, since $\gamma$ is located near the past light cone of the relevant cut of $\scri^+$, we also avoid the causality issues described in section \ref{sec:PSCausality} for the PS proposal. However, we can think of the PS wormholes as a limit of saddle-points which \emph{are} under semiclassical control, where we take $\gamma$ to approach $\evap$. Indeed, this is what will happen to the QES $\gamma$ in the limit $u\to u_\evap$. This provides some justification for using the PS wormholes after evaporation (i.e., for $u>u_\evap$) as a reasonable extrapolation of controlled calculations at earlier times.

In summary, we have considered a context where an asymptotically flat black hole radiates to $\scri^+$, with a focus on the region $\scri_u \subset \scri^+$ before retarded time $u$.  We have then studied the expectation value of cyclic permutation operators $\tau$ on $n$ copies of the radiation in $\scri_u$. This models the actual results of measurements made by a sophisticated experimentalist\footnote{For a black hole above the Planck scale, the experimentalist must be very sophisticated indeed, since the relevant expectation values are exponentially small.  As a result, distinguishing between the two branches requires exponentially many copies of the $n$-replica system.  Indeed, distinguishing between two possible values of (R\'enyi) entropy for an unknown state generally requires a number of copies which is exponential in the smaller candidate entropy. More sophisticated methods improve the coefficient of the exponential over that associated with the simple swap test, but the best algorithm still requires exponentially many copies; see e.g.\ \cite{acharya2019measuring}.} who allows $n$ identically-prepared black holes in the same universe to evaporate, captures the radiation emitted up to corresponding retarded times, and then measures the action of the corresponding permutation. The experimentalist might then use her measurements to deduce the von Neumann entropy of the radiation on $\scri_u$; we interpret the $n\to 1$ limit of the swap calculation as a prediction for the result. With the new QES, following from the replica wormholes, the result reproduces the Page curve, affirming the predictions of BH unitarity.

The expectation is that the replica wormholes exist also for integer $n>1$, and give similar results for the expectation value of permutation operators.  This would allow the above experimentalist to avoid the awkward step of taking the $n\rightarrow 1$ limit.  It remains to establish this in full, though see \cite{Mirbabayi:2020fyk} for analogous numerical $n=2$ constructions in Euclidean signature, and see \cite{CDMRW} for discussions and explicit examples of classical real-time integer $n$ replica saddles (without back-reaction from quantum fields).

\subsection{Replica wormholes for other observables}
\label{sec:RWother}

So far, we have considered the contribution of replica wormholes to the expectation values of permutation operators $U_\tau(\scri_u)$ and thus  swap entropies \eqref{eq:SwapEntn}. It is now natural to ask whether such topologies can also contribute to other observables. This question was also discussed in \cite{Stanford:2020wkf}, which inspired many of the considerations in this section.

From one perspective,  such contributions seem inevitable. We can write the expectation value of $U_\tau(\scri_u)$ as a sum over matrix elements of the density matrix $\rho^{(n)}(u)$ that describes radiation on $n$ copies of $\scri^u$ from $n$ evaporating black holes.  This gives
\begin{equation}\label{eq:rhoElements}
	\Tr \left( U_\tau(\scri_u) \rho^{(n)} \right) = \sum_{i_1,\ldots,i_n} \langle i_n,i_1,\ldots,i_{n-1}|\rho^{(n)}(u)|i_1,i_2,\ldots,i_n\rangle,
\end{equation}
where $i$ labels a complete set of boundary conditions on $\scri_u$. The result \eqref{eq:rhoElements} is simply a reorganisation of the path integral studied above in which we first perform the path integral with fixed matter fields on $\scri_u$ to compute matrix elements of $\rho^{(n)}(u)$, and only then integrate over all possible such boundary values of matter fields with appropriate identifications to perform the sum shown on the right-hand-side. Since the left-hand-side receives replica wormhole contributions, this must be true of the right-hand-side as well, and thus of the $n$-evaporation radiation density matrix $\rho^{(n)}(u)$. One should thus expect generic observables involving $n$ copies of $\scri_u$ to be modified by replica wormholes as well.

However, this argument leaves open whether the required contributions to matrix elements or other observables are large enough to appear at the semiclassical level, and thus also whether replica wormholes need to make an explicit appearance as saddles in their semiclassical computation.
Since the right hand side of \eqref{eq:rhoElements} has a sum over exponentially many terms, a semiclassical description of the sum need not tell us anything about the individual terms.  Nonetheless, we argue below that the conclusion is plausible, and that replica wormholes may well give saddle-points for matrix elements or for the expectation values of simple operators. Our arguments will be rather heuristic and suggestive, so a more detailed study is required to establish this carefully; \cite{Stanford:2020wkf} goes a long way towards this aim.

To illustrate the point, we first consider a very simple observable, namely the product of expectation values of simple operators inserted on different copies of $\scri^+$:
\begin{equation}\label{eq:2pt}
	\Tr(\mathcal{O}_1(u)\mathcal{O}_2(u) \rho^{(2)} ).
\end{equation}
Here $\mathcal{O}_r(u)$ is a simple local operator such as the value of particular field mode on $\scri^+$ at retarded time $u$, and $r$ denotes which `replica' of the Hawking radiation on which it acts. In contrast to our studies of the swap operator above (which mixes boundaries associated with different values of $r$), since we now compute the expectation value of a product of operators that each act on a single boundary the corresponding boundary conditions for the path integral do not include any connections between the two asymptotic regions. Nevertheless, as explained below we anticipate saddle-points in the gravitational path integral which dynamically connect the boundaries via replica wormholes.

The reason for this is in fact much the same as for expectation values of the swap operator. There, the existence of a replica wormhole saddle relied on an interplay between the gravitational action (through contributions associated with the area of the surface $\gamma$) and the matter effective action, in that case the matter R\'enyi entropy. In computing \eqref{eq:2pt}, the role of the matter effective action is played instead by the logarithm of the two-point function $\langle\mathcal{O}_1\mathcal{O}_2\rangle$ evaluated in the replica wormhole geometry. But at the qualitative level this behaves  much the same as the matter R\'enyi entropy. In particular, it has a logarithmic singularity as the surface $\gamma$ approaches null separation from the retarded time $u$ on $\scri^+$, as occurs near the apparent horizon of the black hole sufficiently far in the past. The interplay between the classical area and such a logarithmic singularity was precisely what allowed for the existence of a nontrivial QES above. It is therefore reasonable to expect that there may similarly exist a semiclassical replica wormhole saddle for \eqref{eq:2pt}.

However, the effect of this saddle should be much smaller for \eqref{eq:2pt} than for  expectation values of the swap operator.  In the latter case, replica wormholes dominate the late-time answer because the matter entropy in the Hawking saddle is naturally `extensive' in the sense that it grows linearly with time.  As a result, for  expectation values of the swap operator the one-loop-corrected action of the Hawking saddle becomes larger than the action (associated with the area of $\gamma$) for the replica wormhole.  But there is no such extensive effect for \eqref{eq:2pt}, and no corresponding late-time suppression of contributions from the Hawking saddle.  So one expects replica wormholes to contribute as subdominant saddles, and thus to give corrections which are suppressed exponentially in $S_\mathrm{BH}$.\footnote{An exception to this would occur if the Hawking saddle gives an extremely small answer (or exactly zero) for some other reason. For example, \eqref{eq:2pt} may receive its leading contributions from replica wormholes if the one-point function $\Tr(\mathcal{O}(u)\rho^\mathrm{Hawking})$ vanishes due to a symmetry.} The suppression by exponential factors agrees with our heuristic argument from \eqref{eq:rhoElements}, as the sums on the right-hand-side of \eqref{eq:rhoElements} should run over exponentially many terms so that small corrections of this order in each off-diagonal matrix element lead to the desired leading-order corrections on the left-hand-side of \eqref{eq:rhoElements}.

In practice it may be rather challenging to check for a replica wormhole saddle-point for quantities like \eqref{eq:2pt}, since it would seem to require finding a  back-reacted (and presumably complex) solution to the gravitational equations sourced by the quantum effective action, just as for integer R\'enyi entropies. It may be directly tractable in simple models of gravity (as in \cite{Stanford:2020wkf}), or by studying some appropriate family of quantities with an $n\to 1$ limit analogous to the von Neumann entropy, to evade the complications of back-reaction.

\section{The Hilbert space of baby universes}
\label{sec:BUetc}

The result reviewed above, showing that replica wormholes suffice to make the swap entropy of Hawking radiation follow a Page curve, is satisfying in many ways.  In particular, it gives a completely semiclassical computation that supports Bekenstein-Hawking unitarity.  Moreover, it does so by computing a quantity that is experimentally accessible, at least in principle.

However, it also raises many questions.  While  we now have a path-integral derivation of the Page curve, our new ingredients do not affect the computation of expectation values of observables for the Hawking radiation from a single black hole. The density matrix of radiation is still $\rho_\text{Hawking}(u)$ as computed as in section \ref{sec:semiclassical}, and which still comes just from the saddle-point pictured in figure \ref{fig:rhoEvaporatingSC}.  In particular, the swap entropies obtained in \ref{sec:replicas} are not equal to the R\'enyi entropies of $\rho_\text{Hawking}(u)$. How are these results to be reconciled?

The simple answer is that the density matrix $\rho^{(n)}(u)$ on $n$ copies of radiation is not simply equal to the tensor product $\rho_\text{Hawking}(u)^{\otimes n}$. But this means that
the results of independent and widely separated experiments are correlated, and thus give rise to a violation of cluster decomposition. How are we to interpret predictions of the theory in such a situation? What form can these correlations between experiments take? And what is the Hilbert space interpretation of these results?

In this section we answer these questions by cutting open the path integrals described so far, to obtain a Hilbert space interpretation of the correlations between boundaries from a sum over intermediate states. Before diving in we briefly preview the central ideas, which are much the same as in \cite{Coleman:1988cy,Giddings:1988cx,Polchinski:1994zs}. The intermediate states in question are states of closed `baby' universes which propagate between distinct asymptotic boundaries.  But the resulting correlations are restricted to be purely classical, so that we may regard expectation values of asymptotic observables as random variables selected from some probability distribution. The reason is that such asymptotic observables can be regarded as a mutually commuting set of operators acting on the Hilbert space of baby universes, which can be simultaneously diagonalised into superselection sectors. It therefore appears that semiclassical gravity predicts results for asymptotic observers which are consistent with BH unitarity, though the precise dynamics is not uniquely determined but chosen from an ensemble. That ensemble depends on a choice of the initial state of closed (baby) universes.

\subsection{From path integrals to Hilbert spaces}

To set the stage, we begin by briefly reviewing the relationship between the path integral computations of quantum amplitudes and their Hilbert space formulation, emphasizing features relevant to gravitational theories.

A Hilbert space appears when we cut a path integral into pieces, writing the integral over the cut as a sum over intermediate states. Before incorporating dynamical gravity, let us discuss this for a QFT path integral on a fixed background spacetime $\mathcal{M}$, and cut the geometry along a Cauchy surface $\Sigma$ of our choice. This cut manifold has two new boundaries $\Sigma_-$ and $\Sigma_+$, the past and future sides of $\Sigma$ respectively. We can now perform the path integral on this manifold with boundary, imposing boundary conditions that the fields $\phi$ approach $\phi_\pm$ on the boundaries $\Sigma_\pm$ (for example), and integrating over $\phi$ elsewhere. To obtain the original path integral on $\mathcal{M}$, we ensure continuity of the fields at $\Sigma$ by setting $\phi_+=\phi_-=\phi_\Sigma$, and then integrate over all field values $\phi_\Sigma$ on $\Sigma$.

This cutting and gluing has a Hilbert space interpretation as the insertion of a resolution of the identity, $\id = \int \mathcal{D}\phi_\Sigma \, |\phi_\Sigma\rangle \langle\phi_\Sigma|$. We have a Hilbert space $\hilb_\Sigma$, formally spanned by field eigenstates $ |\phi_\Sigma\rangle$ labelled by field configurations on $\Sigma$, where the inner product $\langle\phi_+|\phi_-\rangle$ is given by a functional delta-function setting $\phi_+=\phi_-$. In a semiclassical approximation, where the path integral is computed by fluctuations around a saddle-point with approximately Gaussian weighting, this Hilbert space $\hilb_\Sigma$ becomes a Fock space for linearised fluctuations about the saddle.

This is somewhat complicated by the inclusion of dynamical gravity, when we also sum over the topology and geometry of spacetime. As in QFT we can cut the path integral along some Cauchy surface $\Sigma$, and include the geometry of $\Sigma$ in the sum over intermediate states. But diffeomorphism invariance makes this more subtle. We have many choices of slice $\Sigma$ that all lead to the same Hilbert space as long as they agree asymptotically (where the geometry is fixed by boundary conditions). These different choices are related by the Hamiltonian constraint or Wheeler-DeWitt equation. We will not need the technical details here, but we do note that this modifies the inner product on the Hilbert space associated with the cut.  Due to gauge invariance, it is natural that distinct field configurations on $\Sigma$ (now including the induced metric) need not define orthogonal states.  But something stronger is true here,  as the inner product is determined by the dynamics. Indeed, we will see below that the effect of replica wormholes can be described as a dynamical modification of the inner product on the Hilbert space at the cut.

In the context of evaporating black holes, we saw that the semiclassical path integral was helpful for computing observables on $\scri^+$ before some retarded time $u_\evap$ at which the black hole becomes Planckian and the semiclassical treatment is no longer valid. For this situation we are most naturally led to describe a gravitational Hilbert space describing the states on a partial Cauchy slice $\Sigma_u$ which, as part of the boundary conditions, is required to meet $\scri^+$ at time $u$. This would describe a system with a boundary.  However,  it will be conceptually simpler and cleaner to consider instead a Hilbert space of closed universes {\it without} boundary. As will be described in section \ref{sec:HSHPS} below, the simplest way to pass from the former to the latter is by making complete measurements on $\scri^+$.  However, as in section \ref{sec:PS} this comes at the cost of requiring some assumptions about the evaporation.  We will thus at first revive the `PS assumption' of  section \ref{sec:PS} in order to explain the main ideas involving in passing to a description in terms of closed universes.  We will use this assumption for the next few sections, though in section \ref{sec:DropPS} we will describe the modifications required to avoid it, and in fact to avoid using any assumption outside the domain of semiclassical physics.

\subsection{Hilbert spaces for Hawking and Polchinski-Strominger}
\label{sec:HSHPS}

Rather than going directly to the replica wormholes of most interest, we will warm up by discussing the Hilbert space interpretation of the Hawking and Polchinski-Strominger calculations of sections \ref{sec:semiclassical} and \ref{sec:PS}. In particular, for now we will make use of the `PS assumption' introduced in section \ref{sec:PS} to simplify the discussion.

We begin with Hawking's calculation using a single black hole and computing the expectation value of some operator $\mathcal{O}$ on $\scri^+$. The Hawking wormhole computing this expectation value consists of bra and ket copies of the black hole spacetime joined along some final Cauchy slice. Using The PS assumption, we may choose this joining Cauchy slice to consist of $\scri^+$ and $\Sigma_\mathrm{int}$, a Cauchy surface for the black hole interior. To obtain a Hilbert space interpretation, we can first cut this geometry along $\scri^+$, where we obtain the Hilbert space of `out states' $\hilb_{\scri^+}$. We choose an orthonormal basis $\left\{|i\rangle_{\scri^+}\right\}_i$ for this space. However, cutting only along $\scri^+$ is not sufficient to write the expectation value as a sum over states, since the geometry is still connected through the black hole interior. We must thus also slice the geometry along $\Sigma_\mathrm{int}$, obtaining a Hilbert space  $\hilb_\mathrm{int}$ with orthonormal basis $\left\{|a \rangle_\mathrm{int}\right\}_a$.

\begin{figure}
	\centering
	\raisebox{75pt}{\scalebox{1.5}{$\langle\mathcal{O}\rangle = \displaystyle\sum_{i,j,a}$}}\quad
	\includegraphics[width=.45\textwidth]{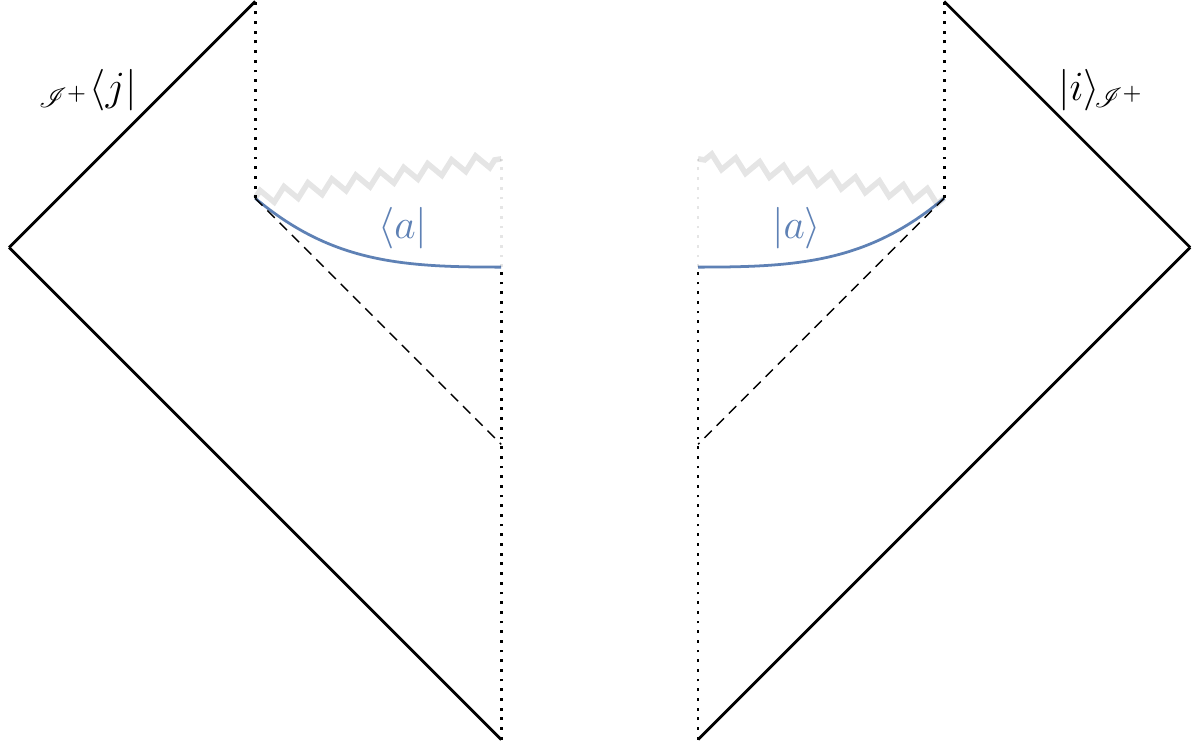}
	\quad
	\raisebox{75pt}{\scalebox{1.5}{$\mathcal{O}_{ij}$}}
		\caption{Cutting the Hawking wormhole computing the expectation value of an operator $\mathcal{O}$ defined on $\scri^+$ (with matrix elements $\mathcal{O}_{ij}={}_{\scri^+}\langle i|\mathcal{O}|j\rangle_{\scri^+}$) to obtain a Hilbert space interpretation. The path integral on the right spacetime, with boundary conditions on $\scri^+$ set by the state $|i\rangle_{\scri^+}$ and on $\Sigma_\mathrm{int}$ by the state $|a\rangle$, computes the wavefunction $\psi_{ai}$ of a pure state in $\hilb_{\scri^+}\otimes \hilb_\mathrm{int}$. The left spacetime computes the conjugate wavefunction, and to obtain the expectation value we sum over all intermediate states, $\langle\mathcal{O}\rangle = \sum_{i,j,a}\bar{\psi}_{aj}\psi_{ai}\mathcal{O}_{ij}$. \label{fig:HawkingHS}}
\end{figure}

Once we have cut along both $\scri^+$ and $\Sigma_\mathrm{int}$, we have decomposed the geometry into disconnected bra and ket copies of the Hawking spacetime as shown in figure \ref{fig:HawkingHS}. The path integral on the ket spacetime with boundary conditions imposed on $\scri^+$ and $\Sigma_\mathrm{int}$ computes the wavefunction $\psi_{a i}$ of a state in $\hilb_{\scri^+}\otimes \hilb_\mathrm{int}$:
\begin{equation}
	|\psi\rangle = \sum_{i,a} \psi_{a i}  |i\rangle_{\scri^+} \otimes |a \rangle_\mathrm{int} \in \hilb_{\scri^+}\otimes \hilb_\mathrm{int}
\end{equation}
The path integral on the conjugate bra spacetime gives the complex conjugate of this wavefunction, which is a state in the dual space $\hilb_{\scri^+}^*\otimes \hilb_\mathrm{int}^*$.

To glue these spacetimes back together along $\Sigma_\mathrm{int}$, we sum over all states of the interior and take the inner product, obtaining the Hawking density matrix for the state on $\scri^+$:
\begin{equation}
	\rho_\text{Hawking} = \sum_{i,j,a,b} \bar{\psi}_{b j}\psi_{a i} \langle b|a \rangle_{\mathrm{int}} \left(|i\rangle\langle j |\right)_{\scri^+}.
\end{equation}
An orthonormal basis on $\Sigma_{int}$ gives $\langle b|a \rangle_{\mathrm{int}}=\delta_{ab}$, and we have
\begin{equation}
	\langle i|\rho_\text{Hawking}|j\rangle = \sum_{a} \bar{\psi}_{a j}\psi_{a i}.
\end{equation}
This is a mixed state because we have traced out the black hole interior, with which the matter on $\scri^+$ is entangled. The Hilbert space on $\scri^+$ is thus insufficient to give a complete description of the original state of the system on $\scri^-$. We must also include information about the state on $\Sigma_\mathrm{int}$, which we may think of as a closed\footnote{We say that the baby universe is closed as the boundary of $\Sigma_\mathrm{int}$ at $\evap$ involves a sphere of zero size. We will comment further on this in section \ref{sec:Disc}. }
 `baby' universe, born from the black hole formed in the `parent' universe.  In introducing this terminology, we warn the reader that there are two slightly different notions of baby universe in the literature.   When required for clarity (mostly in section \ref{sec:DropPS}), we will use the term `PS baby universe' to distinguish the above notion from others that may arise.

So far, this is a fairly conventional description of information loss. But we will go beyond this by considering the computations of section \ref{sec:PS} that involve $n$ copies of the black hole. The Polchinski-Strominger wormholes consist of multiple copies of the Hawking wormhole, so to obtain a Hilbert space interpretation we can again slice them along $n$ copies of $\scri^+$, where we have $n$ copies of the asymptotic Hilbert space $\hilb_{\scri^+}^{\otimes n}$, and along $n$ copies of $\Sigma_\mathrm{int}$. After cutting them open in this way, for each term in \eqref{eq:PSsumperm} the resulting $n$ `ket' spacetimes are identical.  In particular, they compute the wavefunction of the state
\begin{equation}
\label{eq:psi(n)}
	|\psi^{(n)}\rangle  = \sum_{\substack{{i_1,\ldots,i_n}\\{a_1,\ldots,a_n}}} \psi_{a_1 i_1}\cdots \psi_{a_n i_n}  |i_1\rangle_{\scri^+_1} \otimes\cdots \otimes |i_n\rangle_{\scri^+_n} \otimes |a_1,\ldots,a_n \rangle_{\mathrm{BU}}
\end{equation}
in $\hilb_{\scri^+}^{\otimes n} \otimes \hbu$, where $\hbu$ is the Hilbert space of closed (baby) universes. We obtain the density matrix on  $\hilb_{\scri^+}^{\otimes n}$ by tracing out the baby universes,
\begin{equation}\label{eq:PSstate}
	\langle i_1,\ldots,i_n|\rho^{(n)}|j_1,\ldots,j_n\rangle = \sum_{\substack{{a_1,\ldots,a_n}\\{b_1,\ldots,b_n}}} \psi_{a_1 i_1}\bar{\psi}_{b_1 j_1}\cdots \psi_{a_n i_n}\bar{\psi}_{b_n j_n} \langle b_1,\ldots,b_n|a_1,\ldots,a_n \rangle_{\mathrm{BU}}.
\end{equation}

Since we have obtained $\hbu$ by cutting along $n$ copies of $\Sigma_\mathrm{int}$, it is tempting to identify $\hbu$  with the $n$-fold tensor product of $\hilb_\mathrm{int}$.  In that case its inner product would factorize into $n$ copies of the inner product on $\hilb_\mathrm{int}$. But if we made this identification, the state \eqref{eq:psi(n)} would be simply the $n$-fold tensor product $|\psi\rangle^{\otimes n}$, and the density matrix $\rho^{(n)}$ in \eqref{eq:PSstate} would be the tensor product $\left(\rho^{(1)}\right)^{\otimes n}$.  In particular, we would not find the sum over permutations in \eqref{eq:rhon2}.  We will resolve this tension below by not making assumptions about the inner product on  $\hbu$, but instead by computing the inner product induced by PS wormholes.  Replica wormholes lead to similar modifications to the inner product that we will discuss in section \ref{sec:RWBUI}.

Specifically, the correct inner product on $\hbu$ must be obtained by comparing \eqref{eq:PSstate} with \eqref{eq:PSsumperm}.  Since the  PS wormholes involve pairing the $n$ `ket' copies of $\Sigma_\mathrm{int}$ with the $n$ `bra' copies in any of the $n!$ possible ways, this inner product involves  a sum over permutations:
\begin{equation}\label{eq:PSIP}
	\langle b_1,\ldots,b_n|a_1,\ldots,a_n \rangle_{\mathrm{BU}} = \sum_{\pi\in \Sym(n)} \delta_{a_1 b_{\pi(1)}}\cdots \delta_{a_n b_{\pi(n)}}
\end{equation}
Note that this is the inner product on the $n$-fold symmetric product $\Sym^n\hilb_\mathrm{int}$. As a result, in the Polchinski-Strominger proposal, the baby universe Hilbert space contains a Bosonic Fock space built on the `one-universe states' $\hilb_\mathrm{int}$:
\begin{equation}
\label{eq:PSIP5}
	\text{Polchinski-Strominger:}\quad \hbu \supset \bigoplus_{n=0}^\infty \Sym^n\hilb_\mathrm{int}.
\end{equation}
We have written `contains' ($\supset$) here, since this is not in fact quite the full baby universe Hilbert space. As we will see below, it is natural to extend $\hbu$ to be a Fock space built on $\hilb_\mathrm{int}\oplus \hilb_\mathrm{int}^*$, with both baby universes and time-reversed `anti baby universes'. The second summand $\hilb_\mathrm{int}^*$ (the dual space of $\hilb_\mathrm{int}$) gives the states of a single anti-universe.

\subsection{Baby universes and ensembles}\label{sec:BUensembles}

The physical predictions that follow from the state \eqref{eq:PSstate} defined by  the inner product \eqref{eq:PSIP} may not immediately be clear.   We will describe this in some detail below, taking advantage of the fact that the Polchinski-Strominger proposal is simple enough to allow explicit results.  The result will remain useful when we later move beyond the Polchinski-Strominger proposal (and leave behind its challenges), as many of the lessons learned here will remain true for replica wormholes, and also  for gravitational path integrals more generally (under certain weak assumptions).

\subsubsection{The PS Fock space of baby universes}

The predictions of the Polchinski-Strominger proposal can be made manifest by using the familiar representation of the Bose inner product \eqref{eq:PSIP} as a Gaussian integral:
\begin{gather}
	\langle b_1,\ldots,b_n|a_1,\ldots,a_n \rangle_{\mathrm{BU}} =\Big\langle \alpha_{a_1}\cdots \alpha_{a_n} \bar{\alpha}_{b_1}\cdots \bar{\alpha}_{b_n}\Big\rangle_{\mathrm{BU}} \, , \\
	\text{where } \Big\langle F[\alpha,\bar{\alpha}] \Big\rangle_{\mathrm{BU}} := \frac{1}{\mathfrak{Z}}\int \prod_{a} d\alpha_a d\bar{\alpha}_a \; e^{-\sum_a \alpha_a \bar{\alpha}_a}  F[\alpha,\bar{\alpha}]. \label{eq:GaussianBU}
\end{gather}
The normalisation $\mathfrak{Z}$ is chosen so that $\big\langle 1 \big\rangle_{\mathrm{BU}} =1 $. The integration variables are complex\footnote{We use complex fields so that we only count contractions between `kets' and `bras', not between two kets, for example. This distinguishes baby universes from `anti' baby universes.} `baby universe fields' $\alpha_a$ labelled by an orthonormal basis of states $|a\rangle$ on $\Sigma_\mathrm{int}$. In place of the labels $a$ we could instead use field configurations $\phi$ for matter fields on $\Sigma_\mathrm{int}$, so that $\alpha$ is a functional of these fields. Then the Gaussian weighting can be written as an exponential of  $\int \mathcal{D}\phi \, \alpha[\phi]\alpha^*[\phi]$, where we integrate over all field configurations $\phi$. We then compute amplitudes $\big\langle \cdot \big\rangle_{\mathrm{BU}}$ by integrating over all functionals $\alpha[\phi]$ with this weighting.\footnote{Passing from the gravitational path integral to this integral over functionals $\alpha$ is mathematically analogous to the passage from particle dynamics to a description of quantum field theory as a path integral over Eucliden field configurations.  In this analogy, $\phi$ would label points in spacetime, $\alpha$ would be a quantum field (a function of spacetime), and the Gaussian weighting (for a free QFT) is given by the action. The kinetic term in the QFT action can then be understood as arising due to the Hamiltonian constraint on particle worldlines, which we have implemented rather implicitly in \eqref{eq:GaussianBU} by diagonalising the physical `single-universe' inner product.
However, the reader should see \cite{CMMR} for comments and warnings about using this analogy to interpret the physics.}

With this representation, we can write the components of the $n$-evaporation density matrix \eqref{eq:PSstate}  on $n$ copies of $\scri^+$ as
\begin{gather}
\label{eq:boundarycondtionamplitudes}
	\langle i_1,\ldots,i_n|\rho^{(n)}|j_1,\ldots,j_n\rangle = \Big\langle \bar{\Psi}_{j_1}\ldots \bar{\Psi}_{j_n} \Psi_{i_1}\ldots \Psi_{i_n} \Big\rangle_{\!\mathrm{BU}},\\
	\text{where}\quad \Psi_i = \sum_a \alpha_a\psi_{ai}.
\end{gather}
If we for now ignore the integral over $\alpha$ associated with $\big\langle \cdot \big\rangle_{\mathrm{BU}}$ and instead simply fix each $\alpha_a$ to some specific value, then the expression completely factorises between the $n$ copies, and also between `bra' and `ket' indices:
\begin{equation}
	\begin{gathered}
	\langle i_1,\ldots,i_n|\rho^{(n)}|j_1,\ldots,j_n\rangle \xrightarrow{\quad \text{fix }\alpha\quad} \Psi_{i_1}^\alpha \bar{\Psi}_{b_1}^\alpha \cdots \Psi_{i_n}^\alpha \bar{\Psi}_{b_n}^\alpha.
	\end{gathered}
\end{equation}
Here we have included a superscript $\alpha$ to emphasise that $\Psi^\alpha_i$ is now to be regarded as a fixed complex number depending on our choice for each $\alpha_a$. This factorisation means that, for a given value of $\alpha_a$, the Hawking radiation can be described by a pure state $|\Psi^{\alpha}\rangle\in\hilb_{\scri^+}$:
\begin{equation}
	\rho^{(n)} \xrightarrow{\quad \text{fix }\alpha\quad} (|\Psi^\alpha\rangle\langle\Psi^\alpha|)^{\otimes n}, \quad \langle i|\Psi^\alpha \rangle = \Psi_i^\alpha =  \sum_a{\alpha_a\psi_{ai}} \; .
\end{equation}

Now, the above potential factorisation property is spoiled by the fact that we must still integrate over $\alpha$ with some weighting. In other words, the parameters $\alpha_a$ are not fixed, but instead selected from a probability distribution. Note, however, that the same choice of $\alpha$ parameter pertains to all asymptotic observers at all boundaries. In particular, for $n$ black holes the state at $\scri^+$ is obtained by a {\it single} integral over $\alpha$,
\begin{equation}\label{eq:rhonalpha}
	\rho^{(n)} = \int d\mu(\alpha) \left(|\Psi^\alpha\rangle \langle \Psi^\alpha|\right)^{\otimes n},
\end{equation}
for some measure $d\mu(\alpha)$ (which in the PS paradigm is given by the Gaussian \eqref{eq:GaussianBU}).

As a result, any given set of actual measurements\footnote{We remind the reader that quantum mechanics measurements are associated with projection operators and that, while expectation values can be inferred from the relative frequencies of the outcomes associated with projections, a direct measurement of quantum mechanical expectation values would violate the linearity of quantum mechanics.} of the Hawking radiation states on $\scri^+$ from multiple black holes are correlated in such a way that they are always compatible with $n$ copies of some pure state $|\Psi^\alpha\rangle$.  But the theory does not give a specific prediction for $|\Psi^\alpha\rangle$.  Instead, it gives a probabilistic one. For an asymptotic observer, the black hole formation and evaporation can thus be described in terms of an S-matrix taking pure states to pure states, but with an unknown S-matrix selected from an ensemble.

This should be contrasted with the result obtained in the absence of PS wormholes for which the $n$ copies are uncorrelated.
Since the inner product on $\hilb_{\mathrm{int}}$ can also be written as an integral with respect to the same Gaussian measure\footnote{Since $\hilb_{\mathrm{int}}$ gives the $n=1$ term in \eqref{eq:PSIP5}, we may regard
$\hilb_{\mathrm{int}}$  as a subspace of $\hbu$ and use the same inner product.} $d\mu(\alpha)$, we may write this result using  independent integrals over $\alpha$ for each evaporation:
\begin{equation}
\label{eq:comparerho}
	\rho^{(n)}_{\mathrm{Hawking}} = \left(\int d\mu(\alpha) |\Psi^\alpha\rangle \langle \Psi^\alpha|\right)^{\otimes n} = \rho_\mathrm{Hawking}^{\otimes n}.
\end{equation}
We dub this the Hawking result for the $n$-fold experiment, as the predictions of \eqref{eq:comparerho} for experiments at $\scri^+$ are given by a Hawking $\$$-matrix \cite{Hawking:1976ra}.
But with PS wormholes we find instead \eqref{eq:rhonalpha}, which is a classical mixture of $n$ copies of a pure state as described above.

Note that the measure $d\mu(\alpha)$ arising from the Polchinski-Strominger proposal gives a complex multivariate Gaussian probability distribution for the components $ \Psi_i = \langle i|\Psi\rangle$ of the Hawking radiation wavefunction. The mean is zero, and the covariance matrix is given by the Hawking density matrix $\rho_\mathrm{Hawking}$.\footnote{This gives non-normalised wavefunctions. While the normalisations can be absorbed into the measure, doing so appears to introduce a mild $n$-dependence for $d\mu(\alpha)$. One might also consider the possibility of additional involving wormholes connecting the path integral for $\rho^{(n)}$ to the normalising denominator $\Tr\rho^{(n)}$, which should be expected to remove this $n$ dependence. It would be worthwhile to understand this issue in detail, but such a treatment is beyond the scope of this work.}

To understand this from the perspective of the baby universe Hilbert space, we can instead represent the Bose inner product \eqref{eq:PSIP} in terms of a Fock space generated by baby universe creation and annihilation operators,
\begin{gather}
\label{eq:caops}
	|a_1,\ldots,a_n, \bar b_1, \ldots, \bar b_m\rangle = A_{a_1}^\dag \cdots A_{a_n}^\dag B_{b_1}^\dag \cdots B_{b_m}^\dag |\HH\rangle, \\
	[A_a,A^\dag_b] =[B_a,B^\dag_b] = \delta_{ab}, \quad [A_a,A_b]=[B_a,B_b]=[A_a,B_b]=[A^\dag_a,B_b]=0\\
	A_a|\HH\rangle = B_b|\HH\rangle = 0.
\end{gather}
Here, $A_a$ and $A^\dag_a$ annihilate and create a baby universe in the state $a$. Similarly $B_b$ and $B^\dag_b$  annihilate and create a time-reversed object that may be called an `anti' baby universe (an anti-BU)\footnote{These are much the same as the baby universe creation/annihilation operators of \cite{Coleman:1988cy,Giddings:1988cx,Giddings:1988wv}, though those references worked in a real basis.  There may also be minor differences associated with subtleties discussed in \cite{CMMR}.}. Although anti-BUs did not appear in our discussion above, they naturally form the intermediate states if we considered the time-reverse of our boundary conditions (associated with a white hole that explodes to form a smooth $\scri^+$ with classical matter and quantum fields in the vacuum state but with time-reversed Hawking radiation at $\scri^-$).   In \eqref{eq:caops}, we have used $|\HH\rangle$ to denote the oscillator vacuum in order to think of it as a Hartle-Hawking state for reasons that we will explain momentarily.

Now, the Gaussian integral \eqref{eq:GaussianBU} is nothing but the expectation value of (complex) `position operators' $\hat{\alpha}_a$ in the oscillator vacuum $|\HH\rangle$:
\begin{equation}\label{eq:BUHH}
	\Big\langle F[\alpha,\bar{\alpha}] \Big\rangle_{\mathrm{BU}} = \big\langle\HH\big|F[\hat{\alpha},\hat{\alpha}^\dag]\big|\HH\big\rangle,
\end{equation}
where
\begin{equation}
	\hat{\alpha}_a = A^\dag_a +B_a.
\end{equation}
All the operators $\hat{\alpha}_a$ and $\hat{\alpha}^\dag_a$ mutually commute, so we can write the Hilbert space in terms of the position eigenbasis $|\alpha\rangle_\mathrm{BU}$ labelled by a set of complex eigenvalues $\alpha_a$:
\begin{equation}
\label{eq:alphadef}
	\hat{\alpha}_a|\alpha\rangle_\mathrm{BU} = \alpha_a|\alpha\rangle_\mathrm{BU}.
\end{equation}
We obtain the Gaussian integral \eqref{eq:GaussianBU} by inserting the completeness relation
\begin{equation}\label{eq:alphacomplete}
	\id  = \int d\alpha d\bar{\alpha}\, |\alpha\rangle\langle\alpha|
\end{equation}
into the right hand side of \eqref{eq:BUHH}, and using the Gaussian wavefunction of the oscillator vacuum
\begin{equation}
\label{eq:HHwavefunction}
	\langle \alpha|\HH\rangle \propto e^{-\tfrac{1}{2} \sum_a |\alpha_a|^2}.
\end{equation}

\subsubsection{Lessons and comments}
\label{sec:lessons}

Having completed our derivation of \eqref{eq:BUHH} from this Hilbert space point of view, let us now pause to extract some useful lessons.  The first lesson is that the appearance of only a single integral over $\alpha$ in the $n$-evaporation state \eqref{eq:rhonalpha} follows from the fact that the states \eqref{eq:alphadef} simultaneously diagonalize the operators $\hat{\alpha}_a$ and $\hat{\alpha}^\dag_a$. The latter statement is a consequence of a more primitive fact that will remain true when we go beyond the PS proposal, in that the boundary conditions for computing expectation values of asymptotic observables will continue to define simultaneously-diagonalizable operators acting on the Hilbert space $\hbu$ of closed universes.

Let us first illustrate this rather abstract-sounding statement by recalling that,
in the present case, we have boundary conditions $\Psi_i$ specifying both an initial state of matter  on $\scri^-$ which will collapse to a black hole as well as a final state  $|i\rangle_{\scri^+}$  of Hawking radiation  on $\scri^+$. The corresponding operator
\begin{equation}
\label{eq:BCop}
	\hat{\Psi}_i  = \sum_a\hat{\alpha}_a\psi_{ai}
\end{equation}
on $\hbu$ either creates a baby universe in some internal state or annihilates a time-reversed baby universe. The path integral computes an expectation value of a product of such operators, one for each separate boundary.\footnote{One may thus refer to $\hat{\Psi}$ as a `boundary-inserting operator'.  Indeed, it is tempting to refer to these as `boundary-creating' operators.  But one should realize that both $\hat \Psi$ and its adjoint $	\hat{\Psi}_i  = \sum_a\hat{\alpha}_a^\dagger \bar \psi_{ai}$ create boundaries in this sense.  These are thus not standard creation-annihilation operators, and in particular differ from the baby universe creation and annihilation operators $A,B, A^\dagger, B^\dagger$.} It is manifest from \eqref{eq:BCop} that the operators are all built from the (complex) `position' operators
$\alpha_a$, and in particular that creation operators $A_a^\dag$ never appear alone.  Similarly, the time-reversed boundary conditions $\bar \psi_j$ would define operators involving $\alpha_a^\dag$, which thus also commute with \eqref{eq:BCop}.

Although we used explicit results for the $\hbu$ inner product to derive this result, as argued in \cite{Marolf:2020xie} it in fact follows from fundamental properties of the gravitational path integral.  The point is simply that we may regard \eqref{eq:boundarycondtionamplitudes} as an amplitude computed by the quantum gravity path integral with the specified boundary conditions built from $\Psi_i, \bar \Psi_j$.  Since the path integral sums over \emph{all} bulk spacetimes compatible with the stated boundary conditions, the result is independent of how the boundary conditions might have been ordered.  As a consequence, the associated operators $\hat{\Psi}_i$, $\hat{\bar{\Psi}}_j$ commute.\footnote{Equation \eqref{eq:boundarycondtionamplitudes} describes the inner product of two states that involve only baby universes and not anti-BUs, or in other words states created from $|\HH\rangle$ by acting with the $\hat{\Psi}_i$ and not that $\hat{\bar{\Psi}}_j$.  Had we used the latter in the ket-state as well, there would have been additional entries of the $\bar{\Psi}_j$ on the right-hand-side.  But if we had used the latter in the bra-state, we would instead find additional copies of the $\Psi_i$ on the right.  In general, the rule is that the amplitude contains both the bra boundary conditions and the the CPT-conjugates of the ket boundary conditions. This requires the $\hat{\Psi}_i$ to be the adjoint of $\hat{\bar{\Psi}}_i$, so that the above mutual-commutativity means that the operators can be simultaneously diagonalized as claimed; see further discussion in \cite{Marolf:2020xie} and \cite{CMMR}.\label{foot:normal}}

Another lesson that becomes clear from the perspective of the baby universe Hilbert space is a sense in which our predictions depend what we may call the `initial state of baby universes' given by \eqref{eq:HHwavefunction}.  Indeed, the correlations between different copies of Hawking radiation are mediated by the exchange of baby universes, and we have seen that each set of asymptotic boundary conditions modifies the state of $\hbu$ through the action of $\hat{\Psi}_i$ or $\hat{\bar{\Psi}}_j = \hat{\Psi}_j^\dagger$. In the  previous sections we thus have implicitly chosen some initial state for closed universes. But recall that our amplitudes were entirely specified by boundary conditions with the experiments to be performed by our asymptotic observer, and that nothing more was added to adjust the baby universe state.  As a result, our choice of baby universe state must have been specified by the {\it absence} of additional asymptotic boundaries.  It is for this reason that we call it a Hartle-Hawking no boundary state $|\HH\rangle$. Note that we do not use this term for a state of a single connected closed universe, but a state that can include any number of universes (connected components of space) including zero; indeed, the universe number is not even diffeomorphism invariant if universes can split, join, or appear from nothing. Instead, it is defined according to the spirit of \cite{Hartle:1983ai} by the absence of boundaries in the path integral which determines the wavefunction.

Had we instead chosen the baby universe initial state to be e.g.\ $\hat{\Psi}_i |\HH\rangle$, expectation values in this state would be adding additional boundaries with boundary conditions $\Psi_i$ (from the ket) and $\bar{\Psi}_i$ (from the bra). Since we can again expand $\hat{\Psi}_i |\HH\rangle$ in terms of the same basis of  $\alpha$-states, we would  again find the $n$-evaporation density matrix $\rho^{(n)}$ to be a classical mixture of the same pure states $|\Psi^\alpha\rangle$ described above.  However, we will find a different mixture in which the probability distribution for $\alpha_a, \bar \alpha_b$ is defined by the new wavefunction $\langle \alpha |\hat{\Psi}_i |\HH\rangle$, which will generally differ from \eqref{eq:HHwavefunction}.

Finally, before proceeding to replica wormholes, we pause to note that the Hilbert space interpretation on which we have concentrated thus far is not unique. It arises from one particular way of cutting the path integral, regarding $n$ sets of boundary conditions as forming a `ket' state, and taking an overlap with the $n$ conjugate boundary conditions forming the `bra' state. The same path integral can also be cut in several different ways, giving rise to different Hilbert space interpretations -- though always involving the same baby universe Hilbert space $\hbu$. Any such cut splits asymptotic boundaries into two sets, depending on which side of the cut they lie. One set defines a `ket' state and the other defines the `bra', with the overlap between the two being obtained by a sum over intermediate baby universe states joining the two sets.

The different interpretations are readily be described in the operator language by writing  an amplitude as the expectation values of products of $\hat{\alpha}_{a}$, $\hat{\alpha}_{a}^\dag$ in the Hartle-Hawking state $|\HH\rangle$ (or in another state). Since these operators all commute, we can move any subset of them to the right where they act on the `ket' state and move the remainder to the left to act on the `bra.' And we can finally insert a complete set of baby universe states between them.

We illustrate this with a simple example computing the $n=2$ amplitude,
\begin{equation}
	\langle i_1,i_2|\rho^{(2)}|j_1,j_2\rangle = \big\langle\HH\big|\hat{\Psi}_{j_1}^\dag\hat{\Psi}_{j_2}^\dag  \hat{\Psi}_{i_1}\hat{\Psi}_{i_2} \big|\HH\big\rangle.
\end{equation}
We interpreted this earlier as the overlap between the states $\hat{\Psi}_{i_1}\hat{\Psi}_{i_2} \big|\HH\big\rangle$ and $\hat{\Psi}_{j_1}\hat{\Psi}_{j_2} \big|\HH\big\rangle$, so that the intermediate states consisted of two baby universes. Alternatively, we could reorder the boundary-inserting operators $\hat \Psi$, $\hat \Psi^\dagger$ to write the amplitude as the overlap between states $\hat{\Psi}_{i_1}\hat{\Psi}^\dag_{j_1} \big|\HH\big\rangle$ and $\hat{\Psi}_{i_2}\hat{\Psi}^\dag_{j_2} \big|\HH\big\rangle$. The intermediate states are then $|\HH\rangle$ (corresponding to the trivial contribution where the black hole interiors are not swapped) and states $|a,\bar{b}\rangle$ of one baby universe and one anti-universe (corresponding to the nontrivial PS wormhole):
\begin{equation}\label{eq:BUantiBU}
	\hat{\Psi}_{i}\hat{\Psi}^\dag_{j} \big|\HH\big\rangle = \sum_{a,b}\psi_{ai}\bar{\psi}_{bj} (\delta_{ab}|\HH\rangle + |a,\bar{b}\rangle).
\end{equation}

This interpretation \eqref{eq:BUantiBU} is not the most natural one if we wish to describe an intermediate state in real time.  Indeed,  it is somewhat analogous to
describing intermediate states exchanged in the T-channel of some QFT scattering process, which would naively be associated with
a Hilbert space for the QFT associated to a timelike surface that splits space into two parts (as opposed to the usual Hilbert spaces associated with spacelike Cauchy surfaces). However, in the operator description above the intermediate states continue to lie in the same baby universe Hilbert space $\hbu$.  This Hilbert space description will prove useful in the context of replica wormholes.  In particular, it will be straightforward to adapt this description to incomplete measurements at $\scri^+$ by taking a partial sum over the indices $i,j$ to trace out the unobserved piece of the state.

\subsection{Replica wormholes as baby universe interactions}
\label{sec:RWBUI}

We now incorporate the replica wormholes introduced in section \ref{sec:replicas} into our discussion of baby universes. We can think of the Polchinski-Strominger proposal discussed above as a theory of `free' baby universes, in the sense that $\hbu$ is a Bosonic Fock space.  The replica wormholes then modify the inner product on $\hbu$ by incorporating `interactions' between baby universes.

For now, we will continue to make the PS assumption that allows us to treat the union of $\scri^+$ and $\Sigma_\mathrm{int}$ as a Cauchy surface for an evaporating black hole, where we remind the reader that $\Sigma_\mathrm{int}$ runs from the regular origin below the black hole singularity out to the endpoint of evaporation $\evap$. This is a useful crutch to simplify the exposition but, as we will explain later, we will be able to upgrade the argument so as to remove this assumption.  In addition, for simplicity here we will only consider replica wormholes such that the island $\island$ on which we join the replicas lies inside the event horizon. This is well-motivated, since a QES is guaranteed to lie behind the event horizon under the assumption of the quantum focussing conjecture \cite{Bousso:2015mna} (though this does not directly apply to replica wormholes for $n>1$). In such a case, we can choose our Cauchy surface $\Sigma_\mathrm{int}$ for the black hole interior to pass through $\gamma = \partial\island$.

Now, just as the Polchinski-Strominger wormholes induced extra terms in the inner product \eqref{eq:PSIP} by pairing baby universes with permutations, the replica wormholes introduce new terms with a permutation acting only on the associated island.  We thus write
\begin{equation}\label{eq:RWIP}
	\langle b_1,\ldots,b_n|a_1,\ldots,a_n \rangle_{\mathrm{BU}} \supset  (\langle b_1|\otimes \cdots \otimes \langle b_n|) U_\pi(\island) (|a_1\rangle\otimes \cdots |a_n \rangle),
\end{equation}
where the notation $\supset$ means `contains terms of the following form'.
The states and inner products on the right hand side of this equation are taken in the tensor product of $n$ copies of the black hole interior Hilbert space, $\hilb_\mathrm{int}^{\otimes n}$. The operator $U_\pi(\island)$ acts as the permutation $\pi$ on those parts of the $n$ copies of $\hilb_\mathrm{int}$ associated with the island $\island$. If we take $\island=\Sigma_\mathrm{int}$, we recover the terms in \eqref{eq:PSIP}. We can be a little more explicit by choosing a basis of states $|a\rangle = |a',a''\rangle$ for $\hilb_\mathrm{int}$ which factorises between an orthonormal basis of states $|a'\rangle$ for the island $\island$ and a corresponding basis  $|a''\rangle$ for its complement:
\begin{equation}\label{eq:RWIP2}
	\langle b_1,\ldots,b_n|a_1,\ldots,a_n \rangle_{\mathrm{BU}} \supset \delta_{b_1'a_\pi(1)'}\delta_{b_1''a_1''}\cdots  \delta_{b_n'a_\pi(n)'}\delta_{b_n''a_n''}.
\end{equation}

Note that adding analogous terms to the inner product in a continuum quantum field theory would naturally given a vanishing contribution.  Indeed, in direct parallel with our discussions on $\scri^+$, for $n$ copies of a given state they would compute $e^{-S_n^{QFT}[\island]}$ where,
$S_n^{QFT}[\island]$ is the R\'enyi entropy of $\island$.  Such contributions are then exponentially suppressed by the area of $\gamma$ in units of the cutoff.\footnote{This fact is deeply related to the fact that the Hilbert space of quantum field theory does not factorize into a tensor product of a Hilbert space for $\island$ and another Hilbert space for its complement.} However,
as we discovered by computing amplitudes with the path integral in section \ref{sec:QES}, making gravity dynamical naturally  leads to finite contributions from the terms on the right-hand-side of \eqref{eq:RWIP} or \eqref{eq:RWIP2}.
Thus we should think of the states $|a\rangle$ as encoding not only the matter state, but also geometrical degrees of freedom (perhaps including the location of $\gamma$ in the split $|a',a''\rangle$).

Note that if we interpret the sum over states $|a\rangle$ in a natural way as a sum over real Lorentzian geometries, the saddle-point replica wormholes discussed in section \ref{sec:replicas} do not appear directly since they are complex. The direct sum over states $|a\rangle$ will be a sum of highly oscillatory phases, which (as is familiar from steepest descent integrals) can be evaluated by deforming the contour. For further discussion see \cite{CDMRW}.

Using language analogous to that of the Feynman diagrams of perturbative QFT,
we can think of the contribution \eqref{eq:RWIP} to the baby universe inner product as an interaction, giving a `vertex' for $n\longrightarrow n$ `scattering' of baby universes.  Indeed, we can borrow standard techniques from perturbative QFT to compute the associated effect on the expression
in \eqref{eq:GaussianBU} for the inner product in terms of integrals over $\alpha$. To include a replica wormhole, we insert a product of $n$ $\alpha$ fields and $n$ $\bar{\alpha}$ fields, summing over indices to induce the required connections. For example, for $n=2$ we insert a term
\begin{equation}
	\sum_{a'_1,a''_1,a'_2,a''_2} \alpha_{a'_1,a''_1}\bar{\alpha}_{a'_2,a''_1} \alpha_{a'_2,a''_2}\bar{\alpha}_{a'_1,a''_2}
\end{equation}
into the integrand on the right-hand-side of \eqref{eq:GaussianBU}. Summing over all possible combinations of replica wormholes exponentiates this factor (and similar terms for all $n$) so that it modifies the original measure $d\mu(\alpha)$ from $e^{-\sum_a\alpha_a\bar{\alpha}_a}$ to a rather complicated non-Gaussian measure.

On the other hand, aside from this modification of the inner product (and the corresponding changes to the wavefunction of the Hartle-Hawking state and the algebra of universe creation and annihilation operators), there are no further changes to either the arguments or the conclusions of section \ref{sec:BUensembles}. In particular, the expression
\begin{equation}
	\rho^{(n)} = \int d\mu(\alpha) \left(|\Psi^\alpha\rangle \langle \Psi^\alpha|\right)^{\otimes n}
\end{equation}
for the $n$-evaporation density matrix given in equation \eqref{eq:rhonalpha} remains true, with the modified measure $d\mu(\alpha)$ described above.  The details of this measure are not of primary importance for us, except that the modified measure is dominated by states $|\Psi^\alpha\rangle$ of radiation which follow the Page curve (see section \ref{sec:summary} for justification).

\subsection{Baby universes with semiclassical control: dropping the PS assumption}
\label{sec:DropPS}

The above sections developed the notion of PS baby universes and the associated $\hbu$ using the PS assumption.  This allowed us to give a very explicit treatment of the `saddle-point geometries', the associated amplitudes, and the resulting inner product on $\hbu$.  However, it turns out that the most important lessons from the Hilbert space interpretation do not rely on the PS assumption.  These lessons include i) the existence of a baby universe Hilbert space $\hbu$, ii) that the inner product on $\hbu$ is determined  by the path integral, and iii) the fact that asymptotic quantities define simultaneously diagonalizable operators on $\hbu$ and the associated existence of superselection sectors.

As we now show, all of these results can be derived using physics that remains fully under semiclassical control.  However,  the arguments are necessarily more abstract than those using the PS assumption above.  Some readers may thus choose to skip this section on a first reading of this paper.

To proceed, we follow the same basic strategy as in our study of R\'enyi entropies in section \ref{sec:replicas}.  Indeed, we will obtain a Hilbert space interpretation by slicing open the path integrals and the associated replica wormhole saddles discussed in section \ref{sec:repQES}. We thus specify the state on $\scri^+$ only on the subset $\scri_u$, choosing $u < u_\evap$ so that $\scri_u$ does not intersect the future light cone of $\evap$.
We will then sum over all boundary condition on the rest of $\scri^+$.  We may then expect the relevant saddles to remain under semiclassical control as desired.

In particular, since we impose boundary conditions only on $\scri_u$, we may cut the path integral along Cauchy surfaces $\Sigma_u$ which extend to meet the asymptotic boundary $\scri^+$ at the associated retarded time $u$.  We may then further choose $\Sigma_u$ to be well to the past of both the singularity and $\evap$.

In a replica setting, we require several such cuts. The resulting Hilbert spaces $\hilb_{n}$ associated with such cuts are labelled by the number $n$ of boundaries on which these cuts end.\footnote{As described for the Euclidean context in \cite{Marolf:2020xie}, the Hilbert spaces are in fact labelled by the asymptotic geometry of the slices $\Sigma_u$.  For simplicity we restrict to the case where the asymptotic regions are defined by $n$ spheres.  We also mention that, in the current context, there is a notion of `anti-boundary' or `conjugate boundary' (associated with the anti-baby universes discussed below), such that the most general Hilbert space $\hilb_{n,\bar n}$ for the case of sphere boundaries is associated with $n$ boundaries and $\bar n$ anti-boundaries.
(There is a natural linear isomorphism from the dual space $\hilb_{n,\bar{n}}'$ to $\hilb_{\bar{n},n}$.)  Finally, while one might at first expect the Hilbert space to also depend on the advanced times $u$ associated with the location of these spheres on $\scri^+$, choosing a notion of time-translation on $\scri^+$ allows one to canonically identify Hilbert spaces with different values of $u$.} Although in the Hawking saddle it arises from $n$ copies of some given $\Sigma_u$, we emphasize that
 $\hilb_n$ is not just the product $\hilb_1^{\otimes n}$ due to contributions from replica wormholes. 
 In particular, as we discuss below, the Hilbert space $\hilb_0$ without boundaries is not the trivial Hilbert space, but should  instead be the space $\hbu$ of closed universes.

With this small change in boundary conditions, most of the considerations above will continue to hold. We simply take $|i\rangle$ to label a basis of states on $\scri_u$ rather than on the entirety of $\scri^+$, and  we take $|a\rangle$ to be a basis of states on a Cauchy surface $\Sigma_u$ meeting the boundary at time $u$. If we consider the path integral for any single copy of the spacetime, truncate it at $\Sigma_u$, and impose boundary conditions for the quantum fields on $\scri_u$ and $\Sigma_u$, the result computes a wavefunction $\sum_{i,a}\psi_{ai}(u)|a\rangle\otimes |i\rangle$ on $\Sigma_u\cup \scri_u$ for a state in $\hilb_1 \otimes \hilb_u$ (with $\hilb_u$ the Hilbert space on $\scri_u$).

Furthermore, we can write states on the $n$-boundary Hilbert space $\hilb_n$ as linear combinations of a basis $|a_1,\ldots,a_n\rangle$. The notation here is similar to that used above for $n$ baby universes, but there is a crucial difference.  Because we treat asymptotic boundaries as distinguishable, the order of the $a_i$ {\it is} important. The states
$|a_1,a_2\rangle$ and $|a_2,a_1\rangle$ are not the same, and  $\hilb_n$ does not exhibit Bosonic statistics.  This is associated with the fact that
 we specify the asymptotic identifications between Cauchy slices $\Sigma_u$ as part of the boundary conditions, so  there can be no terms in the inner product that permute copies of $\Sigma_u$ in their entirety.

Nonetheless, we still find contributions to the inner product from replica wormholes.  Such contributions again permute island regions $\island$ just as in  equations \eqref{eq:RWIP}, \eqref{eq:RWIP2}:
\begin{align}\label{eq:RWIP3}
	\langle b_1,\ldots,b_n|a_1,\ldots,a_n \rangle &\supset  (\langle b_1|\otimes \cdots \otimes \langle b_n|) U_\pi(\island) (|a_1\rangle\otimes \cdots |a_n \rangle) \\
	&\supset \delta_{b_1'a_\pi(1)'}\delta_{b_1''a_1''} \cdots \delta_{b_n'a_\pi(n)'}\delta_{b_n''a_n''}.
\end{align}
In the second line we have split the index $a$ in two, so the state $|a\rangle = |a',a''\rangle$ on $\Sigma_u$ is labelled by the state $a'$ on the island $\island$ and $a''$ on its complement $\Sigma_u\setminus \island$, where $\Sigma_u\setminus \island$ now extends to infinity. Translating the discussion above to this notation, the boundary conditions require that the $a''$ indices must be paired without permutation, while replica wormholes give rise to the permutation $\pi$ acting on the $a'$ indices.

Again we may use $\Psi_i$ to denote the boundary condition that fixes both matter at $\scri^-$ that collapses to form the black hole and  a  state $|i\rangle$ on the Hilbert space of state on $\scri_u$.  And again we may take $\Psi_i$ to define an operator $\hat{\Psi}_i$ on the Hilbert spaces $\hilb_n$.  But now this operator adds a boundary, increasing the value of $n$.  Thus we write $\hat{\Psi}_i(u): \hilb_n \to \hilb_{n+1}$.  Using our bases, the action of this operator takes the form
\begin{equation}\label{eq:Psiiu}
	\hat{\Psi}_i(u) |a_1,\ldots,a_n\rangle = \sum_a \psi_{ai}(u)|a,a_1,\ldots,a_n\rangle \,.
\end{equation}
Because boundaries are distinguishable, it is important that we added the extra label $a$ to the first slot (we could, if desired, define other versions of $\hat{\Psi}_i(u)$ which choose a different ordering). In particular, it means that these operators no longer commute.  And in any case we cannot talk about diagonalizing them since they map between different Hilbert spaces. Intuitively, this is because the Hilbert spaces $\hilb_n$ carry information not just about the closed baby universes, but also about the state that escapes to $\scri^+$ after time $u$. We would thus like to `trace out' this extra information, leaving only the piece of the state truly associated with baby universes and which mediates the correlations on $\scri_u$ and gives rise to the Page curve.

\subsubsection{Replica Wormhole Baby Universes}

A convenient but abstract method of avoiding the extra information involves using the adjoint operators $\hat{\Psi}_i^\dag(u)$.  By definition, the adjoint operators map between Hilbert spaces in the opposite direction to $\hat{\Psi}_i$, so we have $\hat{\Psi}_i^\dag(u): \hilb_{n+1}\to  \hilb_{n}$. The compositions $\hat{\rho}_{ij}(u) := \hat{\Psi}_j^\dag(u)\hat{\Psi}_i(u)$ are then operators that map $\hilb_{n}$ to itself for any $n$, and for $n=0$ in particular act within the closed universe Hilbert space $\hilb_0 = \hbu$.

More concretely, the adjoint operator $\hat{\Psi}_j^\dag(u)$ acts by inserting a conjugate boundary (with boundary condition labelled by $j$) and gluing it to  an asymptotic boundary associated with the state on which it acts (the first such boundary, since in \eqref{eq:Psiiu} we defined $\hat{\Psi}_i(u)$ to add a boundary in the first slot).   As in our discussion above, this gluing of asymptotic boundaries requires a corresponding gluing of the respective spacetimes on the asymptotic part of $\Sigma_u$.  But again we allow all possible gluings deeper in the bulk, and in particular we allow nontrivial replica wormholes.

\begin{figure}
	\centering
	\raisebox{75pt}{\scalebox{1.5}{$\hat{\rho}_{ij}(u) \longrightarrow$}}\quad
	\includegraphics[width=.45\textwidth]{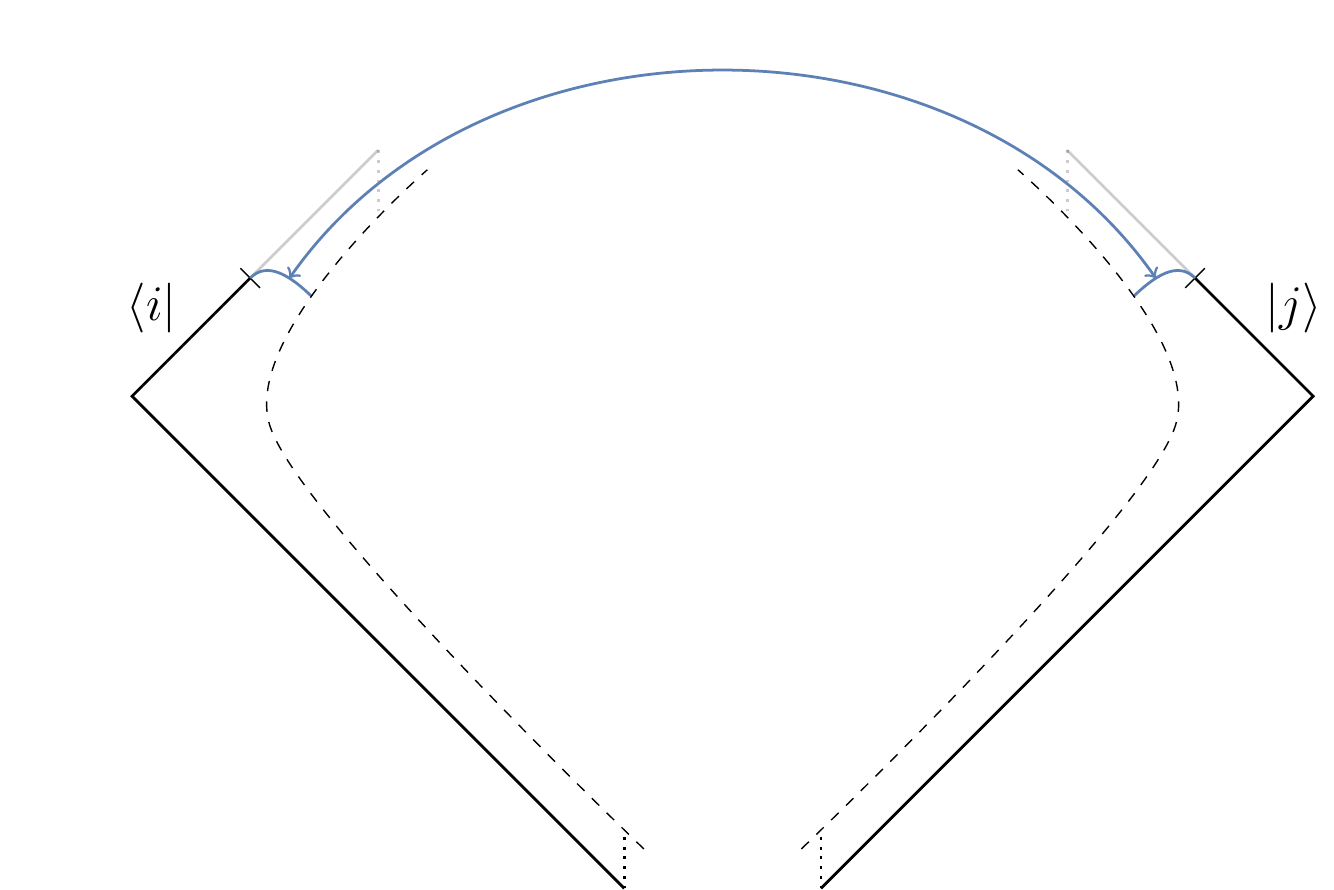}
		\caption{The boundary conditions corresponding to the operator $\hat{\rho}_{ij}(u)$. We have flat asymptotic regions as pictured, with matter boundary conditions at $\scri_u$ specified by the states $|i\rangle,|j\rangle$. The Cauchy slices meeting $\scri^+$ at time $u$ must be identified as shown in the asymptotic region. We do not specify what happens to the spacetime away from the asymptotic region, inside the dashed curved. In particular, this allows the spacetime to connect with other boundaries, and the future Cauchy slices need not be identified in the same way in their entirety. For example, the spacetimes in figure \ref{fig:replicarho2} involve two copies of such boundary conditions, computing $\Tr(\rho^{(2)}(u)^2) = \sum_{ij}\langle\HH|\hat{\rho}_{ij}(u)\hat{\rho}_{ji}(u)|\HH\rangle$. \label{fig:rhouBC}}
\end{figure}

The composition  $\hat{\rho}_{ij}(u)= \hat{\Psi}_j^\dag(u)\hat{\Psi}_i(u)$ thus acts by inserting a boundary condition corresponding to a complete in-in contour, as shown in figure \ref{fig:rhouBC}. This is the boundary condition one would choose for computing the components of the density matrix of Hawking radiation on $\scri_u$, which justifies the choice of notation.

Using  the general argument from \cite{Marolf:2020xie} reviewed in section \ref{sec:lessons} above, it follows that the operators $\hat{\rho}_{ij}(u)$ mutually commute on $\hbu$. Furthermore, they can be simultaneously diagonalised by a basis of $\alpha$-states, giving rise to superselection sectors and ensembles as before. In a given superselection sector (an $\alpha$-state),
the eigenvalues $\rho^\alpha_{ij}(u)$ of the $\hat{\rho}_{ij}(u)$  are interpreted as the components of the density matrix for the Hawking radiation in that superselection sector that emerges before time $u$.

As before, the superselection sectors mean that Hawking radiation emerging from one black hole evaporation is correlated with that emerging from another.  These are classical correlations, described by a classical a probability distribution determined by the decomposition into $\alpha$-states of the specified baby universe state from $\hbu$ (which in the cases discussed above is the Hartle-Hawking no-boundary state $|\HH\rangle$).

In this language, the swap R\'enyi entropies computed in section \ref{sec:replicas} were amplitudes of the form
\begin{equation}
\label{eq:corrho}
	 \sum_{i_1,\ldots,i_n}\langle\HH|\hat{\rho}_{i_1i_2}(u)\hat{\rho}_{i_2i_3}(u)\cdots \hat{\rho}_{i_{n}i_1}(u)|\HH\rangle.
\end{equation}
That is, they were correlation functions in the Hartle-Hawking state of products of the $\hat{\rho}_{ij}(u)$. The replica wormholes gave particular contributions to \eqref{eq:corrho} and, as explained in section \ref{sec:RWother}, they should also contribute to more general amplitudes
$\hat{\rho}_{ij}(u)$.

If we insert intermediate states of $\hbu$ between insertions of $\hat{\rho}$ in \eqref{eq:corrho}, the resulting Hilbert space interpretation generalises \eqref{eq:BUantiBU} above. We give some interpretation of the intermediate states due to replica wormholes in a moment. But this is not the most natural way to describing intermediate states of baby universes in a real-time process, between consecutive experiments on different black holes. For a Hilbert space description that achieves this aim, see appendix \ref{sec:RWBUE}.

\subsubsection{What is a baby universe?}
\label{sec:Whatbaby}

We now pause to more carefully explore the notion of baby universe associated with this replica wormhole construction of $\hilb_n$ and $\hbu$.  We will refer to the result as a replica wormhole baby universe (RWBU), to contrast it with the Polchinski-Strominger notion of baby universe (PSBU) discussed in sections \ref{sec:HSHPS}-\ref{sec:RWBUI}.

\begin{figure}[h]
\centering
\begin{subfigure}[t]{.9\textwidth}
	\centering
	\raisebox{-10pt}{
	\includegraphics[width=.45\textwidth]{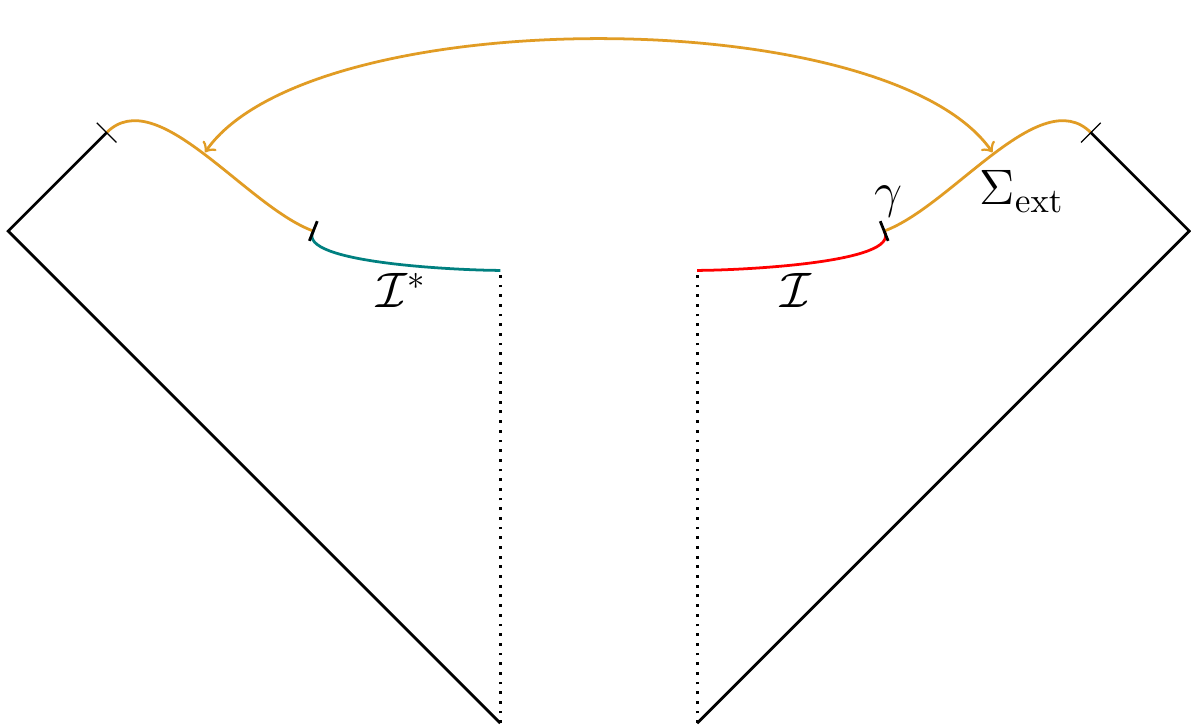}}
	\raisebox{55pt}{\scalebox{1.5}{$\longrightarrow$}}\quad
	\includegraphics[width=.35\textwidth]{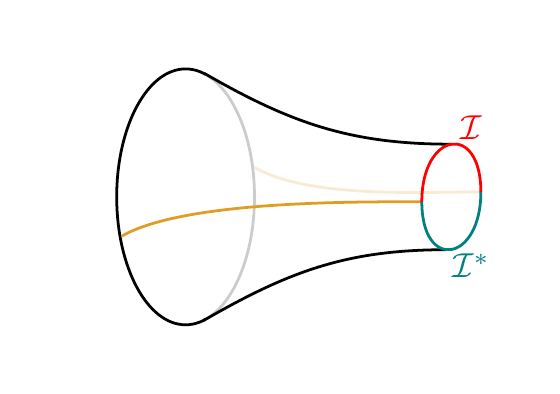}
	\caption{Each of the $n$ replicas (as shown on the left) making up a replica wormhole has topology $S^{D-1}$ times an interval, with the asymptotic boundary at one end of the interval, and two conjugate copies of the island ($\island$ and $\island^*$ from `ket' and `bra' spacetimes respectively) joined along their common $S^{D-2}$ boundary $\gamma$ at the other. A Euclidean continuation resembles the geometry on the right, which could be described in terms of propagation of a RWBU with topology $S^{D-1}$, of which $\island$ and $\island^*$ make up the Northern and Southern hemispheres respectively.\label{fig:PSBUa}}
\label{fig:PSrho2a}
\end{subfigure}
\begin{subfigure}[t]{.9\textwidth}
	\centering
	\raisebox{30pt}{
	\includegraphics[width=.35\textwidth]{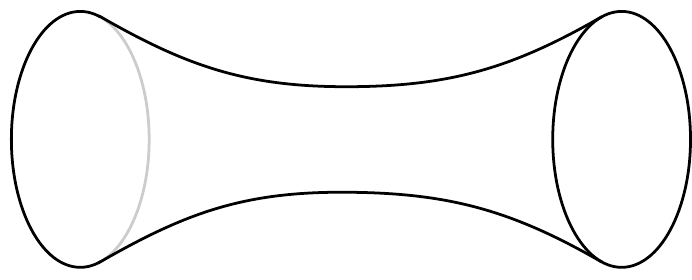}}
	\qquad\qquad
	\includegraphics[width=.3\textwidth]{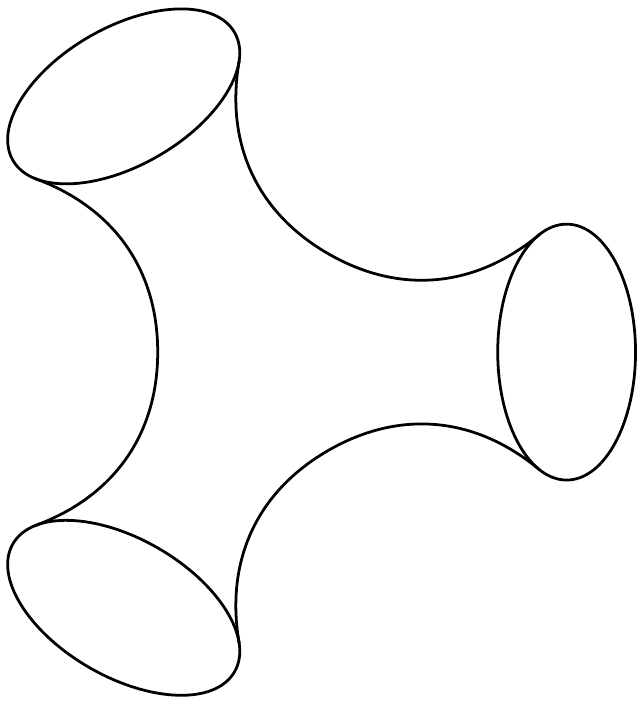}
		\caption{To build a replica wormhole, we sew $n$ of the constituents above together along $\island$ and $\island^*$. We picture the resulting Euclidean spacetimes for $n=2,3$, which we can think of as propagation of a RWBU ($n=2$, left), or an interaction of RWBUs ($n\geq 3$, right).\label{fig:PSBUb}}
\label{fig:PSrho2b}
\end{subfigure}
\caption{We may describe the correlations between sets of Hawking radiation arising from replica wormhole spacetimes as mediated by exchange of `replica wormhole baby universes' (RWBUs) appearing in intermediate states such as $\hat{\rho}_{ij}(u)|\HH\rangle\in\hbu$. Such a state is somewhat unusual in Lorentzian signature, but has a natural Euclidean continuation.\label{fig:PSBU}}
\end{figure}

In particular, let us consider the  intermediate states $\hat{\rho}_{ij}(u)|\HH\rangle \in \hbu$ that mediate the correlations between boundaries in \eqref{eq:corrho}. Since this is the natural object in a replica wormhole discussion, we will call it an RWBU.  Similar states were considered at the end of section \ref{sec:lessons}, where using the PS assumption we interpreted them containing both a PS baby universe and a PS anti-baby universe.  This conclusion must be slightly modified  when we consider replica wormholes instead of PS wormholes, and in particular when we wish to avoid universes with Planckian curvature such as those that end at $\evap$.  However, we can still think of our intermediate state as naturally containing two parts.  The first part may be called a  `partial baby universe', consisting of only the island region on some $\Sigma_u$, while the second part is a partial anti-baby universe.  The two much match at the boundary of the island, and they are joined at $\gamma=\partial \island$.  We could thus perhaps refer to this RWBU as a `BU---anti-BU bound state'. See figure \ref{fig:PSBUa} (left), where the RWBU consists of the red slice (labelled $\island$) and the teal slice (labelled $\island^*$), joined at $\gamma$.

As described after \eqref{eq:BUantiBU}, this notion of `intermediate state' may seem unnatural in Lorentz signature.  Nevertheless, it is the correct Lorentzian continuation of a natural Euclidean notion of intermediate state.  To see this, note that a Euclidean version of the boundary condition $\hat{\rho}_{ij}(u)$ is a spacetime which asymptotes to a closed Euclidean manifold $\mathcal{B}$. For our case of a black hole formed from collapse $\mathcal{B}$ has the topology $S^{D-1}$ (for a $D$-dimensional spacetime), with the two hemispheres of $S^{D-1}$ corresponding to `ket' and `bra' segments of the boundary, joined along an asymptotic spatial $S^{D-2}$. Figure \ref{fig:PSBUa} (right) shows the resulting Euclidean continuation of each replica. A replica wormhole will join $n$ such boundaries as shown in figure \ref{fig:PSBUb} for $n=2,3$.

For example, for $n=2$ the topology of spacetime is $\mathcal{B}$ times an interval, with a boundary lying at each end of the interval. It is then very natural to describe this cylinder in terms of a baby universe with topology $\mathcal{B}$ propagating between the two boundaries. For example, \cite{Almheiri:2019qdq} constructed replica wormholes in a two-dimensional spacetime for a two-sided black hole, with topology of a cylinder for $n=2$, a pair of pants for $n=3$ and so forth (as in figure \ref{fig:PSBUb}). From the Euclidean perspective, it is natural to think of such wormholes as describing interactions between closed universes. And when we analytically continue to Lorentzian signature in the correct sense to describe our density matrix boundary conditions, we arrive at precisely the situation described above. The  `island'  from any ket part of the Lorentzian replica wormhole spacetime is then just half of $\mathcal{B}$, with the other half being the island from a bra part of the replica wormhole.

\section{Discussion}
\label{sec:Disc}

\subsection{Summary}\label{sec:summary}

This work has focussed on describing semiclassical expectations for experiments performed on Hawking radiation collected at $\scri^+$ in an asymptotically flat spacetime. To formulate and perform the relevant computations, we used the Lorentz-signature gravitational path-integral, which in the semiclassical limit involves a sum over saddle-points.  In gravity, as in field theory, classifying all possible saddles tends to be rather difficult, so in practice one works to identify interesting saddles and hopes that they dominate the amplitudes of interest.  We thus began with the familiar Hawking saddle described in the form of figure \ref{fig:rhoEvaporatingSC}, which can be used to compute a density matrix $\rho(u)$ for the Hawking radiation arriving at $\scri^+$ before some retarded time $u$, and hence expectation values of operators acting on that radiation or associated (R\'enyi) entropies. If $u$ is sufficiently to the past of the future lightcone of the endpoint of evaporation $\evap$, the geometry of the Hawking saddle is weakly curved, and all perturbative corrections are small. Of course, at late retarded times the resulting entropies of the Hawking radiation far exceed the Bekenstein-Hawking entropy of the remaining black hole.  We refer to this phenomenon as a violation of Bekenstein-Hawking unitarity (BH unitarity).

No part of our later discussion caused any direct modification of the above conclusions. However, we noted that observers who possess only a single copy of a system cannot experimentally measure its entropy.  We thus imagined experiments to verify the above violation of BH unitarity that involved forming and evaporating $n$ identical black holes, collecting the decay products of each, and identifying the `early' subset of each collection that was emitted before some particular retarded time. We then asked our experimenter to measure a swap operator that acts as a permutation among these $n$ early subsets, but which leaves the remaining late subsets fixed. Such observations performed on many copies of identical-but-independent quantum states give a direct way of measuring entropies, and we accordingly refer to the associated expectation values as `swap entropies'.

In the limit where the $n$ black holes are well separated, we may approximate each black hole formation and evaporation as occurring in a separate asymptotically flat region of spacetime. The boundary conditions on our gravitational path integral then involve $n$ separate asymptotic boundaries. But in performing the sum over all geometries with such boundary conditions we allowed for so-called `spacetime wormholes', which we define as geometries which connect distinct asymptotic boundaries. Such geometries introduce correlations between the $n$ sets of early Hawking radiation, so that the state $\rho^{(n)}(u)$ of these $n$ sets is not in fact equal to the tensor product $\rho(u)^{\otimes n}$. By this mechanism, one might hope that observables such as the swap entropies will be nevertheless be compatible with BH unitarity.

Such an approach was advocated by Polchinski and Strominger in \cite{Polchinski:1994zs}. They considered including in the path integral a class of `PS wormhole' spacetimes shown in figure \ref{fig:PSrho2}, where the various interior connections between copies of $\rho(u)$ are `swapped' in all possible ways relative to the Hawking saddle. We reviewed this proposal in section \ref{sec:PS} to introduce the idea without the technicalities of replica wormholes, finding swap entropies that share certain features with BH-unitarity-compatible Page curves. However, on closer inspection this model continues to violate BH unitarity (and perhaps also causality), as well as requiring us to regard the PS wormholes, which contain a singular and strongly-curved region $\evap$, as saddle-points.

Nevertheless, we saw in section \ref{sec:replicas} that there are other saddles which resolve these issues.  In particular, our swap experiments receive contributions from replica wormholes analogous to those described in \cite{Almheiri:2019qdq,Penington:2019kki}, which are closely associated with the quantum extremal surfaces studied in \cite{Almheiri:2019psf,Penington:2019npb}.  We thus briefly reviewed results and arguments from those references, translating them to our Lorentz-signature asymtptotically-flat setting.  The result is then that our swap $n$-R\'enyi entropies -- at least in the $n\rightarrow 1$ limit where calculations are more tractable -- perfectly reproduce the Page curve associated with BH unitary. This is powerful evidence in favor of the idea that there is an operational sense in which the Bekenstein-Hawking entropy is indeed the black hole density of states.

Finally, section \ref{sec:BUetc} incorporated these ideas into a conceptual framework for the gravitational path integral in the presence of spacetime wormholes, building on the insights of Coleman and Giddings and Strominger \cite{Coleman:1988cy,Giddings:1988cx,Giddings:1988wv}. A particular goal was to reconcile the operational verification of the Page curve described above with the apparent failure of BH unitarity associated with R\'enyi entropies of the Hawking-saddle Hawking radiaton.  To do so, we sought a Hilbert space interpretation by cutting open the relevant contributions to the path integral. This led us to slice open spacetime wormholes along surfaces which do not meet asymptotic boundaries, and which we associated with a Hilbert space of closed `baby' universes $\hbu$. The correlations between multiple sets of Hawking radiation can then be understood as arising from a sum over intermediate states of these baby universes.

It is crucial that the correlations between sets of Hawking radiation can be described are both strict and classical; i.e., any observer who forms and evaporates identical black holes will find identical sets of Hawking radiation, but the particular radiation state obtained may be thought as of as being chosen from a classical probability distribution.  To explain this feature, we considered the expectation values or matrix elements of asymptotic observables in particular states and other asymptotic quantities that one might expect to be c-numbers.  We found that such quantities in fact yield {\it operators} acting on $\hbu$, defined by inserting the relevant boundary conditions in the gravitational path integral. For example, there is an operator $\hat{\rho}_{ij}(u)$ on $\hbu$ for each $ij$ component of the density matrix of Hawking radiation before time $u$. But as argued in \cite{Marolf:2020xie}, all such operators can be simultaneously diagonalized.  In particular, it is easy to show that they
mutually commute.  Since we defined the path integral to sum over all topologies with the required boundary conditions, the output of the path integral cannot  depend on any ordering of the multiple disconnected boundaries.  See also footnote \ref{foot:normal} for the argument that these operators are normal, in the sense that they also mutually commute with their adjoints.  This means that $\hbu$ splits into superselection sectors for the algebra of asymptotic observables.  In other words,  there is a basis of simultaneous eigenstates $\big |\alpha \big \rangle$ of $\hbu$ for all such operators. The correlations between multiple sets of Hawking radiation can then be described as classical correlations from a probability distribution of superselection sectors.

Explicitly, applying this to calculations of the density matrix of Hawking radiation we can write
\begin{align}
	\rho^{(n)}_{i_1\ldots i_n j_1\ldots j_n}(u) &= \big\langle \HH\big| \hat{\rho}_{i_1j_1}(u) \cdots \hat{\rho}_{i_nj_n}(u) \big|\HH\big\rangle \\
	&= \int d\mu(\alpha) \, \rho_{i_1j_1}^\alpha (u) \cdots \rho_{i_nj_n}^\alpha (u) \\
	&= \Big\langle \rho_{i_1j_1} (u) \cdots \rho_{i_nj_n} (u) \Big\rangle \,.
\end{align}
The first line writes our $n$-copy density matrix as an expectation value in the `no-boundary' state $\big|\HH\big\rangle\in\hbu$ of baby universes, which was an implicit choice in our earlier calculations. By inserting a complete set of $\big |\alpha \big \rangle$ states, we write this as an average over superselection sectors, with probability measure $d\mu(\alpha) = |\langle\alpha|\HH\rangle|^2 d\alpha$ (where $d\alpha$ is defined so that the completeness relation $\int d\alpha \, |\alpha\rangle\langle \alpha| = \id$ holds). This defines the notation of the final line, where we write this as an expectation value of random variables $\rho_{ij}(u)$ selected from the ensemble defined by the measure $d\mu$. And while the set of all possible $\alpha$-values will be determined by state-independent considerations involving the algebra of our operators on $\hbu$, the formulae above make manifest that our results depend on the choice of state $|\HH\rangle\in\hbu$ through the measure $\mu(\alpha)$.  A different choice of state results in a different measure, with an extreme example being an $\alpha$-state giving $\delta$-function measure.

This framework finally allows us to reconcile the entropy results described above. So long as the initial baby universe state is $|\HH \rangle$,  the state of Hawking radiation from any given black hole is $\rho_\mathrm{Hawking}(u) =\big\langle \rho(u) \big\rangle$. Its entropy grows with $u$ and fails to follow the Page curve due to entanglement with baby universes.  But, as is always the case for superselection sectors, this entanglement is unobservable. Evaporating additional black holes induces further entanglement with the same baby universe states, correlating the decay products so that measurements designed to deduce the entropy produce the Page curve with the help of replica wormholes.\footnote{For much the same reason, the energy conservation critique \cite{Banks:1983by} does not apply.  There is no experimentally-accessible `dollar matrix'.  Indeed, as described long ago in \cite{Coleman:1988cy,Giddings:1988cx,Giddings:1988wv},  the situation is more similar to that discussed in \cite{Unruh:2012vd} as the baby universes which provide decoherence
carry no energy or momentum.}  In particular, the swap test provides a measurement of $\big\langle \Tr(\rho(u)^n)\big\rangle$; it does not measure  $\Tr\left(\big\langle \rho(u)\big\rangle^n\right)$. Thus as emphasized in \cite{Giddings:2020yes}, replica wormholes do not compute the `true' von Neumann entropy of the state of the radiation; instead they give the entropy of the state projected to a typical superselection sector \cite{Marolf:2020xie}.

It is important that the value of $\Tr(\rho(u)^n)$ in almost any given $\alpha$-sector will be exponentially close to the average value $\big\langle \Tr(\rho(u)^n)\big\rangle$ computed by replica wormholes.  The Page curve is therefore a robust prediction, accurate up to exponentially small corrections. We say that $\Tr(\rho(u)^n)$ (or the entropy) is `self-averaging,' meaning that any given sample from the ensemble is parametrically likely to be parametrically close to the mean. A given quantity $X$ (which we take to be complex in general) is self-averaging if its ensemble variance
\begin{equation}
	\operatorname{Var}(X) =\big\langle X\bar{X} \big\rangle - \big\langle X\big\rangle\big\langle\bar{X} \big\rangle = \big\langle X\bar{X} \big\rangle_\mathrm{connected}
\end{equation}
is much smaller than its mean squared, $\operatorname{Var}(X) \ll \left|\big\langle X\big\rangle\right|^2$. Since the variance is the connected two-point correlator, it is computed gravitationally from a path integral with boundary conditions $X\bar{X}$ over spacetimes which connect the $X$ boundaries to the $\bar{X}$ boundaries. $X$ is self-averaging if these connected contributions are dominated by the disconnected spacetimes.

Now, without any obvious reason to exclude replica wormholes from the gravitational path integral, and in the absence of as-yet-unknown additional contributions (either semiclassical or invoking new physics), we are compelled to consider the following scenario for semiclassical gravity. The scenario is that it predicts observations to always be compatible with unitarity, and with the density of black hole states being given by the Bekenstein-Hawking entropy. But it does not predict the detailed unitary dynamics. Instead, semiclassical gravity gives definite predictions for coarse-grained questions, such as simple observables acting on the Hawking radiation or entropies, but it declines to provide a definite prediction for measurements of fine-grained quantities such as the off-diagonal elements of the density matrix of Hawking radiation.

We have focussed here on black holes in asymptotically flat spacetimes.  But analogous comments can be made in many other contexts as well.  Indeed, the original works \cite{Penington:2019npb,Almheiri:2019psf,Penington:2019kki,Almheiri:2019qdq,Marolf:2020xie} reviewed above (and on which much of this work was based) were performed with asymptotically AdS boundary conditions.  While it is common to study AdS settings with reflecting boundary conditions, one can also couple the AdS system to an auxiliary non-gravitational system that can absorb Hawking radiation and remove it from the asymptotically AdS spacetime.  We may view this as an analogue of the experimental processes described in the current paper, with the auxiliary system playing the role of the `quantum memory' into which our experimenter uploads the Hawking radiation's quantum state.  Studying the action of various swap operators on the non-gravitating auxiliary system then leads precisely to the replica wormholes and associated entropies studied in \cite{Penington:2019kki,Almheiri:2019qdq}.

\subsection{What have we gained?}
\label{sec:disc2}

We now we reflect on the position in which we are left after drawing such conclusions. In particular, it is natural to ask for a complete description of the physics in a particular superselection sector. Even for a single evaporation event, such a description must yield a density matrix that reproduces the Page curve, and for a single evaporation must do so without the help of replica wormholes.   We might then ask: what have we gained from the above considerations?\footnote{As has often been stressed to us by Steve Giddings, no clear answer was provided by the discussions of wormholes and baby universes from the 1980's and early 1990's.} To explore this question and the associated physics of superselection sectors, we will introduce some ideas that are not directly apparent from our considerations so far, but which were studied in more detail in \cite{Marolf:2020xie}.

First, however, we should briefly discuss the predictive status of a framework involving superselection sectors.  One issue is that, as pointed out by \cite{Susskind:1994cv}, the correlations between successive experiments mean that we cannot use a strict frequentist interpretation for the `probability' of getting a particular state of the radiation. Any given observer will decohere onto a branch of the wavefunction with a baby universe state tightly concentrated around some particular superselection sector. But it is natural to instead interpret `probabilities' of $\alpha$-states as minimal Bayesian priors, assigning credences to different possible superselection sectors and thus to particular states for the Hawking radiation.  Now, for general baby universe initial states this perspective allows us to make definite predictions only when certain features are common to all allowed superselection sectors, or when we consider self-averaging observables with parametrically sharply peaked probability distributions.  But this is also the only sense in which frequentist probability makes definite predictions for standard systems, though in that context one may use the number of experiments as a parameter controlling the width of the distribution.

It is important to emphasize that measuring the actual state of Hawking radiation is tantamount to experimentally determining, at least in part, the $\alpha$-sector in which we live. The situation is thus much the same as when working with a theory with unknown free parameters (and indeed, we could identify these parameters with coefficients in the effective action giving the S-matrix for black hole formation and evaporation). Alternatively, we can view the $\alpha$-state as determined by the initial conditions of baby universes as described above.  It thus has the same logical status as any other measurement of initial conditions, a situation which has been much discussed in cosmology and to which many of the same words will necessarily apply.

With the above as prologue, we now point out an important difference between the wormholes studied here and those studied in the late 1980's \cite{Coleman:1988cy,Giddings:1988cx,Preskill:1988na}. The earlier works primarily studied the effect of microscopic wormholes, and in particular of wormholes much smaller than any macroscopic scale of interest. In that case, they can be `integrated out', and the resulting ensemble of $\alpha$-states is describable as providing a distribution of random couplings for terms (such as the cosmological constant term) in a local effective action; see \cite{Hebecker:2018ofv} for a recent review. Each member of the ensemble is thus a local theory on the scale of interest. But since the typical scale of replica wormholes is that of the event horizon of the black hole undergoing evaporation, integrating out such wormholes will \emph{not} provide such a local effective theory on black hole scales.\footnote{One could think of it as defining a local effective theory on scales larger than the black hole, but then the black hole itself would simply be treated as a particle with a large but finite number of internal states.} So in our context the effect of $\alpha$ states cannot be absorbed into a shift of local coupling constants in a useful way. Indeed, this is just the sort of non-locality required for the scenarios discussed in \cite{Giddings:2011ks,Giddings:2012gc}.

It thus appears that we will not obtain a local semiclassical description of superselection sectors by integrating out topology changing processes. On the other hand, we might still ask if one can find a local semiclassical description that retains such processes, but which explicitly includes an initial $\alpha$-state for the baby universes.   The answer to this question will hinge on whether
$\alpha$-states lie in the regime of semiclassical validity.\footnote{Unless, perhaps, we introduce new objects to resum certain contributions: see section \ref{sec:ebranes}.}

This seems unlikely, and  semiclassical physics seems similarly unlikely to determine the precise spectrum of possible superselection sectors. The basic reason is that writing $\alpha$-states in the occupation number basis leads to large weights for terms involving very large numbers of baby universes. But for exponentially large occupation numbers, the `interactions' of baby universes (i.e., the topology changing processes which split and join universes) become important at leading order because the exponential suppression of any particular interaction is compensated by the number of possible such interactions. In this regime, there is no guarantee that $\hbu$ has any useful semiclassical description, since it is no longer even approximately a Fock space of single universes, and we do not obtain a good approximation by truncating the path integral to any finite number of topologies.  In particular, if we try to sum over the large number of semiclassical terms involved, small corrections to the semiclassical approximation in each term may accumulate to yield large corrections to the final answer.

This issue is exemplified by toy models of black holes which are so simple that we may perform the path integral exactly, namely Jackiw-Teitelboim (JT) gravity \cite{Saad:2019lba,Penington:2019kki} and the even simpler topological model introduced in \cite{Marolf:2020xie}. In the exact solution of these models, the superselection sectors have features expected of unitary quantum systems, but which are remarkable when appearing from a gravitational path integral: they have a discrete spectrum of black hole microstates,\footnote{While there are no propagating degrees of freedom in these theories, we may nonetheless model the black hole interior by `end-of-the-world branes' with a large number of internal states, perhaps much greater than $e^{S_\mathrm{BH}}$.} bounded in number by the Bekenstein-Hawking entropy. But this is not manifest from the semiclassical approximation, where we expand in a small parameter of order $e^{-S_\mathrm{BH}}$ which suppresses more complicated spacetime topologies. If we truncate that expansion at any finite order, we see no restriction to superselection sectors with the above features. In fact, the precise spectrum of $\alpha$-states turns out to be sensitive to \emph{doubly} nonperturbative effects.  The effects are not merely of order $e^{\# S_\mathrm{BH}}$ as for subleading saddle-point geometries, but are of order $e^{\# e^{S_\mathrm{BH}}}$.  This strong suppression is associated with their arising from an infinite sum of exponentially-suppressed geometric saddles.\footnote{Moreover, these effects may not be determined uniquely from the semiclassical expansion since (as is the case in JT gravity) the sum over topologies describes only an asymptotic expansion that does not converge. For JT, there is an extremely natural completion of the sum over topologies defined by Hamiltonians selected from an ensemble of random matrices, since the topological expansion precisely fits the rigid structure required by such a completion. For more realistic models it is unlikely that we will be so lucky as to identify an obvious completion.} From these considerations it would appear that $\alpha$-states involve a regime where quantum fluctuations of spacetime topology are untamed. See section 5 of \cite{Marolf:2020xie} for a more detailed discussion.

It would thus appear that we can say little about individual superselection sectors using only semiclassical physics, and that we can only access averaged or other simple statistical properties.\footnote{By focusing on clever averaged quantities, semiclassical calculations can nevertheless give more indirect hints at the structure of $\alpha$ states. For example, \cite{Saad:2018bqo,Saad:2019lba} show that a single topology produces the `ramp' in the spectral form factor that is characteristic of long-range eigenvalue repulsion and hence indicative of a discrete spectrum with statistics resembling that of a random matrix. However, the feature of the spectral form factor which more directly signifies a discrete spectrum (the `plateau') appears to require summation of all topologies or going beyond a geometric description.} Nonetheless, we can make much stronger statements by taking an axiomatic approach and making use of consistency conditions. Specifically, let us assume that the Hilbert spaces of intermediate states considered in section \ref{sec:BUetc} are well-defined, and that they each have a positive semidefinite inner product. For example, while the replica wormholes discussed above showed that the entropy of Hawking radiation is consistent with BH unitarity on average, general consistency arguments show something much stronger, requiring consistency with BH unitarity for \emph{every} superselection sector. More precisely, section 4 of \cite{Marolf:2020xie} showed that the number of linearly independent pure states below a given energy (say, prepared by forming and partially evaporating a black hole and projecting the Hawking radiation onto various possible states) is bounded in every superselection sector by the thermodynamic entropy (defined by the inverse Laplace-transform of a Gibbons-Hawking type path integral \cite{Gibbons:1976ue} with periodic Euclidean boundary conditions).
Since old black holes have large interiors and thus naively give rise to many more internal states than allowed by the Bekenstein-Hawking entropy (see e.g.\ \cite{Rovelli:2017mzl}), such a bound requires surprising linear relations between such states (equivalently, some linear combinations of states must be unexpectedly `null', with vanishing inner product with every other state, and so must be set equal to the zero state). This was seen very explicitly in the toy model of \cite{Marolf:2020xie} and generalisations \cite{Balasubramanian:2020jhl}. These relations rely on the same doubly-nonperturbative physics as discussed above in relation to $\alpha$-states. In \cite{Marolf:2020xie}, following \cite{Jafferis:2017tiu}, we interpreted such relations as novel nonperturbative manifestations of diffeomorphism invariance.

As a result of these considerations, we have no reason to expect semiclassical physics to be a good approximation in the interiors of old black holes for an individual superselection sector. While this has of course been suggested before  the situation is now much improved because the semiclassical approxiation itself suggests principled reasons to doubt its validity.   The approximation predicts its own break-down as it should.

However, the attentive reader will still want to be assured that we have not thrown out the baby with the bathwater. If semiclassical physics is inadequate to describe old black holes in a given superselection sector, what ensures that we may still trust it in weakly gravitating regimes? In the language of \cite{Mathur:2009hf}, what is the `niceness condition' which ensures that we may neglect topology changing processes involving replica wormholes or interactions with large baby universes in contexts where BH unitarity was not in danger? The key observation in this regard is that replica wormholes become important only when the matter entropy is so large that the sum over internal states can compensate for the usual exponential suppression of topology change. We therefore need to consider these effects only when we have a region with entropy exceeding the area of its perimeter in Planck units; i.e., when $S\gtrsim \frac{A}{4G}$.

\subsection{Further open questions}
\label{sec:discussions}

We now close with some open questions and further comments.

\subsubsection{AdS/CFT and the factorisation problem}

A potential concern with the above conclusions is the strong tension with the traditional understanding of the AdS/CFT correspondence. The point is that this correspondence provides us with examples of theories of quantum gravity with a nonperturbative, UV complete description in terms of a dual conformal field theory, but in which there is no sign of the superselection sectors that we inferred from the existence of replica wormholes.

To be specific,
in the asymptotically AdS context, our considerations point to the idea that semiclassical gravity should be dual not to a single unitary CFT, but should instead be dual to an ensemble of such theories, with a different theory for each superselection sector. While examples of such dualities have been recently discovered for simple two-dimensional models of gravity \cite{Saad:2019lba,Stanford:2019vob}, the more well-established examples of gauge/gravity duality (such as the paradigmatic duality between $\mathcal{N}=4$ super Yang Mills and type IIb string theory in AdS$_5\times S^5$) involve a unique dual theory.

This tension is not entirely new; rather, it brings to the fore an old puzzle, touched upon in section \ref{sec:SW}, which has become known as the factorization problem \cite{Rey:1998yx,Maldacena:2004rf,ArkaniHamed:2007js}. The AdS/CFT correspondence equates gravitational amplitudes with fixed asymptotically AdS boundary conditions to the partition function of a CFT, with background geometry determined by the conformal boundary of the gravitational `bulk' spacetime. If that boundary is disconnected, locality of the CFT immediately implies that the result should factorize as the product of partition functions on each connected component. But this result is surprising from the gravitational point of view:
contributions from bulk spacetimes that connect different boundary components appear to spoil the above factorization property, but it seems arbitrary to exclude such spacetimes from the gravitational path integral.  From the point of view of the baby universe Hilbert space discussed in section \ref{sec:BUetc}, factorization requires that $\hbu$ is one-dimensional, so that all states of baby universes are somehow equivalent \cite{Marolf:2020xie,McNamara:2020uza}.

There has not been any entirely satisfactory resolution to this puzzle.  It thus remains to be seen whether e.g.\ type IIb string theory in AdS$_5\times S^5$  has a one-dimensional $\hbu$ (perhaps due to the proper inclusion of various stringy objects and features that go beyond semi-classical supergravity), or whether this bulk theory is in fact dual to an ensemble of field theories with only one member of the ensemble being given by $\mathcal{N}=4$ super Yang Mills using the standard bulk-to-boundary dictionary.\footnote{As described in \cite{Marolf:2020xie}, there is a possibility that $\mathcal{N}=4$ super Yang Mills is the unique dual, but that different bulk superselection sectors map to this dual using distinct dictionaries.  In effect, the different dictionaries would then be related by (perhaps non-local) bulk field redefinitions. One might also think of this as the $\alpha$-sectors defining different quantum error correcting codes in the sense of \cite{Almheiri:2014lwa}.}

In the light of replica wormholes, the factorisation problem is directly related to the black hole information problem, since the entropy computations involved wormholes connecting multiple boundaries. We do not immediately require factorisation for the entropies, since the boundary conditions for separate boundary components are correlated in a way which explicitly spoils factorisation. But if we decompose the R\'enyi entropies into quantities which do require factorization, it appears that the wormholes remain and spoil factorization \cite{Stanford:2020wkf}. Somewhat less concretely, as pointed out in section \ref{sec:SW} a mixed state of Hawking radiation represents a failure of factorization: components of the density matrix are computed by a product of two `S-matrix' boundary conditions, and the state is pure exactly when the amplitude similarly factorizes.

Now, it may well be that physics similar to replica wormholes appears naturally for theories with a single unitary dual after some appropriate coarse-graining which explicitly spoils factorization: see e.g.\ \cite{Pollack:2020gfa,Liu:2020jsv,VanRaamsdonk:2020tlr}. But the more pertinent question for us is whether replica wormholes are relevant in a situation where we have performed no such explicit coarse-graining. Paraphrasing \cite{Stanford:2020wkf}, in a situation like the standard AdS/CFT setting having factorization and without superselection sectors, can we nonetheless understand replica wormholes as the first term in a systematically improvable expansion?

\subsubsection{Description of superselection sectors}
\label{sec:ebranes}

In section \ref{sec:disc2}, we were rather pessimistic about describing individual superselection sectors directly in terms of standard semiclassical gravitational physics. Nonetheless, there is still scope for a relatively simple description using a different language. One such idea which has appeared recently in toy models is that of `spacetime D-branes' or  `eigenbranes' \cite{Blommaert:2019wfy,Blommaert:2020seb,Saad:2019lba,Marolf:2020xie}. These are dynamical boundaries for spacetime (analogous to D-branes providing boundaries on which the string worldsheet can end) which have the effect of (perhaps partially) fixing an $\alpha$-state. While these appear to be new objects in the theory, they can also be thought of as an emergent, collective description of a coherent state of baby universes (much like regarding D-branes as a coherent state of closed strings, as opposed to new fundamental objects). Does something similar apply going beyond these toy models, to theories which are rich enough to include evaporating black holes?

In the context of evaporating black holes, the idea of providing boundary conditions for spacetime in the black hole interior to produce a pure state of Hawking radiation is not new: this is essentially the final state proposal \cite{Horowitz:2003he}. Perhaps these ideas can be revisited as an effective description of baby universe $\alpha$-states. Certainly, it remains an outstanding open problem to find a more complete, and perhaps more physical, description of the transfer of information from a black hole to the outgoing Hawking radiation in each superselection sector.

\subsubsection{Contributions from UV physics}

We have been careful to make use only of low-energy physics which is well-established and tested, and in regimes where there is no reason to expect that it fails to be trustworthy. However, we cannot rule out the possibility that the quantities we have studied are sensitive to more exotic physics from the UV completion of the theory. Indeed, this may be required to solve the factorization problem in the AdS/CFT context.

One such set of ideas is the fuzzball proposal (reviewed in \cite{Mathur:2005zp,Bena:2013dka}), which we highlight due to some conceptual similarity with physics of an individual superselection sector discussed above. Specifically, one piece of the fuzzball proposal is that gravitational collapse does not lead to formation of a horizon, but instead there is a tunnelling event to a horizonless configuration. The amplitude to tunnel to any given configuration is small, but this is compensated for by the large number of possible states. We can compare this to the situation for superselection sectors described in \ref{sec:disc2}, where interactions with baby universes were similarly suppressed individually, but compensated for by a large population of baby universes. One might speculate that the fuzzballs replace the baby universes, effectively selecting a distinguished $\alpha$-state. But since this selection depends on fine details of the UV completion, with extra dimensions, strings, branes and so forth, the low-energy gravity is ignorant of the details: it does the best job it can in the face of its ignorance, which is to average over the possibilities. In the hope of making such a connection, we conclude with one comment: while the fuzzball literature suggests that the tunnelling event happens before the horizon forms, from our considerations we see that this is in fact unnecessary to solve the information problem. It suffices if this physics kicks in only after the Page time, when the parametrically large interior can play a role, and when large corrections to the state of Hawking radiation are required.

\subsubsection{Spacetimes with singular causal structure}
\label{sec:SingCausal}

The fact that replica wormholes can provide gravitational saddles strongly suggests that spacetimes with singular causal structures play an important role in the gravitational path integral.  As noted in section \ref{sec:repQES}, the past light cone of any splitting surface $\gamma$ has multiple disconnected parts.  In particular, it has one such part for each of the bra-spacetimes that join at $\gamma$ (and similarly one such part for each of the ket-spacetimes).

This idea that such causal singularities should be included is not new (see e.g.\ \cite{Louko:1995jw,Dowker:1997hj,Borde:1999md}), though its implications remain to be fully explored.  One would like to understand just how general such causal singularities can be, and in particular what singularities arise in saddle-point geometries.  For example, can one find saddles where splitting surfaces for replica wormholes lie outside horizons (and thus in the past of $\scri^+$)?  If so, how are we to understand their effects on measurements performed by asymptotic observers?  Similarly, are there saddles with multiple splitting surfaces that are causally related to each other? See \cite{Chen:2020tes} for an example of timelike separated islands in a cosmological context.  It may be possible to probe the physics of such settings using time-folds, as may be familiar from the study of out-of-time-order correlation functions. That is, instead of each replica being constructed from one branch of forward evolution (`ket') and one of backward evolution (`bra'), we add further forward and backward branches, with the possibility of nontrivial replica-wormhole-like identifications. Such time-folds might be used to connect $\scri^+$ with the past of a splitting surface (where the physics is understood).

Conversely, our work above took as a fundamental assumption that the low-energy gravitational path integral sums over topologies.  While this is a common discussion in treatments of gravitational path integrals, and despite its utility in describing the Hawking-Page transition in AdS space \cite{Hawking:1982dh} and defining the Hartle-Hawking no-boundary wavefunction \cite{Hartle:1983ai}, some readers will ask if there might be formulations of quantum gravity in which it fails to hold.  This important issue also deserves further attention in the future.

\subsubsection{Non-perturbative physics of of Baby Universes}

There also remain certain questions about how non-perturbative corrections will affect our discussion of baby universes.  For example, as described in section \ref{sec:Whatbaby}, the Polchinski-Strominger assumption led to a certain notion of PS baby universe, while our analysis of replica wormholes led to a different notion of RW baby universe.  In particular, the latter can roughly be thought of as a bound state of a PS-baby and a PS-anti-baby universe.  The difference between the two was in part due to the fact that the PS assumption allowed us to discuss the path integral associated with forming a black hole and the performing a complete projective measurement at $\scri^+$.  But did the PS-assumption lead to the correct conclusion?  We presume the full non-perturbative theory to allow such boundary conditions, but what are the results?  Do the resulting baby universes resemble the PS-babies, or does each PS-baby necessarily come attached to an anti-baby so that the result is more like the RW baby universes?  Or is this question fundamentally ill-defined  due to the presence of null states as described in section \ref{sec:disc2}?  And on a similar note, does the non-perturbative theory have a meaningful distinction between universes and anti-universes?

\subsubsection{More details of unitarity}

Our work above focussed on the Page curve.  This is a prominent signature of BH unitarity, but it is not it itself enough to guarantee unitarity for asymptotic observers. Does semiclassical gravity make predictions that are in line with unitarity in other ways, and in more detail?

As an illustration that challenges may lie ahead, we give an example in the context of the Polchinski-Strominger proposal in section \ref{sec:PS}. In section \ref{sec:PSP}, we found this proposal to give predictions consistent with a pure state on $\scri^-$ evolving to a pure state on $\scri^+$ (for example, as probed by the swap test). But unitary evolution also requires that the inner product is conserved, so two orthogonal states on $\scri^-$ should evolve to orthogonal states on $\scri^+$. We can check this using a swap test, except that we now prepare two black holes with orthogonal states at $\scri^-$, perhaps by throwing a particle with two possible internal states into the black hole. Unitarity demands that the expectation value of the swap operator acting on $\scri^+$ for these two black holes is zero. But this is not the case for the PS proposal: the expectation value is exponentially small, but nonetheless positive.\footnote{If the particle's internal state transmits perfectly into the black hole interior in the semiclassical approximation, so that both initial states give rise to the same density matrix $\rho_\mathrm{Hawking}$ of Hawking radiation at $\scri^+$, then the swap expectation value is $\Tr \rho_\mathrm{Hawking}^2$.} Thus the Polchinski-Strominger proposal does not result in a unitary S-matrix.

If we remain within the semiclassical regime, considering only experiments on the radiation before the black hole becomes too small, then we do not have such a sharp contradiction with unitarity. Nonetheless, it provides a warning that more must be checked, and motivates a careful study of the situation when we consider several different initial states.

\subsubsection{Moving away from asymptotics}

We studied black hole formation and collapse in an idealised setting, using states that were prepared and measured at asymptotic boundaries, and using experiments with multiple black holes placed in separate spacetimes. This allowed us to make very clean statements (like commutativity of operators acting on the baby universe Hilbert space), but it can only be an approximation to more realistic settings. Any actual experiment will involve experimenters subject to gravitational physics, even if only weakly.  While it is natural to assume that such real-world experiments would be well-modeled by the idealized ones described above (or involving an auxiliary system coupled to AdS, or involving sharp boundary conditions imposed on finite `cutoff' surfaces as in implicit in e.g.\\cite{Anegawa:2020ezn,Hashimoto:2020cas,Gautason:2020tmk,Krishnan:2020oun}), this remains to be shown in detail. In particular, our concept of cluster decomposition, in the sense that experiments on multiple black holes will approach our `separate universe' idealisation as the separation between them is taken to infinity, is as yet only an expectation.

It is clearly of interest to explore this further, not least in the context of cosmology.  Indeed, in analogy with Everett's treatment of the quantum mechanical `measurement problem' \cite{Everett:1956dja}, the most interesting question would appear to be what form of conceptual framework (if any) would allow a sharp discussion of experiments whose final records -- and not just the intermediate steps -- are subject to quantum gravity effects.

\subsubsection{The experience of an infalling observer}

Our main focus in this paper has been to compute observables defined far from the black hole, in asymptotic regions. We have not directly commented upon the more difficult question of predictions for the observers who enter the black hole. This is more challenging, since it is far from obvious how to give a gauge invariant description of such observers, who are inevitably part of the quantum system of the black hole (a situation familiar from quantum cosmology). We will not say anything definitive on this question, but we make a few comments below.

If the baby universe state is simple (as for the Hartle-Hawking state), our path integrals describing any one black hole are dominated by the usual semiclassical black hole spacetime, with a smooth interior until the singularity. This gives us no obvious reason to doubt the conventional description that an infalling observer will experience no drama at the horizon. The firewall paradox \cite{Almheiri:2012rt,Almheiri:2013hfa} is evaded because, in a technical sense, information is lost: the late radiation is not required to be entangled with the early radiation.

However, the situation is less clear for multiple identically prepared black holes or more complicated baby universe states.  In particular, in the AdS context one could make use of an
auxiliary bath system as in \cite{Almheiri:2019psf} to effectively `measure' the $\alpha$-parameters, thus decohering the different superselection sectors of the gravitating spacetime.  Since infalling observers have no access to the bath, one might expect their experiences to be described by individual superselection sectors.  The firewall problem then arises with full force.  In addition, we must deal with the vast number of null states required by the discussion in section \ref{sec:disc2}. What it means to discuss physics in this context, and how it relates to previous proposed resolutions  remains a fascinating topic for both discussion and further investigation.

\acknowledgments

We thank Steve Giddings and Douglas Stanford for conversations motivating much of this work. We also thank Mukund Rangamani for comments on a draft. We are grateful for support from NSF grant PHY1801805 and funds from the University of California. H.M.~was also supported in part by a DeBenedictis Postdoctoral Fellowship, and D.M.~thanks UCSB's KITP for their hospitality during portions of this work.  As a result, this research was also supported in part by the National Science Foundation under Grant No. NSF PHY-1748958 to the KITP.

\appendix

\section{Further review of the Hawking effect in a fixed spacetime}
\label{sec:MoreRev}

This appendix completes our brief review of the derivation of the Hawking effect in a fixed curved spacetime that was begun in section \ref{sec:Heisenberg}.  The argument below follows \cite{Hawking:1974sw} and \cite{Jacobson:2003vx}.  Recall that we consider a free massless quantum field on the classical spherically symmetric uncharged collapsing black hole spacetime of figure \ref{fig:collapse} (left).  On $\scri^-$, the state $|\psi\rangle$ of the quantum field coincides with the Minkowski vacuum on $\scri^-$.

We wish to characterize the state of the field on $\scri^+$ using the number operators $N(\omega; \scri^+) = a^\dagger(\omega,\scri^+)a(\omega, \scri^+)$ associated with modes of definite positive frequency $\omega$ with respect to some affine parameter $u$ along $\scri^+$. We follow  \cite{Hawking:1974sw} in working in the Heisenberg picture, so we need to evolve the operators $a^\dagger(\omega,\scri^+)$, $a(\omega, \scri^+)$ backwards in time to express them in terms of the corresponding operators $a^\dagger(\omega,\scri^-)$, $a(\omega, \scri^-)$ on $\scri^-$.  This will give a Bogoliubov transformation that will allow us to compute the distribution of occupation numbers at $\scri^+$.

Now, linearity of the quantum fields also means that the desired backwards evolution of the operators $a^\dagger(\omega,\scri^+)$, $a(\omega, \scri^+)$ be can be found by studying the behavior of the corresponding field modes.  And the latter behavior is obtained by solving the classical wave equation.  For modes $L$ localized at late retarded times (large affine parameter $u$ along $\scri^+$), the desired backwards propagation can then be broken into two phases; see the right panel of figure \ref{fig:collapse}.

In the first phase (closest to $\scri^+$), the localization at large $u$ means that the spacetime is very close to that of a stationary black hole as any transient effects associated with the collapse will have either dispersed to distant parts of the asymptotic region (where its gravitational effect is minimal) or will have fallen into the nascent black hole.  In the approximation that the region is exactly stationary, the evolution of a mode of definite frequency $\omega$ amounts to solving a Schr\"odinger-type scattering problem, resulting in a reflected mode $R$ and a transmitted mode $T$; see again figure \ref{fig:collapse} (right).

The reflected mode $R$ reaches $\scri^-$ at late advanced times (i.e., large affine parameter $v$ along $\scri^-$) without leaving the Phase $I$ region where the spacetime remains nearly stationary.  As a result, $R$ has the same positive frequency $\omega$ along $\scri^-$ as does $L$ along $\scri^+$.  This means that $R$ contributes only to what are usually called the $\alpha$ Bogoliubov parameters (which map annihilation operators to annihilation operators) as opposed to the more interesting $\beta$ Bogoliubov parameters associated with mixing between creation and annihilation operators.

On the other hand, the transmitted mode $T$ travels through the region where the spacetime is dynamical.   However, since $L$ is localized at large $u$, the transmitted mode $T$ has high frequency with respect to natural freely-falling observers.  As a result,
the WKB approximation may be used to justify the use of geometric optics in propagating $T$ back to $\scri^-$ and completing the calculation.  This is the 2nd phase of the backwards evolution that was foreshadowed above.   But rather than complete the full calculation, the end result can be seen \cite{Jacobson:2003vx} by noting that in the overlap of the regions corresponding to phases 1 and 2, the $T$ mode is localized in a region close to the horizon that is well approximated by Rindler space.  Furthermore, since the corresponding Rindler time-translation coincides with the (approximate) time translation symmetry outside the black hole, in this region $T$ is purely positive-frequency with respect to Rindler time.  But any smooth state will locally approximate the Minkowski vacuum in this Rindler region.  Thus the occupation numbers of $T$ modes are thermally distributed.

The Hawking effect is thus associated primarily with the transmitted mode $T$.  The fact that it corresponds only to the part of the original mode that was transmitted through the potential barrier into the region near the horizon during the phase 1 evolution introduces the famous ``grey-body'' factors into the Hawking effect.  Here the name comes from the fact that when evolving modes toward the future the corresponding transmitted part would fall into the black hole and be absorbed, and also to the fact that absorption and emission coefficients must agree in thermal equilibrium.  Thus the (squared) fraction of the original mode that remains present in $T$ is naturally interpreted as the coefficient for emission of the original mode by a radiating black hole.

\section{Intermediate states of baby universes}
\label{sec:RWBUE}

In section \ref{sec:DropPS}, we discussed a Hilbert space interpretation of replica wormhole calculations, in particular allowing only measurements on a region $\scri_u$ before the black hole becomes too small. However, this description was not particularly natural from the point of view of consecutive measurements on different sets of Hawking radiation, so in this appendix we give an alternative Hilbert space interpretation.

Consider in particular a real-time process in which $\hbu$ begins in some initial state (perhaps $|\HH\rangle$), and an asymptotic observer creates a black hole before making a complete projective measurement for the Hawking radiation on $\scri_u$. It is then natural to ask for the state of $\hbu$ required for predicting subsequent similar measurements. Since we are leaving the radiation which emerges after time $u$ unobserved, we have necessarily lost some information.  The state of baby universes will thus become mixed due to entanglement with the unobserved part of the asymptotic state.  As a result, this process is best described as a map from density matrices to density matrices (a quantum channel) on $\hbu$.

More generally, we can write down the map from an initial density matrix on $\hbu$ to a final density matrix on $\hbu\otimes \hilb_u$ that describes the state of both the baby universes and the state of the Hawking radiation to which we have access. We can write this map explicitly as
\begin{equation}
\label{eq:QC}
	\rho_\mathrm{BU} \mapsto  \mathcal{N} \sum_{i,j} \Tr_u(\hat{\Psi}_i(u) \, \rho_\mathrm{BU} \hat{\Psi}_j^\dag(u)) \otimes |i\rangle\langle j|,
\end{equation}
where we have introduced the operation $\Tr_u$ (to be defined below), which enacts the partial trace over the unobserved radiation. The first term in the tensor product is an operator on $\hbu$, the second factor $|i\rangle\langle j|$ on the radiation Hilbert space on $\scri_u$, and $\mathcal{N}$ is chosen to normalise the trace to unity.

To explain \eqref{eq:QC}, recall that a density matrix $\rho_\mathrm{BU}$ is an operator on $\hilb_0 = \hbu$. The operator $\hat{\Psi}_i(u)$ maps $\hbu \rightarrow \hilb_1$, and so
$\hat{\Psi}_j(u)^\dagger$ maps $\hilb_1 \rightarrow \hbu$.  Thus $\hat{\Psi}_i(u) \, \rho_\mathrm{BU} \hat{\Psi}_j^\dag(u)$ is a map from the one-boundary Hilbert space $\hilb_1$ to itself. Its matrix elements are computed by a path integral bounded by a pair of Cauchy surfaces $\Sigma_u$ meeting $\scri^+$, one on the `ket' branch and one on the `bra'.  The operation $\Tr_u$ identifies these two branches along $\Sigma_u$ asymptotically, producing an operator on $\hbu$. It is not meaningful to specify a priori how far this identification persists into the interior.  The gravitational path integral sums over all such choices; replica wormholes will lead to semiclassical contributions where the identifications persist to the edge of the associated island.

As is the case for all our discussions, the formula \eqref{eq:QC} simplifies if $\rho_\mathrm{BU}$ is built from asymptotic boundary conditions (so that is part of the superselected algebra of asymptotic observables). In that case one finds $\Tr_u(\hat{\Psi}_i(u) \, \rho_\mathrm{BU} \hat{\Psi}_j^\dag(u)) = \hat{\Psi}_j^\dag(u)\hat{\Psi}_i(u) \, \rho_\mathrm{BU} = \hat{\rho}_{ij}(u) \rho_\mathrm{BU}$. In particular, if the baby universes are in an $\alpha$-state (so that $\rho_\mathrm{BU} = |\alpha\rangle\langle\alpha|$), the map leaves the state of baby universes unchanged, while producing the state $\rho^\alpha(u)$ on $\scri_u$ whose components are given by the $\alpha$-eigenvalues of $\hat{\rho}_{ij}(u)$.

We can also use this same simplification more generally if
we only wish to use \eqref{eq:QC} to compute expectation values of asymptotic observables (thought of as operators on $\hbu$).
By essentially the same argument as above, tracing such observables against \eqref{eq:QC} gives the same result as tracing the observables against
$\hat{\rho}_{ij}(u) \rho_\mathrm{BU}$. The point is simply that we can commute $\hat{\Psi}_j^\dag(u)$ past these obervables and then use the cyclic property of the trace in order to use $\hat{\Psi}_j^\dag(u)\hat{\Psi}_i(u) = \hat{\rho}_{ij}(u)$. However, it should be borne in mind that other states and operators exist and may be of interest, for example to describe the experience of an observer falling into a black hole.

\bibliographystyle{JHEP}
\bibliography{biblio}

\end{document}